\documentclass[prx ,amssymb, aps , twocolumn ]{revtex4-2}

\usepackage{graphicx}
\usepackage{amsmath}
\usepackage{amssymb}
\usepackage{bm}
\usepackage[latin1]{inputenc}

\usepackage[caption = false, labelfont= bf]{subfig}

\usepackage{tabularx}
\usepackage{xcolor}
\usepackage{xspace}
\usepackage{ulem}

\usepackage[squaren,Gray]{SIunits}

\definecolor{linkcolor}{rgb}{0,0,0.6}

\usepackage[pdftex,colorlinks=true,
pdfstartview=FitV,
linkcolor= linkcolor,
citecolor= linkcolor,
urlcolor= linkcolor,
hyperindex=true,
hyperfigures=false]
{hyperref}

\usepackage[final]{changes}

\definechangesauthor[name = validated, color=orange]{OK}

\usepackage{nicefrac}
\usepackage[usestackEOL]{stackengine}

\newcommand {\Hc} {$H_{c2}$\xspace}
\newcommand {\HcT} {$H_{c2}(T)$\xspace}
\newcommand {\Hco} {$H_{c2}(0)$\xspace}
\newcommand {\Hcl} {$H_{c1}$\xspace}

\newcommand {\aaxis}{$a$ axis\xspace}
\newcommand {\baxis}{$b$ axis\xspace}
\newcommand {\caxis}{$c$ axis\xspace}
\newcommand {\UTe}{UTe$_2$\xspace}
\newcommand {\Hm} {$H_{m}$\xspace}
\newcommand {\ucoge} {UCoGe\xspace}

\newcommand {\hftransi} {HF\xspace}
\newcommand {\lftransi} {LF\xspace}
\newcommand {\deltaCT} {\frac{\Delta C}{T}}

\newcommand {\dHc} {$\nicefrac{d H_{c2}}{d T_c}$\xspace}
\newcommand {\dHcT} {$\nicefrac{d H_{c2}}{d T}$\xspace}
\newcommand {\dHcl} {$\nicefrac{d H_{c1}}{d T_c}$\xspace}
\newcommand {\Tc} {$T_{c}$\xspace}
\newcommand {\Cp} {$C$\xspace}
\newcommand {\CT} {$C/T$\xspace}
\newcommand {\diese} {$\#$}
\newcommand {\vF}{$v_{F}$\xspace}

\newcommand{\Hparab}{$H \parallel b$\xspace}
\newcommand{\Hparac}{$H \parallel c$\xspace}
\newcommand{\Hparaa}{$H \parallel a$\xspace}

\newcommand {\llambda}{$\lambda$\xspace}
\newcommand {\llambdaH}{$\lambda(H)$\xspace}
\newcommand {\llambdaHHm}{$\lambda(H/H_m)$\xspace}

\newcommand {\swave}{$s$-wave\xspace}
\newcommand {\pwave}{$p$-wave\xspace}
\newcommand {\dwave}{$d$-wave\xspace}
\newcommand {\fwave}{$f$-wave\xspace}
\newcommand {\dvector}{$\bm{d}$-vector\xspace}

\begin{document}
	\title{Field-induced tuning of the pairing state in a superconductor}
	\author{A. Rosuel$^1$, C. Marcenat$^1$, G. Knebel$^1$, T. Klein$^2$, A. Pourret$^1$, N. Marquardt$^1$, Q. Niu$^{1,3}$, S. Rousseau$^1$, A. Demuer$^4$, G. Seyfarth$^4$, G. Lapertot$^1$, D. Aoki$^5$, D. Braithwaite$^1$, J. Flouquet$^1$, J.P. Brison$^1$ }
		
	\email[Corresponding author: ]{jean-pascal.brison@cea.fr}
	\affiliation{$^1$Univ. Grenoble Alpes, CEA, Grenoble-INP, IRIG, Pheliqs, 38000 Grenoble France}
	\affiliation{$^2$Univ. Grenoble Alpes, CNRS, Institut N\'{e}el, 38000 Grenoble France}
	\affiliation{$^3$Anhui Key Laboratory of Condensed Matter Physics at Extreme Conditions, High Magnetic
Field Laboratory, HFIPS, Anhui, Chinese Academy of Sciences, Hefei 230031, P. R. China}
	\affiliation{$^4$Univ. Grenoble Alpes, INSA Toulouse, Univ. Toulouse Paul Sabatier, EMFL, CNRS, LNCMI, Grenoble 38042, France}
    \affiliation{$^5$IMR, Tohoku University, Oarai, Ibaraki, 311-1313, Japan}
	
	\begin{abstract}
		The recently discovered superconductor \UTe, with a superconducting transition temperature \Tc between 1.5~K and 2~K, is attracting much attention due to strong suspicion of spin-triplet and topological superconductivity.
		Its properties under magnetic field are also remarkable, with field-reinforced (\Hparab) and field-induced (H in the $(b,c)$ plane) superconducting phases. 
		Here, we report the first complete thermodynamic determination of the phase diagram for fields applied along the three crystallographic directions. 
		For field along the easy \aaxis, we uncover a strong negative curvature of the upper critical field very close to \Tc, revealing a strong suppression of the pairing strength at low magnetic fields. 
		By contrast, measurements performed up to 36~T along the hard magnetisation \baxis confirm a bulk field-reinforced superconducting phase. 	
		Most of all, they also  reveal the existence of a phase transition line within the superconducting phase. 
		Drastic differences occur between the low-field and high-field  phases pointing to different pairing mechanisms.
		Detailed analysis suggests a possible transition between a low-field spin-triplet to high-field spin-singlet state, a unique case among superconductors. 	
	\end{abstract}
	
	\maketitle
	
	\date{}
	
\section{Introduction}
    The major breakthrough of the last 40 years in the field of superconductivity has been the discovery of several families of unconventional superconductors: heavy-fermions, organics, high-$T_c$ cuprates, iron pnictides in chronological order. 
    All are controlled by pairing mechanisms dominated by purely electronic interactions instead of the conventional electron-phonon interaction.
    These new pairing mechanisms also lead to new superconducting states, with different possible spin-states (spin-singlet/even parity or spin-triplet/odd parity), and corresponding orbital states, dubbed \dwave, \pwave, \fwave. 
    Spin-triplet superconductivity is presently highly sought-after in the context of quantum engineering, as the required starting point to build robust topologically protected qubits. 
    Artificial heterostructures have been proposed to create this rare state from conventional s-wave superconductors \cite{FuPRL2008}. 
    However, experimental success following this line remains controversial. 
    Hence, bulk spin-triplet superconductors with strong spin-orbit coupling is an appealing alternative.

    Besides superfluid $^3$He \cite{Leggett1975}, a perfect analogue of a neutral spin-triplet \pwave superconductor below 3~mK, most candidates for bulk spin-triplet superconductivity are found among uranium-based heavy-fermion systems. The most famous is undoubtedly UPt$_3$, which is one of the rare systems showing not only one superconducting phase, but three superconducting phases differing by their symmetries \cite{JoyntRMP2002}. 
    Another major series of systems is that of UGe$_2$, URhGe and UCoGe, where the bulk coexistence of ferromagnetism and superconductivity leaves little doubt that they are also  spin-triplet superconductors. 
    They all display in addition uncommon superconducting phase diagrams, with a re-entrance or reinforcement of superconductivity under large magnetic fields  \cite{AokiRevJPSJ2019}.  

    Naturally, these uranium heavy-fermion superconductors are also central in modern condensed matter physics, as potential topological superconductors.
    For example, UPt$_3$ is also considered as a prime candidate for topological chiral spin-triplet superconductivity, like URu$_2$Si$_2$ is a prime candidate for topological chiral spin-singlet ($d$-wave) superconductivity \cite{GhoshJPhysCondMat2020}. 

    However, spin-triplet superconductivity remains a rare state of matter, with no clear understanding of the main reasons driving its appearance.
    In this context, the discovery of superconductivity in \UTe above 1.5~K back in 2018 \cite{RanScience2019} drove much enthusiasm in the condensed matter physics community. 
    \UTe is a stunning system for almost all of its properties, starting with the fact that it is metallic only thanks to strong electronic correlations \cite{Xu2019, IshizukaPRL2019, Shick2021}. Its superconductivity shows extraordinary robustness to very large magnetic fields, \cite{RanScience2019, Aoki2019, Knebel2019, RanNatPhys2019}, a feature 
    also associated to the ferromagnetic superconductors, and requiring at least spin-triplet superconductivity. 
    Still \UTe is a paramagnetic heavy-fermion 
    with no strong evidence for being close to a ferromagnetic instability \cite{AokiJPCM2022}. 
    Among the hardest yet most important questions is to understand what could drive such a system toward a spin-triplet superconducting ground-state? 
    Having better hints of where to look for or how to build spin-triplet superconductors would be of interest far beyond the community of quantum materials.  
    This work brings unprecedented elements directly related to this issue, supporting notably that magnetic field  could drive \UTe from a spin-triplet to a spin-singlet ground state.

    Up to now, evidence for spin-triplet pairing in \UTe, comes from NMR Knight-shift measurements, recently performed along all crystallographic axes \cite{FujibayashiJPSJ2022}, and from the strong violation of the paramagnetic limit by the superconducting upper critical field \Hc \cite{RanScience2019, Knebel2019, RanNatPhys2019, Knafo2021, RanNPJQM2021}.
    The analysis of \Hc is however not straightforward in \UTe, because the field-reinforced phase observed for magnetic fields applied along the hard \baxis strongly suggests a field-enhanced pairing strength \cite{RanScience2019, Knebel2019}. 
    Such a field-dependent pairing is difficult to model theoretically and makes the analysis of \Hc more complex, because it opens other routes for a violation of the paramagnetic limit than spin-triplet pairing \cite{AokiJPCM2022}.
    Moreover, even the simplest question of whether or not the field-reinforced state has a different symmetry than the low-field one is up to now not settled.
    Yet experimentally no thermodynamic phase transition had been observed separating low field and high field reinforced superconductivity.

    Nonetheless, high pressure experiments demonstrate unambiguously that \UTe  does belong to the very select class of unconventional superconductors displaying transitions between different superconducting phases 
    \cite{Braithwaite2019,Lin2020,Aoki2020,ThomasPRB2021}.
    Some samples even show a double transition at ambient pressure in zero field, which could explain how this low symmetry system (\UTe is orthorhombic) could also be chiral. 
    Indeed, the observation of time reversal symmetry breaking below \Tc by polar Kerr-effect measurements \cite{Hayes2021}, or the asymmetric spectrum observed with STM spectroscopy \cite{Jiao2020} suggest such a chiral state.
    Actually, there is now more and more evidence that the double transition at ambient pressure in zero field is not an intrinsic effect \cite{AokiJPCM2022, ThomasPRB2021, Rosa2022, IguchiArXiv2022}.
    
    This rich superconducting state inspired several theoretical works, proposing different scenarios for the possible symmetry states under magnetic field and pressure \cite{IshizukaPRL2019, Shishidou2021, Ishizuka2021}. 
    These works also address the 
    deeper question, of why this system is spin-triplet? 
    It is particularly acute in \UTe because contrary to initial expectations, no ferromagnetic fluctuations 
    have yet been detected. 
      Such magnetic fluctuations, present in the ferromagnetic superconductors \cite{AokiRevJPSJ2019} and necessary to explain the stability of the $A$-phase of superfluid $^3$He \cite{Leggett1975}, are natural candidates for the pairing mechanism of spin-triplet superconductors. 
    By contrast in \UTe, neutron measurements have only revealed incommensurate magnetic fluctuations \cite{Duan2020, Knafo2021, ButchNPJ2022}, and a resonance at finite $\bm{Q}$ below the superconducting transition temperature \Tc \cite{DuanNature2021,RaymondJPSJ2021}. 
    \UTe could be similar to UPt$_3$, where neutron scattering studies also mainly detect antiferromagnetic correlations \cite{Aeppli1988}.  Therefore, some  theories have attempted to explain the observed $E_{2u}$ superconducting ground state of UPt$_3$ with a pairing mechanism based on pure antiferromagnetic fluctuations \cite{NomotoPRL2016}. 

    In \UTe, theoretical models of the spin-triplet superconducting state have explored three main scenarios, starting from band structure calculations: 
    i) ferromagnetic fluctuations, winning over antiferromagnetic fluctuations for specific values of the normal state parameters (exchange constants, Coulomb repulsion...)\cite{Xu2019, Ishizuka2021}; ii) local (intra unit-cell) ferromagnetic correlations between the nearest neighbour uranium ions \cite{Shishidou2021, HazraArXiv2022}; iii) pure antiferromagnetic fluctuations, combined with multi-orbital degrees of freedom \cite{ChenArXiv2021} or peculiar conditions on the $\bm{Q}$-dependent susceptibility and the underlying Fermi-surface \cite{KreiselPRB2022}. 
    
    Our detailed thermodynamic measurements presented here demonstrate that the field reinforcement of \Hc in UTe$_2$ arises from a new superconducting phase. It is driven by a pairing mechanism different from that controlling the low field superconducting phase, imposing new constraint on the possible pairing mechanisms. 
    Conversely, we also find that in the low field phase, pairing is suppressed by magnetic fields applied along the easy magnetisation \aaxis.
    All together, we show that the superconducting symmetry may change from a spin-triplet to a spin-singlet state as a function of magnetic field on approaching the metamagnetic instability transition at \Hm=34.75~T for field along the \baxis. 
    These new features are solid experimental inputs challenging the theoretical scenarios for the superconducting pairing in \UTe.
    
    The present paper is organised as follows: 
    in the next section~\ref{sec:overview}, we give a rapid overview of the main experimental results.
    Section~\ref{sec:experiment} gives the experimental details. 
    The results are presented in section~\ref{sec:results} and analysed in section~\ref{sec:analysis}.
    Section~\ref{sec:discussion} discuss their impact and the open questions. 
    Additional data or details on the methods used for the analysis are reported in appendices.
 
\section{Overview of the main results}
\label{sec:overview}    
    This work reveals a thermodynamic phase transition between the low and high-field reinforced superconducting phases, for magnetic fields applied along the hard magnetisation \baxis. 
    This has been made possible thanks to specific heat experiments up to 36~T and thermal dilatation/magnetostriction measurements up to 30~T. 
    We also uncover a drastic alteration of the specific heat anomaly along \Hc between the two phases, a unique feature, still never observed in any unconventional superconductor, strongly suggesting a change of pairing mechanism between the two superconducting phases.
  
\begin{figure}
	\centering
	\includegraphics[width=1\linewidth]{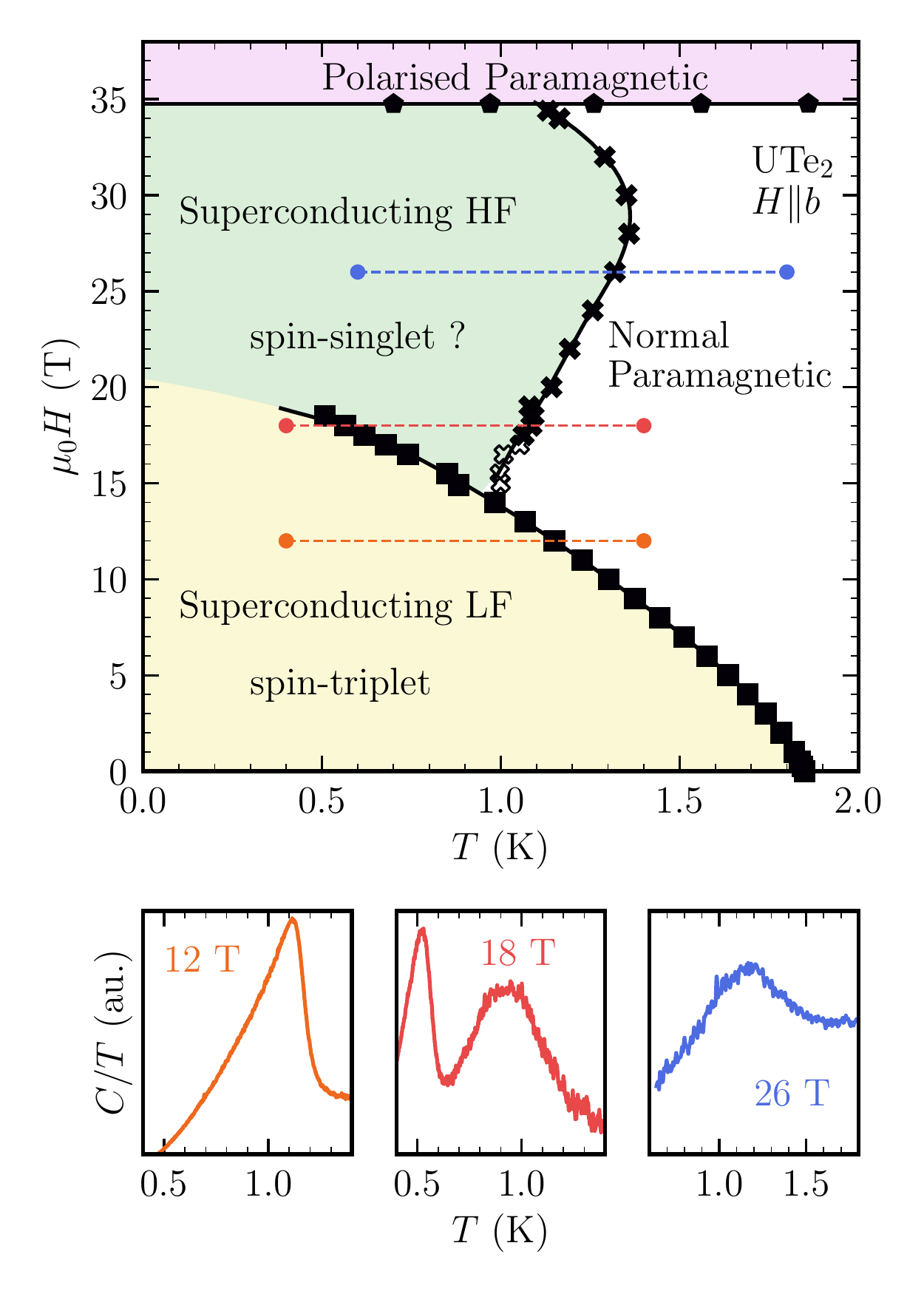}
	\caption{Top: Bulk superconducting phase diagram of \UTe for a magnetic field $H$ along the \baxis. 
	The specific heat measurements reveal a phase transition line between a low-field (\lftransi) (yellow) and high-field (\hftransi) (green) superconducting state. 
	As a function of magnetic field, superconductivity is suppressed at 34.75~T by a first-order metamagnetic transition marking the entrance in a partly polarised magnetic phase (magenta). 
	Bottom: Specific heat divided by temperature as a function of temperature at different magnetic fields: (left) At 12~T a sharp anomaly marks the superconducting transition to the \lftransi phase. (middle) On cooling a broad hump-like transition occurs at the transition to the \hftransi phase above a sharp low temperature transition to the \lftransi superconducting phase. (right) At 26~T only the hump-like transition to the \hftransi superconducting phase is observed.  
	}
	\label{fig:SimplePhaseDiagram}
\end{figure}

    Figure~\ref{fig:SimplePhaseDiagram} is a summary of this main result, showing the phase diagram for field along the \baxis, with the two superconducting phases (labelled \lftransi and \hftransi for "low-field and "high-field"), and the specific heat anomalies at the different phase transitions. 
    Phase transitions between superconducting phases of different symmetries are rare, however such a change in the specific heat anomaly between the different phases is, to the best of our knowledge, unique. 
    Analysis is required to reveal that it suggests different spin-states characterising these two phases, with a \hftransi phase in strong interplay with the metamagnetic transition occurring at \Hm. 
    The result, which is rather counter-intuitive, suggests that the \lftransi phase would be spin-triplet, and the \hftransi phase spin-singlet, most likely triggered by the development of antiferromagnetic correlations on approaching \Hm. 
    Hence, this would be a direct consequence of the competing pairing interactions (or of the change of dominant finite $\bm{Q}$-vector in the magnetic excitation spectrum) predicted to occur in \UTe \cite{Xu2019, Ishizuka2021, KreiselPRB2022}.  

\begin{figure}
	\centering
	\includegraphics[width=.9\linewidth]{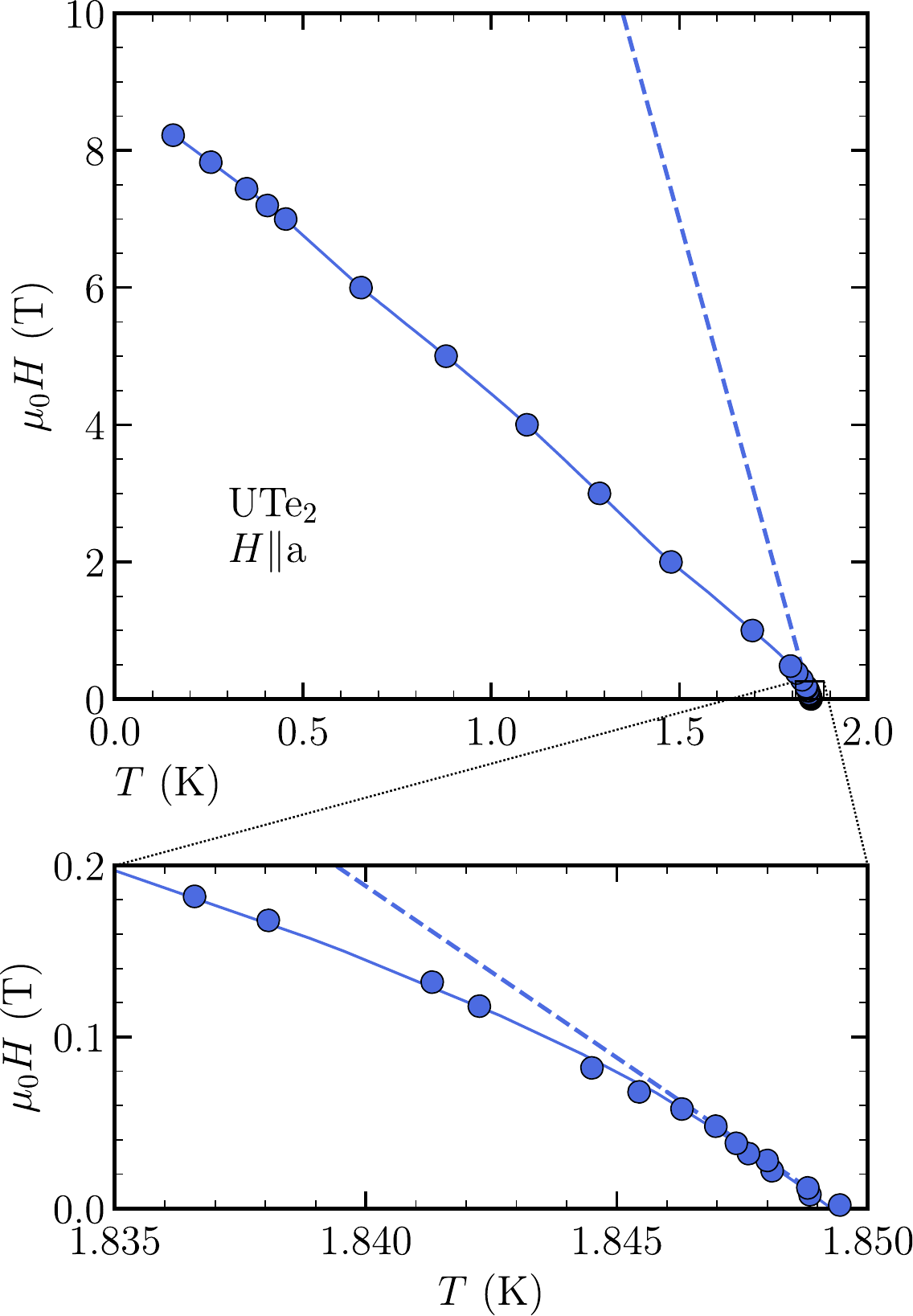}
	\caption{Top: \Hc along the \aaxis. Dashed line : orbital limit deduced from the measured initial slope at \Tc. 
	Bottom: zoom on the very low field behaviour, close to Tc, showing the very strong negative curvature responsible for the deviation of \Hc along $a$ from the orbital limitation.
    This anomalous behaviour signs a strong suppression of the pairing strength for fields applied along the easy magnetisation axis, an effect also observed in the ferromagnetic superconductors.}
	\label{fig:thePbWithHc2_along_a}
\end{figure}

    Another surprise uncovered by these specific heat measurements is that \HcT appears to be anomalous not only along the \baxis, with the field-reinforced \hftransi phase, but also along the $c$ and most importantly along the easy \aaxis. 
    This had been completely overlooked up to now, but determination of \HcT by specific heat reveals a very strong negative curvature of \Hc along the \aaxis very close to \Tc, and an initial slope four times larger than initially thought. 
    Hence, a new mechanism is required to explain the very strong deviation of \HcT from the linear behaviour expected below \Tc/2 (see Fig.~\ref{fig:thePbWithHc2_along_a}): 
    we show that a strong paramagnetic limitation, anyway excluded by recent NMR Knight-shift measurements \cite{FujibayashiJPSJ2022}, would not be sufficient to explain this singular temperature dependence of \Hc. 
    Instead, it points to a severe suppression of the pairing strength along the easy axis. 
    In the case of \UTe, it could arise from at least two different sources. 
    Suppression of (hypothetical) ferromagnetic fluctuations by fields along the easy axis would lead to such a decrease of the pairing strength, a  mechanism similar to that in ferromagnetic superconductors \cite{Hattori2012ucoge, Wu2017}, but also working for paramagnetic systems \cite{Mineev2017}.
    Or the strong sensitivity of finite-$\bm{Q}$ (spin-triplet) pairing \cite{ChenArXiv2021, KreiselPRB2022} to Fermi-surface instabilities, already revealed in \UTe at rather low magnetic fields along the \aaxis \cite{NiuPRL2020}, might also lead to pairing strength suppression.  
    
    Hence, both results along the hard \baxis and the easy \aaxis bring new elements on the possible symmetry states in \UTe and on the competing pairing mechanisms.
    It enlightens the stunning superconducting properties of \UTe, and uncovers key features which should guide future theory developments. 
    Indeed, understanding the mechanisms leading to the strong field dependence of the pairing strength in \UTe is an issue barely touched by current theoretical models, but certainly central for this system and clearly of major interest for the whole field of unconventional superconductivity.

\section{Experimental Details}
\label{sec:experiment}
\subsection{Single crystal growth and samples}

	Different single crystals of \UTe from three different batches have been studied by specific heat and magnetostriction/thermal expansion measurements. 
	All single crystals were prepared by the chemical vapour transport method with iodine as transport medium. 
	A starting ratio of U:Te $=$ 2:3 has been used, and the quartz ampules have been heated slowly up to a final temperature of 1000~$^\circ$C on  one side and 1060$^\circ$C on the other side 	and this temperature gradient was maintained for 18 days. 
	The ampules have been slowly cooled down to ambient temperature during 70~hours. 
 	A sample ($\#$1) with a superconducting transition temperature \Tc of 1.45 K, a mass of 12.3~mg, from the same batch than those of Refs.~\onlinecite{Knebel2019, NiuPRR2020}, has been used for the magnetostriction studies.  
 	Samples, \diese2 and \diese3 with masses of 5.6 mg and 27 $\mu$g respectively, are from another batch with a critical temperature around 1.85~K, and were used for most of the specific heat measurements.  
 	Thermal expansion has been measured on a fourth crystal (sample \diese4), very similar to samples \diese2 and \diese3 regarding the specific heat in the normal state, the specific heat jump at $T_c \approx 1.85$~K as well as the residual value $\gamma_0$:  we found $\gamma_0 \approx 0.03$~J K$^{-2}$mol$^{-1}$ at 0.1~K, or 0.011~J K$^{-2}$mol$^{-1}$ extrapolated toward $T=0$ from above 0.3~K.
 	Note that the entropy balance at \Tc is also well satisfied on these samples (\diese2-\diese4, see Appendix \ref{appendix:Cp normal phase}).
	
\subsection{Specific heat measurements}
	The specific heat of two samples ($\#$1 and $\#$2) has been measured by a quasi-adiabatic relaxation method in a dilution refrigerator up to 15~T in a superconducting magnet and down to 100~mK. 
	Small heat pulses of maximum 1$\%$ of the temperature $T$ (0.5$\%$ in the superconducting transition) were applied to the samples. 
	The specific heat $C$ is extracted from the temperature response of the sample during the whole pulse sequence. 
	Down to the lowest temperatures, only one relaxation time was measured in the exponential decay. 
	The addenda have been measured separately. 
	They represent $8\%$ of the total measured specific heat at 2~K and $2\%$ at 100~mK. 
	Mainly temperature sweeps were performed, but also some field sweeps for the transitions at the lowest temperatures (between 100~mK and 200~mK).

    To align the samples in the magnetic field, we used a piezoelectric rotator allowing a rotation over 90$^{\circ}$ in a plane parallel to the field, and a goniometer allowing a $\pm$3$^{\circ}$ rotation perpendicular to the plane.  
	Furthermore, the set-up is rigid, so that the torque between magnetic field and the anisotropic magnetisation of the sample could not induce a misalignment.
	 
	 Sample (\diese3 - of 27 $\mu$g) has been measured with an ac specific heat technique in a $^3$He refrigerator down to 600~mK, and up to 36~T in the M9 magnet at the high magnetic field laboratory LNCMI in Grenoble. 
	 Details of the specific heat set-up are shown in the Supplemental Materials of Ref.~\onlinecite{MichonNature2019}.
	 For fields up to 18.5~T, the ac calorimetry has been performed using a 20~T superconducting magnet (M2) in combination with a $^3$He refrigerator down to 400~mK. 
	 An attocube piezorotator allowed for a rotation in the ($b$,$c$) plane.

	 The value of the critical temperature \Tc is extracted from the specific heat transition by a fit to an ideal jump broadened by a Gaussian distribution of critical temperatures: \Tc corresponds to the centre of this distribution.
     This model, which reproduced the data very well, allows to extract directly other parameters like the width and jump of the specific heat at the transition (more details in the Appendix \ref{appendix:Gaussian fit} ).

\subsection{Linear magnetostriction and thermal expansion measurements}
    The linear magnetostriction $\Delta L_b/L_b$ of \UTe has been measured on single crystal \diese1. 
    In addition, we measured the linear thermal expansion at constant magnetic field on single crystal \diese4.

    These measurements have been performed using a high resolution capacitance dilatometer \cite{Kuechler2012}. 
	The capacitance has been determined using an Andeen Hagerling capacitance bridge AH2550A. 
	High magnetic field experiments have been performed using the 30~T magnet M10 of the high magnetic field laboratory LNCMI Grenoble. 
	Due to the limited diameter of the magnet it has been only possible to measure the length change $\Delta L_b$ parallel to the magnetic field applied along the $b$ axis of the crystal. 
    The magnetic field has been swept with a maximal rate of 100 G/sec to avoid eddy currents heating.
	The dilatometer was positioned at the end of a silver cold finger in a $^3$He cryostat, with a base temperature near 370~mK. A RuO$_2$ thermometer and a heater were fixed directly on the dilatometer. Temperature sweeps at fixed magnetic field have been performed with maximal heating rates of 0.1~K/min.  
	
	Additional thermal expansion measurements have been performed in a superconducting magnet up to 13~T using a dilution refrigerator in CEA Grenoble.

\section{Results}
\label{sec:results}
	\subsection{Magnetic field \texorpdfstring{\Hparab}{H parallel b}}
    \subsubsection{Specific heat}
\begin{figure}
	\centering
	\includegraphics[width=1\linewidth]{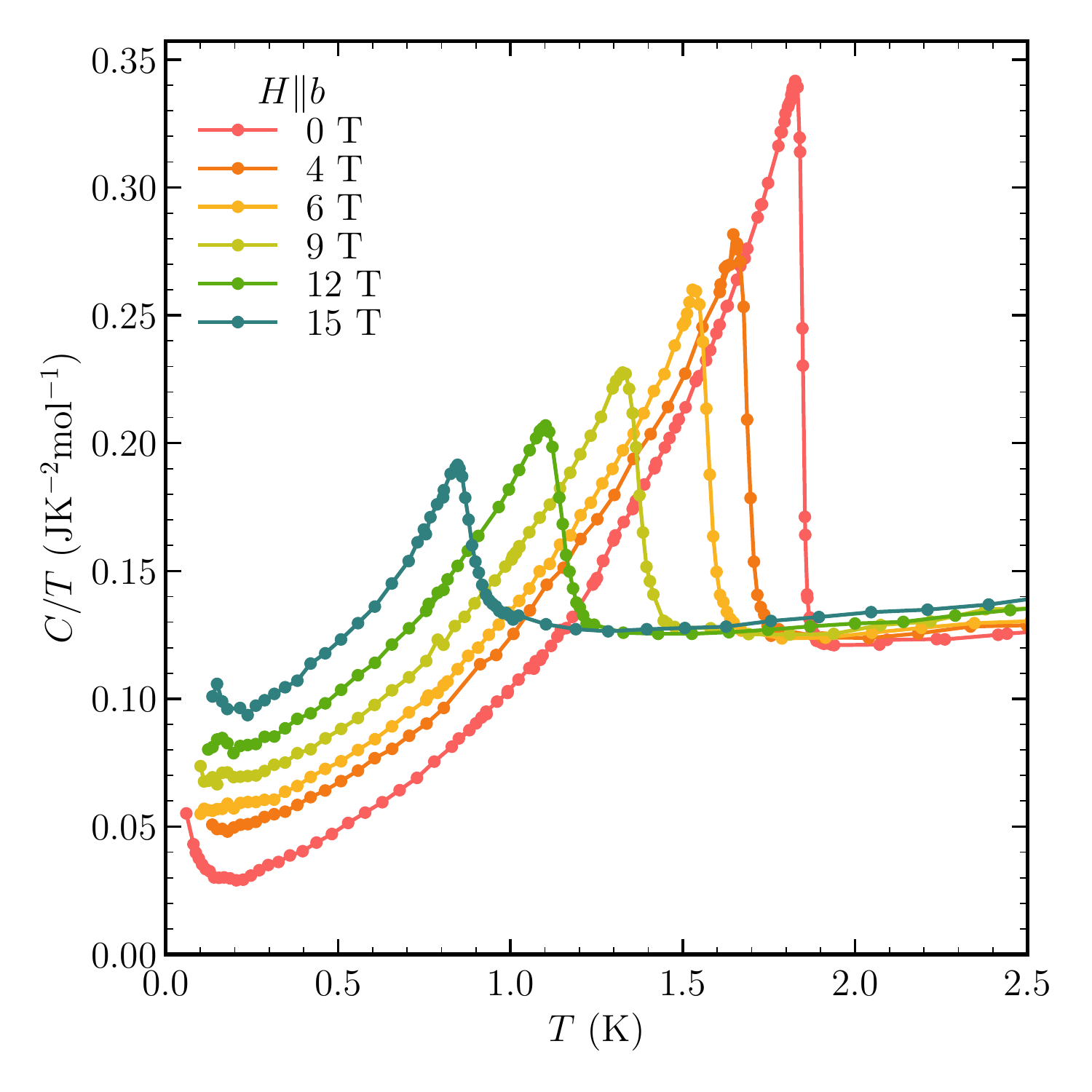}
	\caption{Temperature dependence of the specific heat $C/T$ at different magnetic fields $H \parallel b$ from 0~T to 15~T measured on sample \diese2. }
	\label{fig:RawCpLF}
\end{figure}

	In zero field, all samples studied exhibit a single sharp superconducting transition (width $\Delta T_c \approx 20$ to $38$~mK), with a large jump at the superconducting transition (up to $\Delta$\Cp/\Cp$\approx$~1.85), emphasising the high quality and homogeneity of the samples (Figure~\ref{fig:RawCpLF} for sample \diese2).
	The specific heat measurements on the first \UTe samples displayed an upturn and a large residual term of $C/T$ at low temperatures \cite{Metz2019a}.
	Both became smaller with improved sample quality. 
	Our measurements on crystal \diese2 show indeed only a small residual term, and an upturn shifted to lower temperatures compared to samples with lower \Tc (more details in the Appendix \ref{appendix:Cp superconducting phase}). 
	This agrees with recent works claiming that residual term and upturn are extrinsic to \UTe \cite{Cairns2020,AokiJPCM2022,Rosa2022}.

\begin{figure}
	\centering
	\includegraphics[width=1\linewidth]{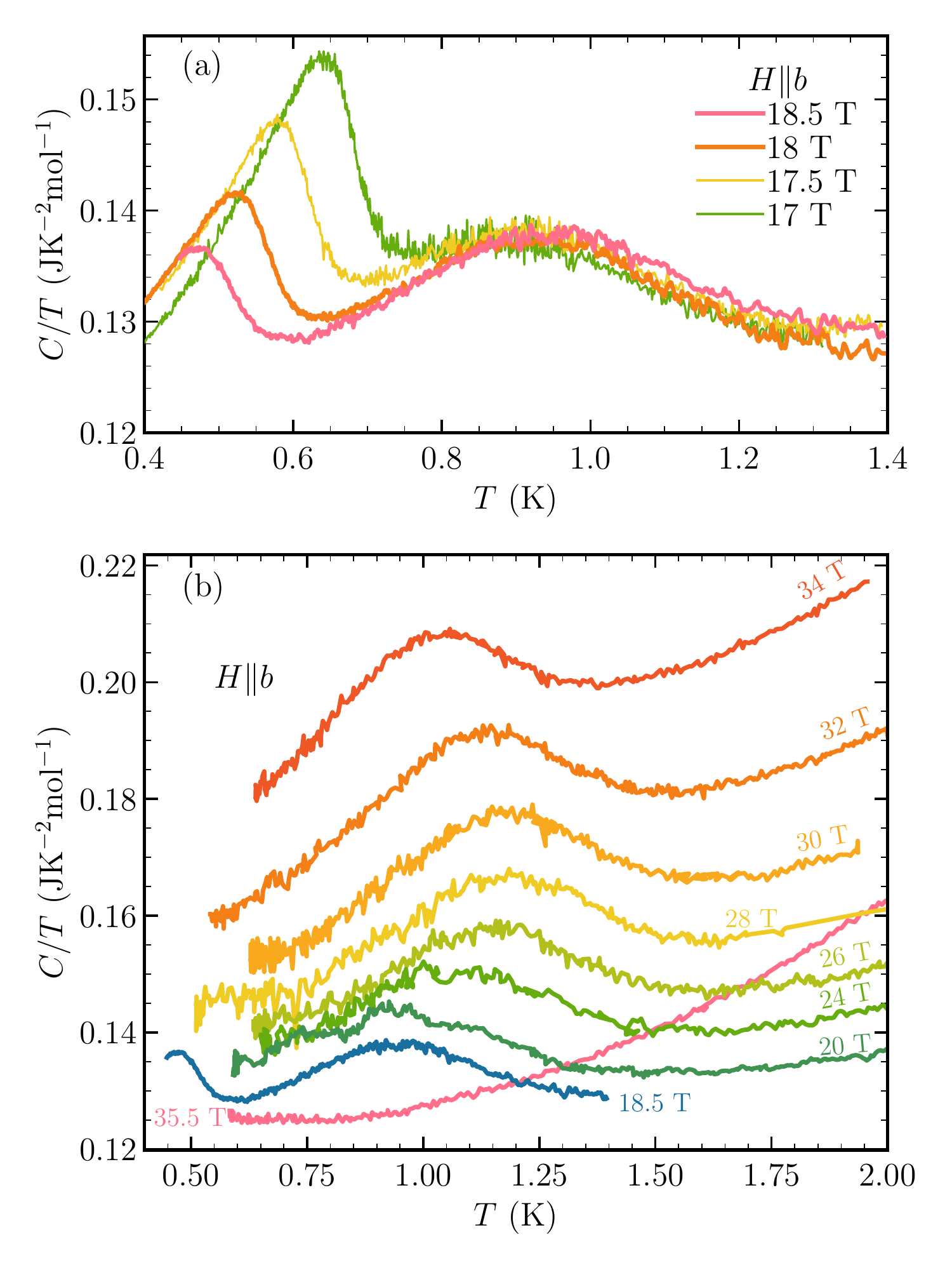}
	\caption{(a) Temperature dependence of $C/T$ measured on sample \diese3 for fields \Hparab from 17 to 18.5~T.  A second wide transition appears above the sharp low temperature transition. (b) $C/T$ for the different magnetic fields up to 35.5~T, which is above the metamagnetic transition. }
	\label{fig:RawCpLF_HF}
\end{figure}

	Figure~\ref{fig:RawCpLF} shows $C/T$ as a function of temperature also for several magnetic fields \Hparab up to 15~T. 
	Under field, the transition remains sharp with a pronounced jump $\Delta C/T_c$ up to 18.5~T (see Fig.~\ref{fig:RawCpLF_HF}(a), so they are easily followed under field. 

	Remarkably for \Hparab, $C/T$ shows two transitions above 15~T: in addition to the marked low temperature transition
    a second wide transition (350~mK width at 18~T) appears above this field (see Fig.~\ref{fig:RawCpLF_HF}(a) and 
    above $H \gtrsim 17$~T, it becomes well detached from the sharp transition. 
	We could follow this second wide \hftransi transition up to the metamagnetic transition,  
	\cite{Knafo2019, MiyakeA2019, RanNatPhys2019}, see Fig.~\ref{fig:RawCpLF_HF}(b).

\begin{figure}
	\centering
	\includegraphics[width=1\linewidth]{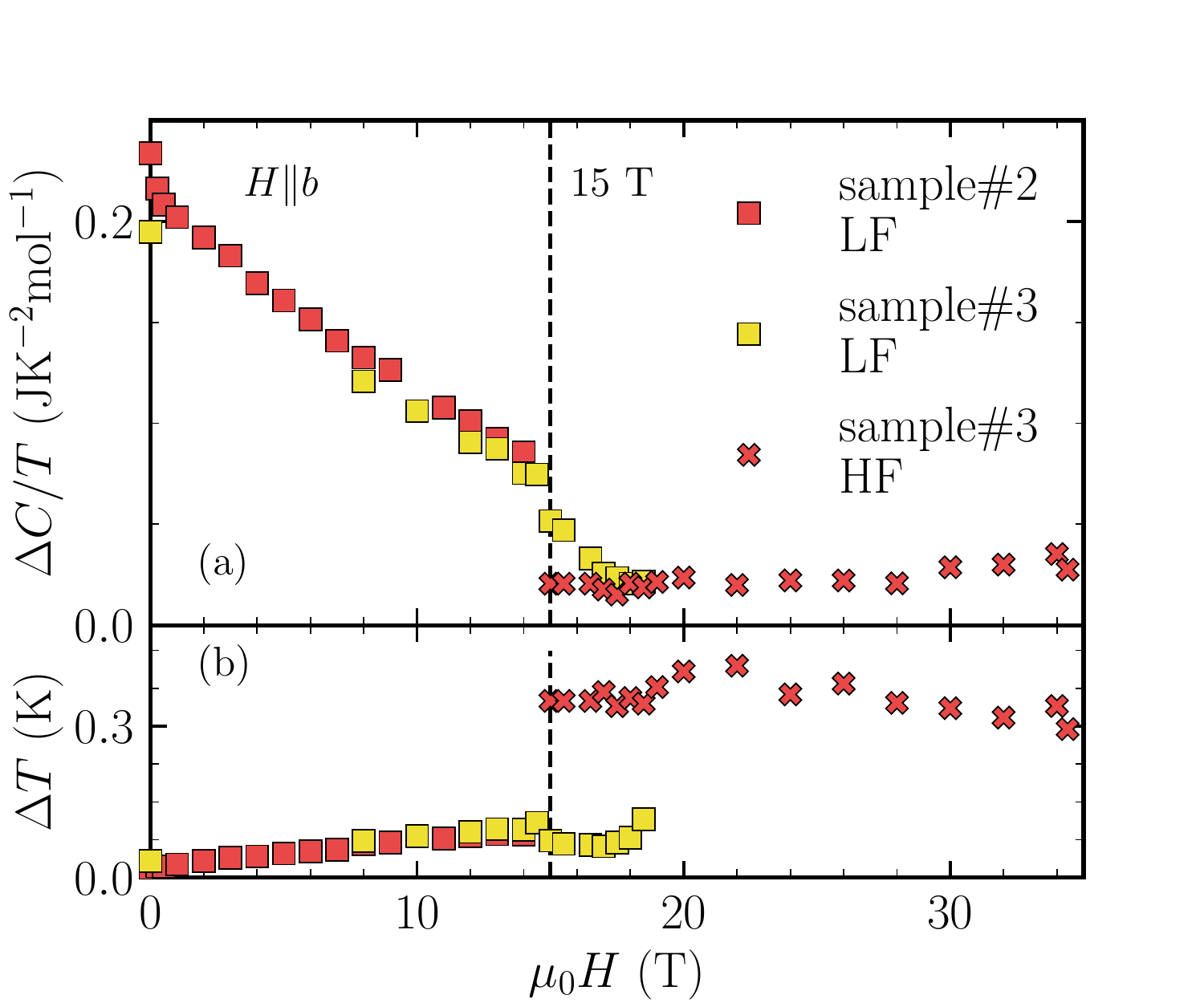}
	\caption{(a)Jump of the transitions $\Delta C/T_c$ as a function of field, for \Hparab, determined on sample \diese2 and \diese3 for the \lftransi and \hftransi transitions.
	(b) Width $\Delta T_c$ of the transitions as a function of fields. $\Delta T_c$ is equal to 2.35 times the standard deviation of the Gaussian distribution of T$_c$ used to fit the transition.}
	\label{fig:Cp_Jump_Width}
\end{figure}

    A Gaussian analysis of the temperature dependence of \CT allows to deconvolute broadening effects and to determine the jump $\Delta C/T$ at \Tc and the width of the transition as a function of magnetic field. They are shown in Fig.~\ref{fig:Cp_Jump_Width}.
    The specific heat jump at the transition from the normal to the \lftransi  phase decreases with field up to 15~T. When the second broad transition appears above 15~T, the jump at the \lftransi superconducting transition displays a marked drop (see also Fig.~\ref{fig:RawCpLF_HF}(a)). 
    Essentially, it goes down to the same level as that of the wide transition of the normal to \hftransi phase, which remains roughly constant up to \Hm .
    Hence, as expected, the emergence of the \hftransi transition goes along with a redistribution of entropy between both phases.
    
	The \Tc of this broad anomaly is increasing with field, except very close to \Hm where the transition temperature decreases slightly.
	This may be due to a slight misalignment of the sample in the high-field experiments or to the torque at the highest field, but it could also be intrinsic.
	Above \Hm, the broad anomaly abruptly disappears. 
	This \hftransi transition is the expected bulk signature of the field-reinforced superconducting phase observed in transport properties for the same field direction \cite{Knebel2019, RanNatPhys2019}. 

\begin{figure}
	\centering
	\includegraphics[width=1\linewidth]{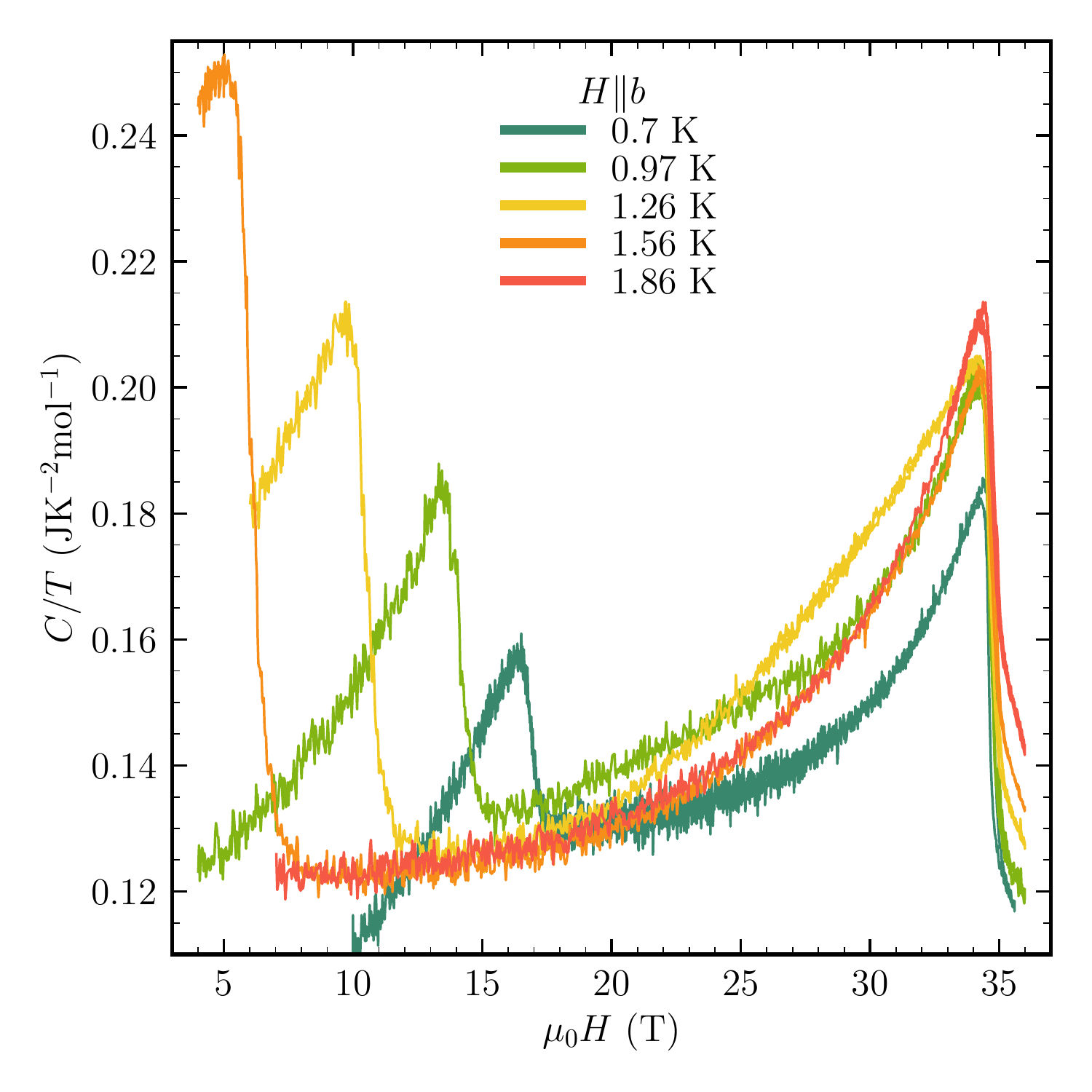} 
	\caption{$C/T$ versus field for \Hparab on sample \diese3 at different temperatures between 0.7~K and 1.86~K. The \lftransi superconducting transition and the peak at \Hm$=34.75$ T are clearly visible, but the \hftransi superconducting transition reported in Fig.~\ref{fig:RawCpLF_HF} appears only as a very broad anomaly.}
	\label{fig:RawCpFieldSweepLF_HF}
\end{figure}

\begin{figure}
	\centering
	\includegraphics[width=1\linewidth]{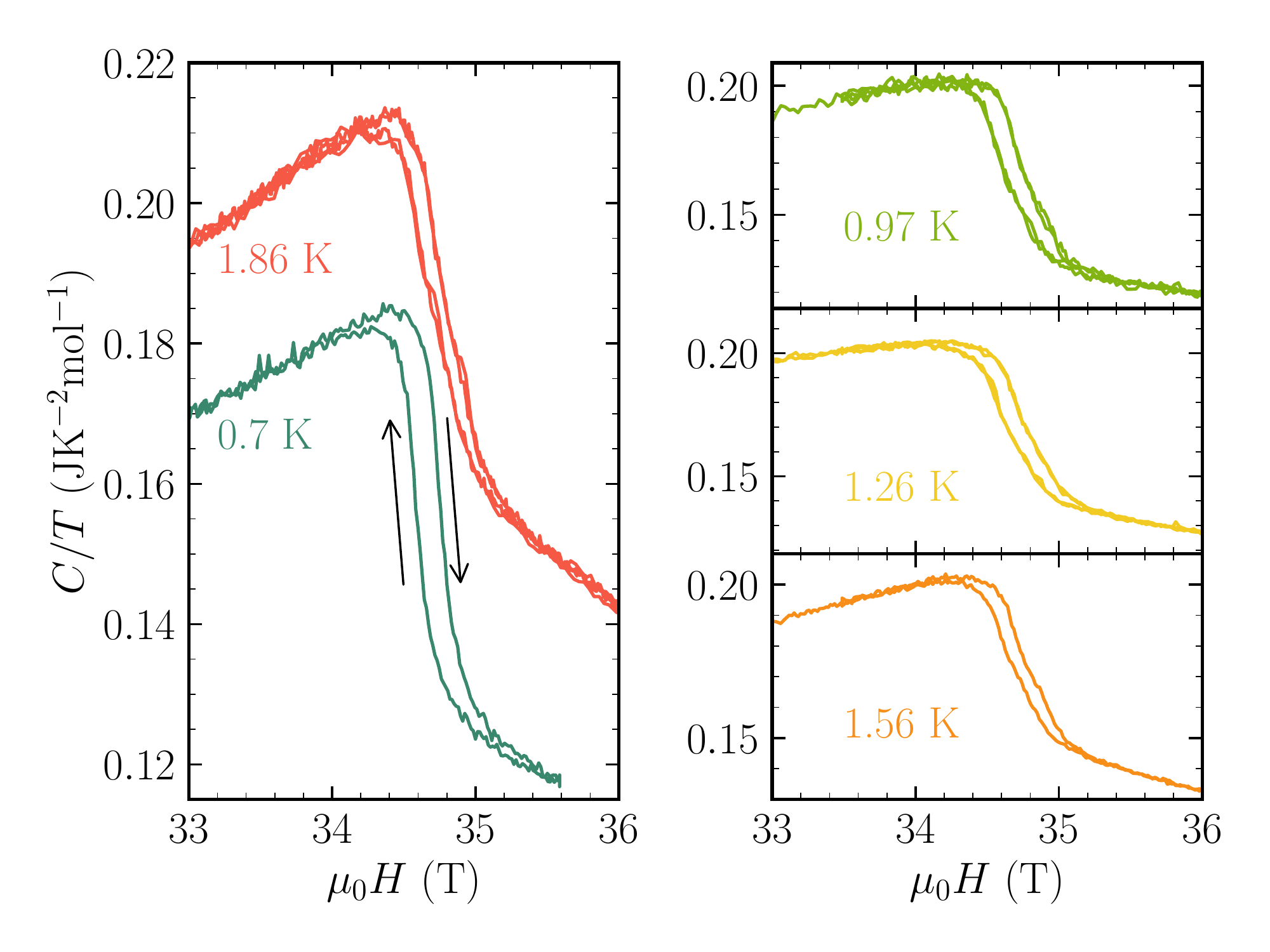} 
	\caption{$C/T$ versus field (\Hparab) at the metamagnetic transition for sample \diese3 at different temperatures between 0.7~K and 1.86~K. The hysteresis is independent of the field sweep rate. On the left panel, the arrows indicate the direction of the field sweeps.}
	\label{fig:RawCpHm}
\end{figure}

	Figure~\ref{fig:RawCpFieldSweepLF_HF} displays the field dependence along the $b$ axis of $C/T$ up to 36~T. 
    The sharp \lftransi transition is well observed on these field sweeps. 
    However, the \hftransi transition observed in temperature scans appears here as a very broad and shallow anomaly, noticeable only by comparison with curves at different temperatures. 
    This arises from the combination of an already large \Tc distribution at fixed field, with an almost vertical \Hc, so that this \hftransi transition appears extremely broad as a function of field (see Appendix \ref{appendix:comparison Hc2 resistivity}).
    
	On approaching \Hm, \CT shows a strong increase, with a large drop (of order 25 \%) at the first-order metamagnetic transition, and a hysteresis independent of the sweep rate, displayed on  Fig.~\ref{fig:RawCpHm}. 
	The drop of the specific heat above \Hm, is sharpest at the lowest temperature, with a width of 0.25~T. 
	However, a possible interplay between the superconducting and metamagnetic transitions at this temperature may influence the shape of the anomaly. 
	The width of the hysteresis decreases linearly with increasing temperature, starting from 0.17~T at 0.7~K. 
    This behaviour of \CT at \Hm agrees qualitatively with previous measurements \cite{Imajo2019, MiyakeA2021} performed in pulsed magnetic fields (see comparison in the Appendix \ref{appendix:SuppCpHm}).

\subsubsection{Complete phase diagram}
\begin{figure}
	\centering
	\includegraphics[width=0.9\linewidth]{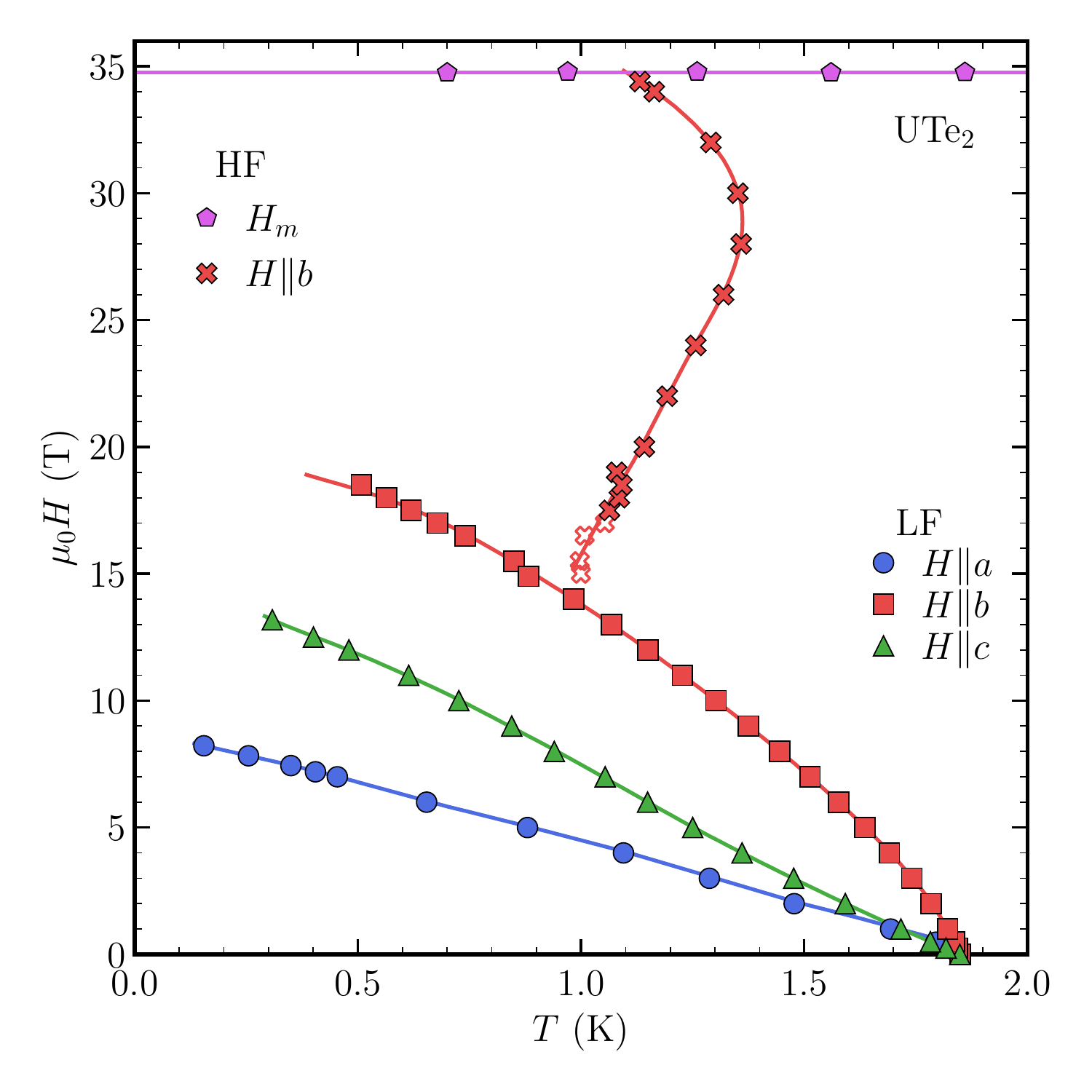}
	\caption{Phase diagram up to 36 T for $H$ applied along the three crystallographic axis, established with specific measurements on samples \diese2 (below 15~T) and \diese3 (above 15~T). 
	Blue circles: \Hparaa, green triangles: \Hparac, red symbols: \Hparab. Red squares are \Tc from the sharp \lftransi transitions, crosses are \Tc from the broad \hftransi transitions.
	For empty crosses, \Tc could only be determined by fixing the width and jump of the \hftransi transition. Magenta hexagons: \Hm determined on sample \diese3 by specific heat from the field-up sweeps.}
	\label{fig:PhaseDiagCp}
\end{figure}

	Since samples \diese2 and \diese3 come from the same batch and have essentially the same \Tc at 0~T, 
	specific heat measurements on these two samples can be used to establish the complete superconducting phase diagrams for all field directions from 0 to 36 T.
	It is shown in Fig.~\ref{fig:PhaseDiagCp}. 
	As underlined already in section \ref{sec:overview}, a most obvious result is that for \Hparab, two superconducting phases are clearly present, with a point around $H=15$~T and $T=1$~K where the three transition lines join.
	
	The limits of the \lftransi superconducting phase correspond to the sharp transition that can be followed from 0~T up to 18.5~T. 
	The emergence of the \hftransi phase is revealed by the broad transition appearing above 15~T, and followed up to \Hm (Figures~\ref{fig:RawCpLF}-\ref{fig:RawCpLF_HF}).
	For magnetic fields slightly above 15~T, the two transitions overlap, thus it is difficult to determine unambiguously \Tc for the \hftransi phase transition. 
	Naturally, this also prevents a precise determination of the nature of the crossing of the \Hc lines of the \lftransi and \hftransi phases: they could merge with a sharp change of slope, or could be tangential.
	Points on between 15 and 17~T (empty crosses in Fig.~\ref{fig:PhaseDiagCp}) have been determined by fixing  the jump height and width of the \hftransi anomaly, using the fact that they both seem to have a negligible field-dependence (Fig.~\ref{fig:Cp_Jump_Width}) 
	(for more details in Appendix \ref{appendix:Gaussian fit}).

	We also found no sign of a fourth transition line inside the \lftransi phase, either on temperature or field sweeps, as would be expected for a crossing of second order phase transitions \cite{YipPRB1991, Braithwaite2019}.
	However, this absence of a fourth line  could also arise from a further broadening of the \hftransi transition inside the \lftransi phase (for field below 15~T). 

	Nevertheless, the phase diagram clearly demonstrates the existence of at least two different superconducting phases for \Hparab: 
	comparison with the resistivity measurements, and the observation of vortex pinning in the \hftransi phase by linear magnetostriction (see next paragraph) show that it is also a superconducting phase.

\subsubsection{Linear Magnetostriction}
\begin{figure}
		\centering
		\includegraphics[width=1\linewidth]{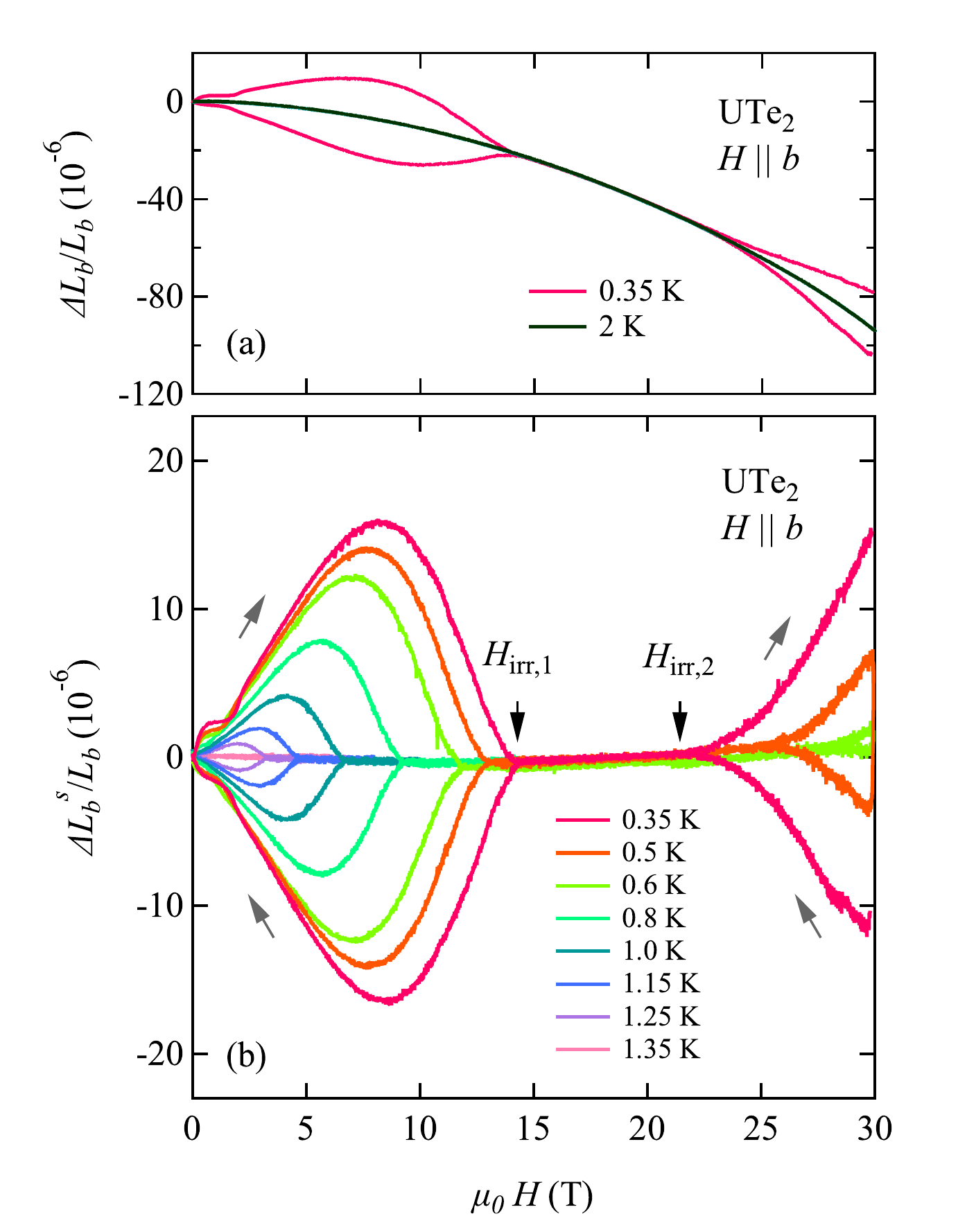} 
		\caption{(a) Longitudinal linear magnetostriction $\Delta L_b/L_b$ of sample \diese1 along the $b$ axis for a field applied along the \baxis, in the superconducting state at 0.35~K and in the paramagnetic state at 2~K. (b) Linear magnetostiction in the superconducting state at different temperatures after subtraction of the paramagnetic contribution measured at 2~K. The grey arrows indicate the direction of the field sweep. Black vertical arrows mark the closing and reopening of the hysteresis as a function of field, which corresponds to the irreversibility field $H_{\rm irr}$ of the vortex motion in the superconductor.}
		\label{fig:RawdL}
	\end{figure}

	The longitudinal linear magnetostriction $\Delta L_b(H)/L_b$ along the \baxis for \Hparab of sample \diese1 is shown in Fig.~\ref{fig:RawdL}.
	In the normal state ($T =2$~K), the linear magnetostriction is negative and shows roughly a $H^2$ field-dependence, as usually observed in paramagnetic metals (Fig.~\ref{fig:RawdL}(a)).
	This is in agreement with the low field measurements of Ref.~\cite{ThomasPRB2021} and with the very recent measurements in pulsed magnetic fields, which show a strong negative jump of the linear magnetostriction at the metamagnetic transition \cite{MiyakeA2022}. 
	Following Maxwell's relations, the negative sign of $\Delta L_b/L_b$  indicates that under uniaxial stress applied along the \baxis, the susceptibility $\chi_b$ along this axis should increase, as observed under hydrostatic pressure \cite{Li2021}.

	In the superconducting state at $T = 0.35$~K, the linear magnetostriction shows a very pronounced hysteretic behaviour.
	In Fig.~\ref{fig:RawdL}(b) we display the additional contribution to the linear magnetostriction $\Delta L_b^s$ which appears in the superconducting state.
	It is obtained from the measured linear magnetostriction at fixed temperature in the superconducting state after subtraction of the paramagnetic contribution measured at 2~K. 
	The linear magnetostriction in the superconducting state is very large and shows a strong hysteresis with a fishtail-like behaviour both below $\approx 15$~T and above $\approx 20$~T. 
	
	This irreversible magnetostriction appears very similar to the behaviour of the magnetisation in the mixed state of type~II superconductors with strong vortex pinning.
    Indeed, in  the critical state model \cite{Bean1964}, magnetic flux penetration or expulsion, when increasing (or decreasing) the field, is impeded by vortex pinning.
    If magnetic flux lines are trapped by the action of pinning forces, equal but opposite forces will act on the lattice, 
    with possible effects on the magnetostriction \cite{Eremenko1999}.
	
\begin{figure}
	\centering
	\includegraphics[width=0.9\linewidth]{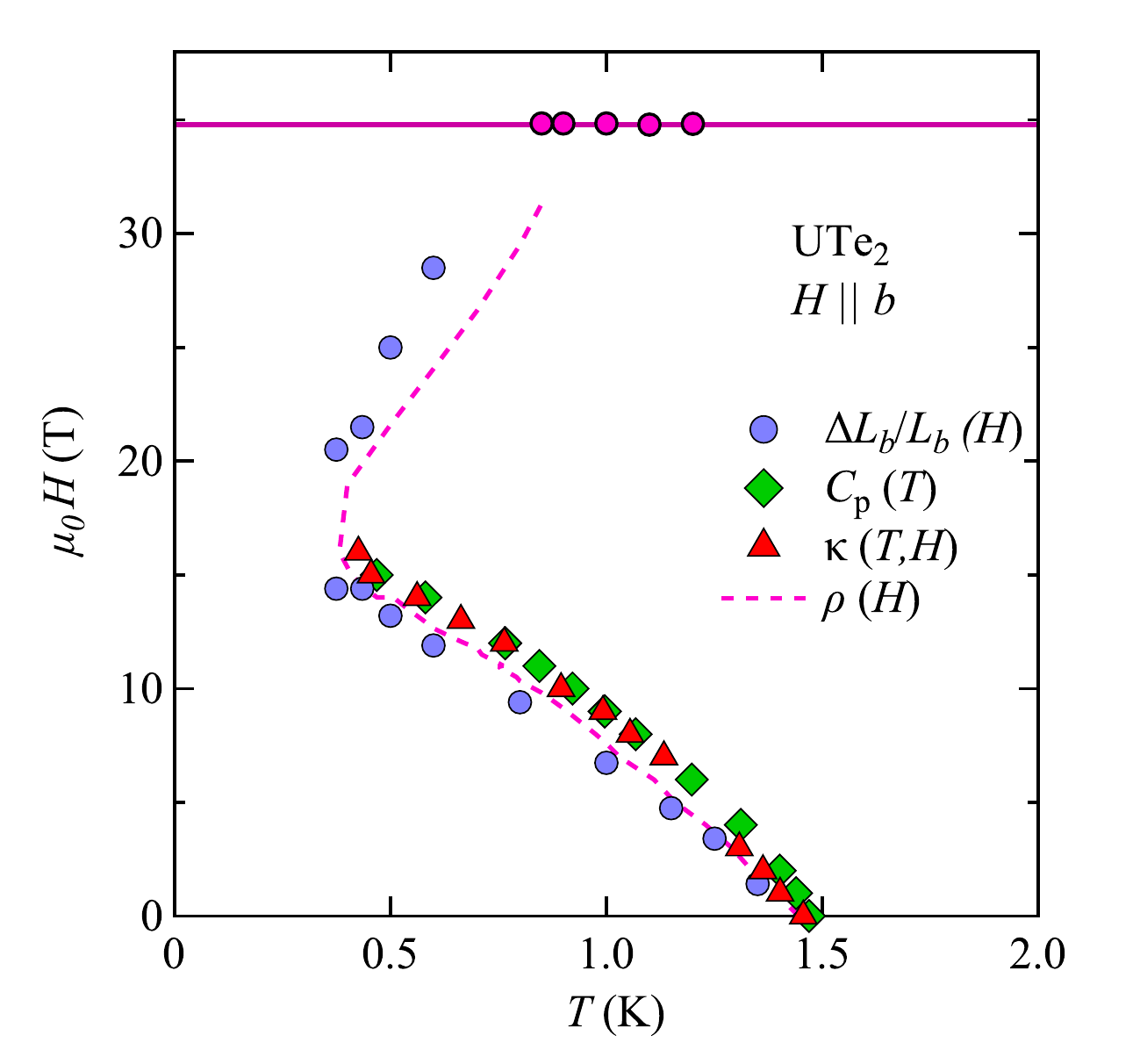} 
	\caption{Phase diagram magnetic field vs temperature for field $H \parallel b$ for sample \diese1 with a \Tc of 1.45~K. 
	Blue circles: irreversibility fields determined from the magnetostriction measurements (see Fig.~\ref{fig:RawdL}); green diamonds: \Hc from specific heat measurements performed on the same sample; red triangles: \Hc from thermal conductivity on a sample from the same batch (appendix \ref{appendix:thermalconductivity}), and (dashed line) \Hc determined from the resistivity $\rho (H)$ of another sample from the same batch 
	(data from Ref.~\cite{Knebel2019}). 
	Magenta line: \Hm determined from resistivity (circles, \cite{Knebel2019}).
	} 
	\label{fig:PhaseDiagdL}
	\end{figure}

    At the lowest temperature, the hysteresis in the linear magnetostriction vanishes above $H_{\rm irr,1} \approx 14.5$~T, and it opens again above $H_{irr,2} \approx 21$~T, being maximal at 30~T, which is the highest field we could reach in the experiment. 
	On warming, the lower field $H_{\rm irr,1}$ decreases, while the upper field $H_{\rm irr,2}$ increases and above 0.6~K, it exceeds the achievable field range. 
	These irreversibility fields are displayed in Fig.~\ref{fig:PhaseDiagdL} and compared 
    with \Hc determined from specific heat measurements (only below 15~T on this sample), with thermal conductivity measurements (appendix \ref{appendix:thermalconductivity}) and also with resistivity measurements on a sample from the same batch with similar \Tc (published in \cite{Knebel2019}). 
    
    In the \lftransi phase, thermodynamic as well as thermal and electrical measurements are in good agreement, with only small quantitative differences on the amplitude of the negative curvature of \HcT in the 0-5~T field range.
    Obviously, flux pinning in the sample at low magnetic fields is rather strong, as found by low field-low temperature magnetisation measurements \cite{Paulsen2021}, and the irreversibility  field follows the upper critical field: $H_{irr,1} \sim H_{c2}$. 
    In the field-reinforced \hftransi superconducting phase above 15~T, the difference between $H_{irr,2}$ and $H_{c2}$ is much more pronounced: there is a broader reversible regime, between \Hc and $H_{irr,2}$, suggesting a decrease of the pinning strength in this phase.
    Actually, the observation of the irreversible magnetostriction due to the flux pinning between 20 and 30~T is a further proof that the \hftransi phase delimited by the broad specific heat anomaly is indeed a bulk, field-reinforced superconducting phase. 
 	
   In Appendix \ref{appendix:thermalexpansion}  we also show the linear thermal expansion $\Delta L_b(T)/L_b$ as a function of temperature measured on sample \diese4, which has a \Tc = 1.82~K similar to that of samples \diese2 and \diese3 studied by specific heat. 
   It confirms the difference of pinning strength between both phases.

\subsection{\texorpdfstring{\Hc}{Hc2} close to \texorpdfstring{\Tc}{Tc}, along all directions}
\begin{figure}[pt]
	\centering
	\includegraphics[width=1.\linewidth]{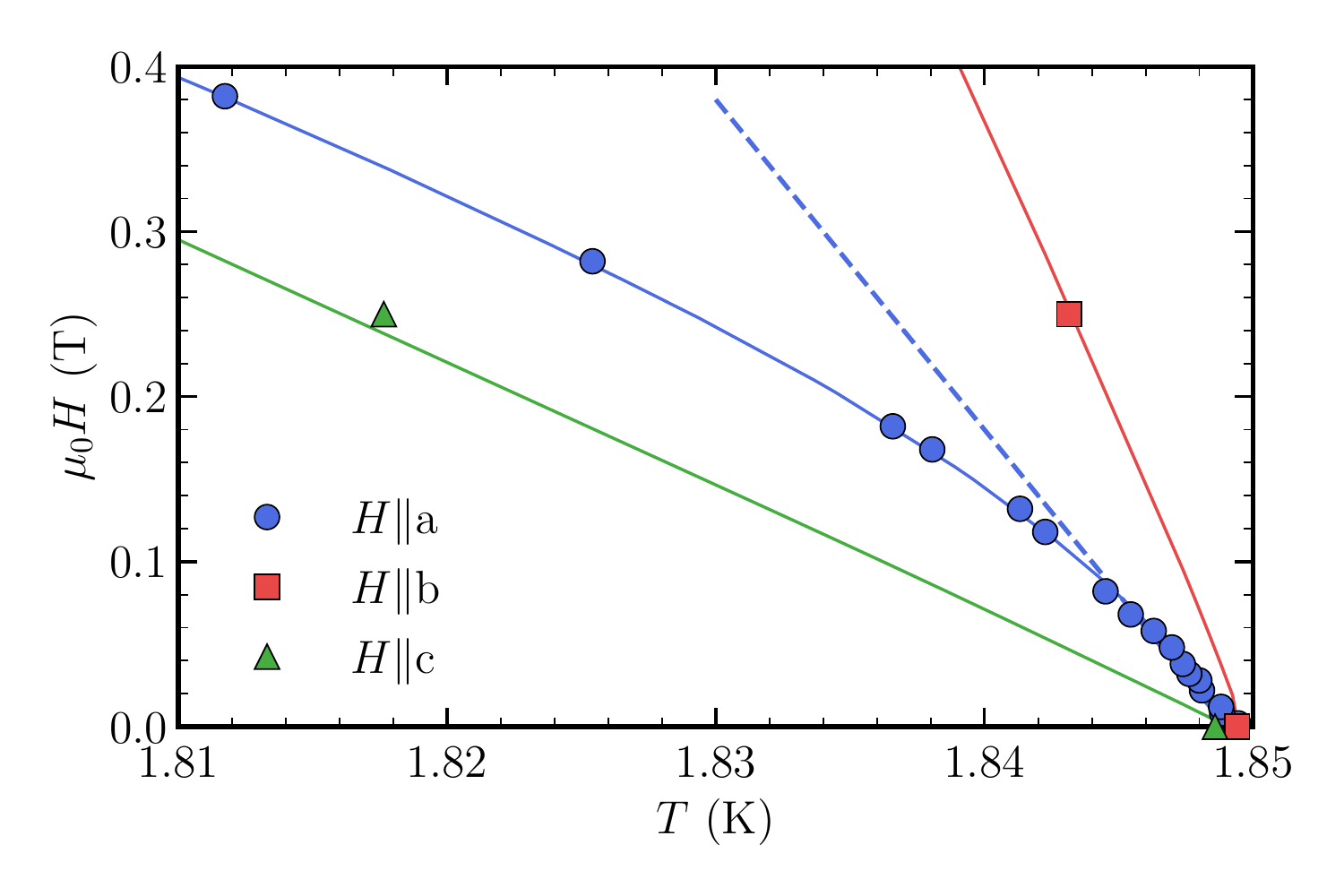}
	\caption{\Hc as a function of temperature determined from specific heat measurements on sample \diese2 for the three axes at very low fields, close to \Tc. Error bars on the determination of transition temperature deduced from the Gaussian analysis of the specific heat anomaly are smaller than the size of the points.
	The dashed line is a linear fit following the initial slope of \Hc along the easy magnetisation \aaxis. 
	It is close to the value along the \baxis, however followed by a very strong negative curvature showing up at very low fields (0.4~T is about $\mu_0 H_{c2}(0)/20$ for \Hparaa).}
	\label{fig:Hc2LF}
\end{figure}

    Figure~\ref{fig:PhaseDiagCp} also shows the upper critical field \Hc of sample \diese2 along the $a$ and $c$ axes. 
    Similar to the previous reports \cite{RanScience2019, Knebel2019}, \Hc is strongly anisotropic and extrapolates to 9~T for the $a$ axis, 15~T for the $c$ axis.
    A more detailed examination reveals that \Hc in \UTe present anomalies not only due to the presence of the two superconducting phases for \Hparab:
    it also has an anomalous temperature dependence for all field directions. 
    The most striking feature appears along the \aaxis.
    Over a large temperature range \HcT along the \aaxis appears linear. 
    However, a closer look very near \Tc yields a contrasted view, showing that contrary to field directions \Hparab and \Hparac, this linear behaviour would extrapolate to a critical temperature 36~mK larger than the experimental value. It can be visualized on Fig.~\ref{fig:Hc2LF}, showing the very low field region ($H<0.4~T$) on an enlarged scale. \Hc along the easy \aaxis displays a very strong negative curvature very near \Tc, following an initial slope of order -20~T/K, much larger than anticipated by the "large scale" linear behaviour.
    It is also much larger than the values determined previously by resistivity measurements:  around -5 or -6~T/K depending on the samples \cite{RanScience2019, Aoki2019}.

    Actually, this effect was already present in Ref.~\onlinecite{Kittaka2020} reporting also specific heat measurements, but was barely discussed.
    We have confirmed this anomalous behaviour (the very large slope followed by a very strong curvature), on three different samples measured also by specific heat (appendix \ref{appendix:Hc2 para a}). 
    Discussion of the possible origin of this very strong curvature along the \aaxis in \UTe is of course central for the question of the symmetry state of the superconducting order parameter in \UTe and of the pairing mechanism (see section \ref{sec:analysis}-A below).
 
    For the other directions, there are no anomalies close to \Tc. The initial slope has its largest value for \Hparab: 
     from our specific heat measurements, we determine \dHc$\approx-34$ T/K, which is larger than the values obtained from electrical transport measurements: \dHc$\approx-25$ T/K \cite{NoteResHc2}.
	This initial slope is lowest along the \caxis: \dHc$\approx-7.5$ T/K\Hc, and \HcT displays an usually large linear regime from \Tc down to 0.5~K.


	\subsection{Normal phase specific heat}
	The very large specific heat jump at \Tc clearly indicates that \UTe is in a strong-coupling regime. 
	The most natural explanation for the field-reinforced superconducting phase is that the superconducting pairing itself is enhanced under fields along the \baxis \cite{RanScience2019, Knebel2019}.
	As discussed already for the ferromagnetic superconductors, in the strong coupling regime, such a field dependence should be reflected also in the normal state Sommerfeld specific heat coefficient $\gamma$ \cite{Wu2017, MineevAnnPhys2020}.
	In \UTe, it is not easy to determine the Sommerfeld coefficient, because \CT remains strongly temperature dependent almost down to \Tc, and there is no simple way to analyse this temperature dependence (appendix \ref{appendix:Cp normal phase}). 
	This arises mainly from a marked anomaly in the specific heat with a maximum at a temperature $T^{\ast} \approx 12$~K, attributed to magnetic fluctuations \cite{Willa2021a}.

	\begin{figure}
	\centering
	\includegraphics[width=1\linewidth]{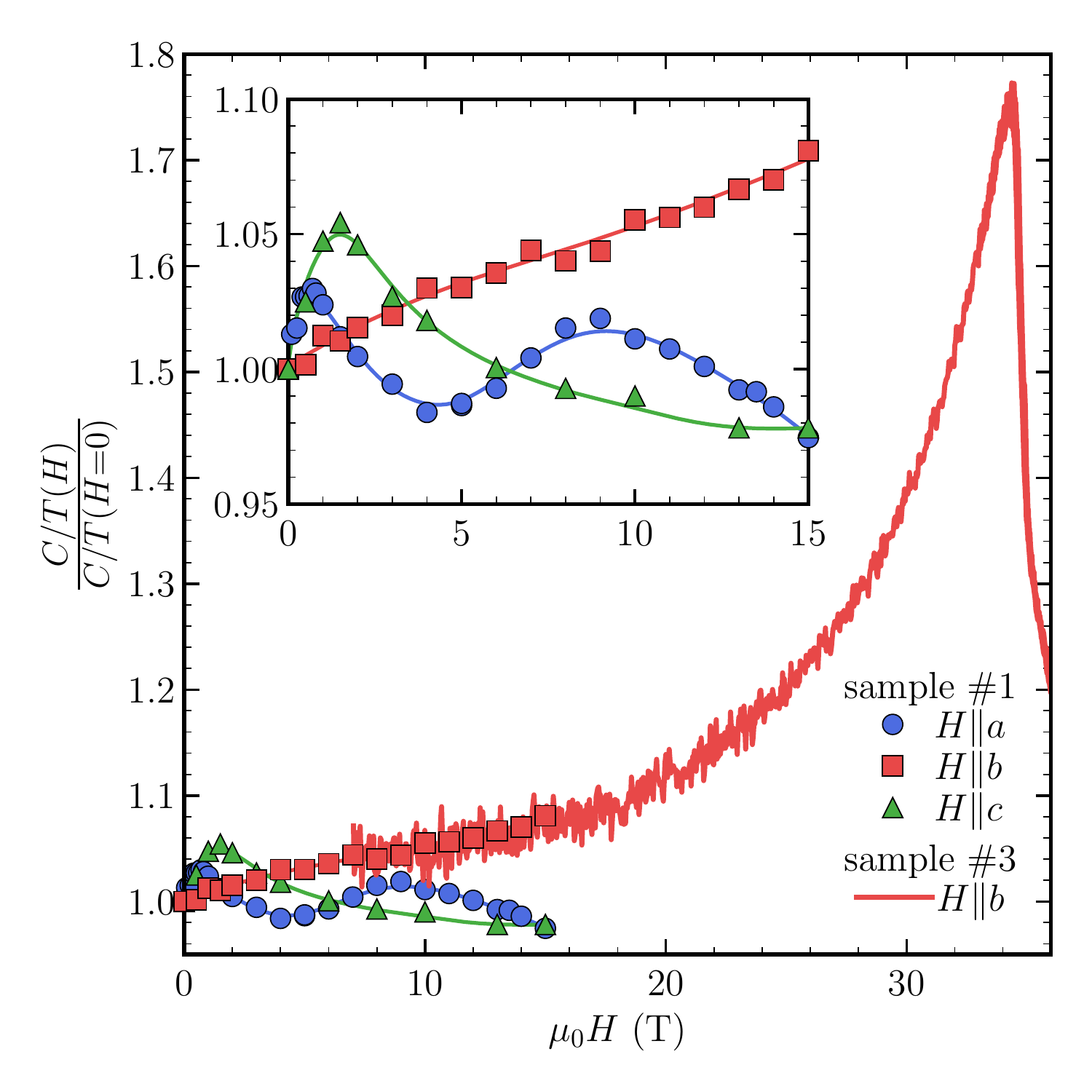}
	\caption{$C/T$ normalised by its value at zero field, as a function of field for $H$ along $a$,$b$ and $c$ axis at 1.86 K. 
	Measurements below 15 T have been done on sample \diese1, by temperature sweeps. Measurements for \Hparab above 15~T have been done on sample \diese3 by a field sweep at 1.86 K. Insert shows a zoom on the measurements below 15 T, lines are guides to the eyes.}
	\label{fig:Cp_T_FieldSweep}
\end{figure}

    We show in Fig.~\ref{fig:Cp_T_FieldSweep} $C(H)/T$ normalised to $C/T$ in zero field, at $T=1.86$~K for fields applied along the three crystallographic directions. 
	As the temperature dependence of \CT becomes weak at this temperature, it can be considered as a reasonable estimation of the $\gamma$ value, at least as long as the applied field is not too close to \Hm.
	Indeed, the specific heat anomaly at 12~K 
	is shifted to much lower temperatures on approaching \Hm for \Hparab \cite{Willa2021a,AokiJPCM2022}.
	Hence for this field direction, magnetic fluctuations are likely to contribute to the large enhancement of \CT close to \Hm, clearly visible in Fig. \ref{fig:Cp_T_FieldSweep}. 
    The subsequent sharp drop of \CT above \Hm might arise from the Fermi surface instability detected at the metamagnetic transition \cite{NiuPhysRevRes2020} and/or from a suppression of magnetic fluctuations in the polarized phase.

	These results for \Hparab are similar to already published data \cite{MiyakeA2019, Imajo2019, MiyakeA2021}, with some quantitative differences notably above \Hm, where the ac technique in static fields probably allows for more precision. 
	For field along the $a$ or $c$ axis, $C/T(H)$ has an even more complex behaviour.
	It is known that along the \aaxis Lifshitz anomalies appear around 5 T and possibly 9~T \cite{NiuPRL2020}. 
	They do appear as extrema of \CT on our measurements (temperature dependence in Appendix \ref{appendix:Cp normal phase}).
	In addition, we also observe  pronounced maxima of \CT along the $a$ and $c$ axes at low field, respectively at 0.5 T and 1.5 T, whose origin is still unclear. 
	Until better understood, it is difficult to rely on the field dependence of \CT in the normal phase to discuss quantitatively the behaviour of the pairing strength with field.

\section{Analysis}
\label{sec:analysis}
	\subsection{\texorpdfstring{\Hc}{Hc2} in the  \texorpdfstring{\lftransi}{LF} phase}
 
    We first discuss the behaviour of \Hc close to \Tc. 
    Indeed, the observed negative curvature of \Hc along the $a$ axes suggesting a severe paramagnetic limitation might seem to contradict the common belief that \UTe is a \pwave superconductor with a \dvector perpendicular to the easy \aaxis.
    
    Such a paramagnetic limitation would be at odds with the value of $H_{c2}(0) \sim 9$~T, which is much larger than the weak-coupling paramagnetic limit of around 3.5~T (for a gyromagnetic factor $g=2$ with \Tc=1.85~K).
    Actually, the negative curvature is so concentrated close to \Tc that it requires a very large value of $g$ ($g \approx 3.2$ in the weak-coupling limit, so even larger in the strong coupling regime) to match the initial deviation from linearity of \Hc along the \aaxis, leading to a saturation of \Hco at 2.25~T at low temperatures (Fig.~\ref{figS:CompHc2_Hc1}(b) in Appendix \ref{appendix:fitParameters}). 
    In other words, \HcT along the easy axis does not follow at all the temperature dependence of an upper critical field solely controlled by paramagnetic and orbital limitations \cite{WerthamerPR1966_3}: 
    paramagnetic limitation is not a satisfying explanation for the strong negative curvature close to \Tc.
     
    Nonetheless, the large value of the initial slope \dHc along the \aaxis obtained when taking account of this strong curvature (Fig. \ref{fig:Hc2LF}) is in excellent agreement with the initial slope \dHcT$=-20.4~T/K$ 
    determined from the lower critical field $H_{c1}$ and the thermodynamic critical field $H_{c}$ (for details see Appendix \ref{appendix:comparison Hc2 Hc1} and Ref.\onlinecite{Paulsen2021}).
    From the resistivity measurements which essentially extrapolate the linear regime up to \Tc, a much smaller value (around -8~T/K) is found, which contradicts the relation with \Hcl and the critical thermodynamic field.
    So, this agreement between the large value of the \dHc and the measurement of \Hcl supports the intrinsic character of the strong curvature of \Hc close to \Tc for H along the \aaxis.
 
    Before examining more quantitatively a possible explanation for this curvature, it is worth analysing the situation along the $c$ and $b$ axes.
    
    Regarding the same comparison of \Hcl with \Hc along the \caxis, the agreement is also very good:
    from the values of \dHcl along the \caxis in Ref. \onlinecite{Paulsen2021}, we expect a value of \dHc of $-7.6$~T/K, again in excellent agreement to the present determination of $-7.5$~T/K. 
    This contrasts with the case for \Hparab. 
    As stated already in Ref. \onlinecite{Paulsen2021}, the anisotropy of \Hcl between the $b$ and $c$ axes at \Tc is very small.
    Hence \dHc should be roughly equal (and of the order of $-8$~T/K) in both directions, whereas the present experiment yield \dHc$=-34$~T/K along the \baxis. 
    In addition, the temperature dependence of \Hc along the \caxis and the \baxis further away from \Tc also cannot be reproduced by any combination of paramagnetic and orbital limitation (appendix \ref{appendix:fitParameters}).

    The most direct way to explain these anomalies regarding the value of the slope at \Tc (for \Hparab) and the temperature dependence of \Hc along all directions and notably  for \Hparaa, is a field-dependent pairing strength.
    This happens also in ferromagnetic superconductors \cite{Wu2017} and has been already proposed for \Hparab in \UTe \cite{RanScience2019, Knebel2019}. 
    If we call $\lambda$ the strong-coupling parameter controlling this pairing strength, it has to be field-dependent in all directions.
    We can rely on \Hcl, which is small enough for the effects of such a field dependence to have negligible impact \cite{Paulsen2021}, to fix the average Fermi velocities controlling \dHc for \Hparab, without the contribution of the field-dependent pairing strength. 
    Along the $a$ and $c$ axes, the agreement between \dHcl and \dHc shows that $\frac{d\lambda(H)}{dH} \approx 0$ in zero field (at \Tc).

    For the estimation of $\lambda(H)$ in the different field directions, there are general constraints which are model-independent. 
    First of all, along the \baxis, the discrepancy between \dHcl and \dHc can only be reconciled with an increase of the pairing strength: 
    increasing \dHc requires $\frac{d\lambda(H)}{dH}(H=0) > 0$.
    Hence, we expect an increase $\lambda(H)$ along the \baxis not only in the \hftransi phase, but  also in the \lftransi phase.
      
    For \Hparac, starting with $\frac{d\lambda(H)}{dH}(0) = 0$, the small positive curvature and the very linear behaviour also requires an increase of $\lambda(H)$, whatever the strong-coupling model and the mechanisms (orbital and/or not paramagnetic limitation).

    The situation is not as straightforward for \Hparaa, however, the most natural explanation is that the deviation from linearity observed very close to \Tc arises from a strong suppression of $\lambda(H)$, with an \Hc otherwise purely orbitally-limited (details in Appendix \ref{appendix:fitParameters} ).
    This scenario is consistent with the NMR results, yielding essentially no change of the Knight-shift at \Tc along the \aaxis  \cite{FujibayashiJPSJ2022}. 

    Following theses NMR results we also assume a negligible paramagnetic limitation in the (\lftransi) phase along the $b$ and $c$ axes. 
    The superconducting order parameter symmetry has an impact on the value of the initial slope \dHc, but influences little the temperature dependence of \Hc due to the orbital limitation.
    Hence, we can use the same strong-coupling model as in \cite{Wu2017,Knebel2019}, which mimics \Hc for a spin-triplet superconductor with a calculation for an \swave superconductor without paramagnetic limitation (taking $g=0$ in the equations). 
    Averaged Fermi velocities for the different field directions have been chosen so that the initial slopes \dHc for a field independent pairing strength have the values calculated from \dHcl (Table~\ref{tabS:ParameterTable} in Appendix \ref{appendix:comparison Hc2 Hc1}).
    
    \begin{figure}
    	\centering
    	\includegraphics[width=1\linewidth]{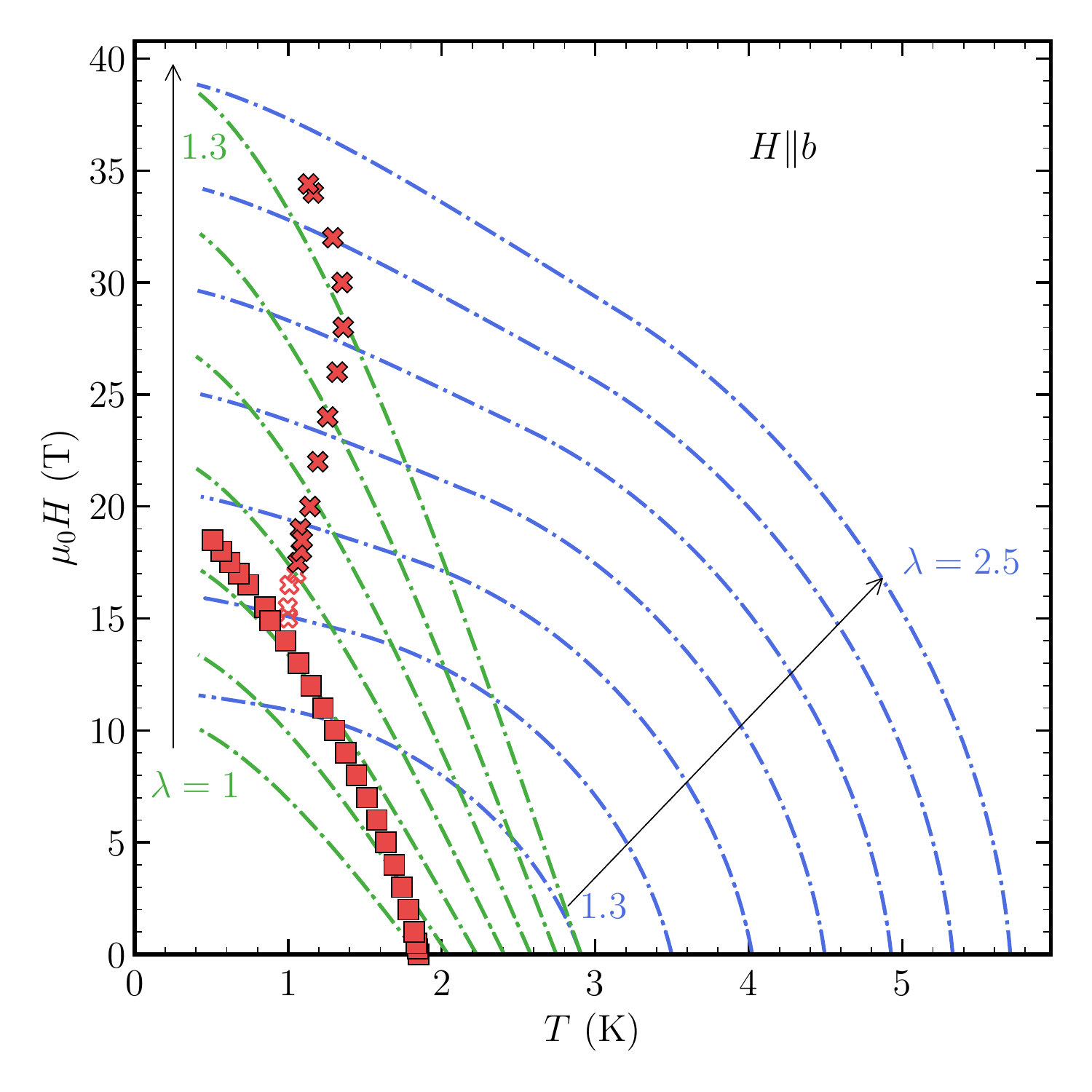}
    	\caption{Crosses and squares are data of \Hc for \Hparab. 
    	Dashed lines are the \Hc calculated for different fixed values of the coupling constant $\lambda$. 
    	The green lines correspond to calculation for $g=0$ by steps of $\Delta \lambda = 0.05$, and the blue lines for $g=2$ by steps of $\Delta \lambda = 0.2$.}
    	\label{fig:FitsHc2LambdaFixed}
    \end{figure}

\begin{figure}
	\centering
	\includegraphics[width=1\linewidth]{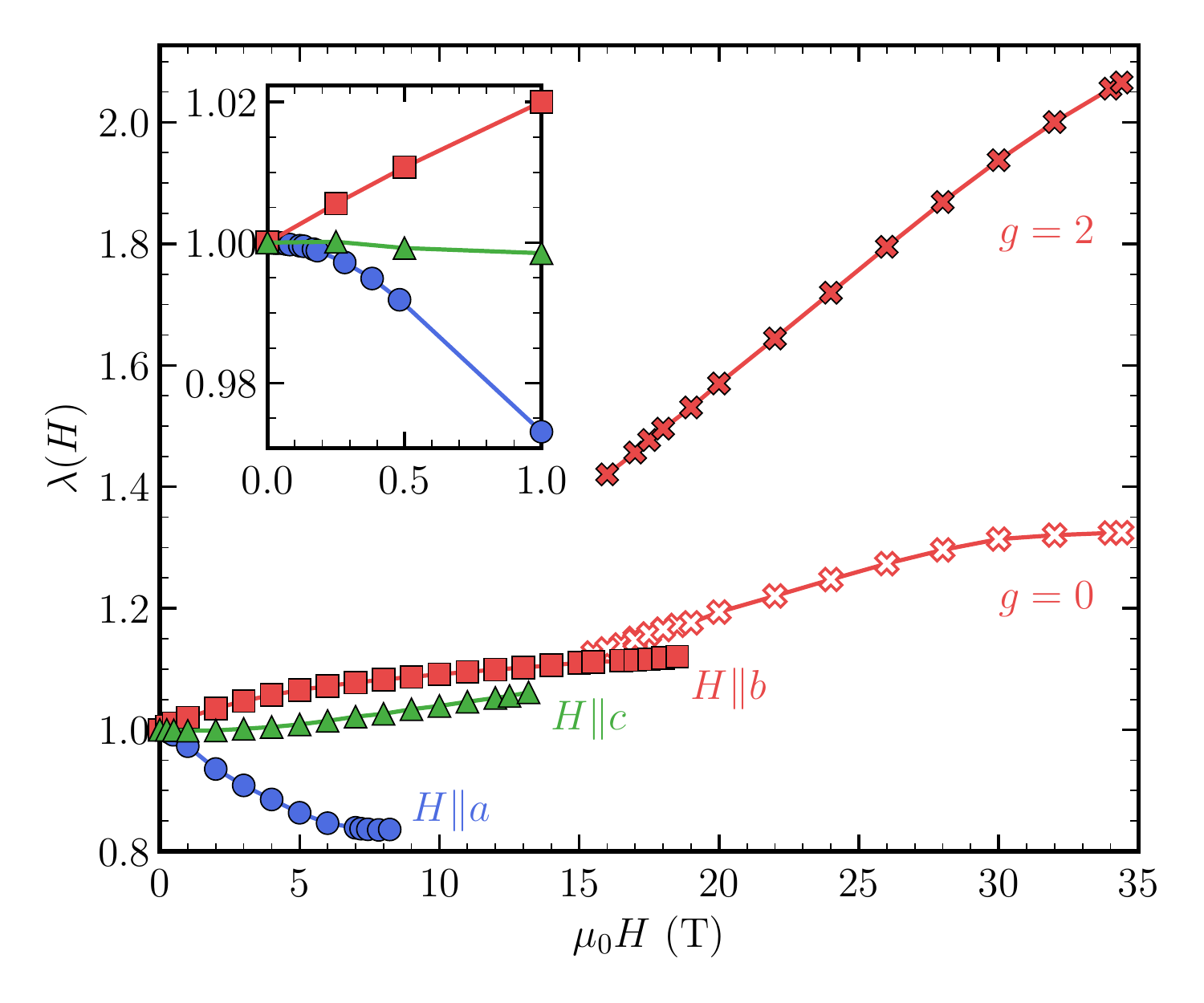}
	\caption{$\lambda(H)$ determined from \Hc along three cristallographic axes, measured on sample \diese2. $\lambda$ was set to 1 at 0~T, $g$ to 0 for the \lftransi phase. For the \hftransi phase, plain red crosses are \llambdaH determined with $g=2$, and empty red crosses are \llambdaH without paramagnetic limit ("$g=0$"). A zoom on fields below 1~T is shown in the insert.}
	\label{fig:Lambda_H}
\end{figure}

    The different calculations of \Hc for \Hparab with the two hypotheses,  spin-triplet superconductivity with no paramagnetic limitation, and spin-singlet with full paramagnetic limit at constant pairing strength, are presented in Fig.~\ref{fig:FitsHc2LambdaFixed}. 
    The large enhancement of the paramagnetic limit due to the correlated increase of the pairing strength $\lambda$ and the corresponding critical temperature in zero field is clearly visible.
    The deduced field dependence of the pairing strength $\lambda(H)$ is reported in Fig.~\ref{fig:Lambda_H} for the three crystallographic directions and the two models for \Hparab. 
    Fit parameters are reported in Table~\ref{tabS:ParameterFitHc2} in the Appendix \ref{appendix:fitParameters}, and the model is described in Appendix \ref{appendix:strong coupling model}.

   The increase of $\lambda$ along the $c$ and even $b$ axes is modest in the (\lftransi) phase, at most 10\%, whereas a strong suppression (factor 2 between \Tc and $T \rightarrow 0$) is required for \Hparaa.
    The inset of Fig.~\ref{fig:Lambda_H} shows that despite the strong suppression of the pairing for \Hparaa, in zero field $\frac{d\lambda}{dH}=0$. 
    This originates directly from our choice of matching the initial slope \dHc at constant $\lambda$ with the values deduced from \Hcl. 

	\subsection{\texorpdfstring{\Hc}{Hc2} in the  \texorpdfstring{\hftransi}{HF} phase}
	
     Theoretical models for the \hftransi phase have proposed a field-induced symmetry change of the order parameter \cite{IshizukaPRL2019, Shishidou2021}.
    The main idea is that for a spin-triplet superconducting state arising from ferromagnetic fluctuations along the easy magnetisation axis, at low fields, the \dvector should be perpendicular to the $a$ axis. Hence a $B_{3u}$ (or more generally $B_{3u}+iB_{1u}$ state under field \cite{AokiJPCM2022}) is favoured at low fields.
    By contrast, for strong fields along the \baxis, a rotation of the \dvector is expected toward a $B_{2u}$ state (or $B_{2u}+iA_{u}$), to minimise the component of the \dvector along the \baxis.
    Such a symmetry change would imply a phase trafsition somewhere between the low and high field regimes, which had not been detected until the present specific heat measurements.
    Nonetheless, this change of \dvector orientation alone could only explain that an initial paramagnetic limitation is exceeded thanks to the new orientation of the \dvector. It will not explain the positive \dHcT observed in the \hftransi phase between 15 and 30~T.
    
    Conversely, empirical explanations of the reinforcement of \Hc focused on field-induced enhancement of the pairing (positive $\frac{d\lambda}{dH}$)  \cite{RanScience2019, Knebel2019}, which is compatible with, but does not require an additional phase transition.
    Indeed, even the rather sharp upturns observed on \Hc extracted from electrical transport could be reproduced with a smooth continuous increase of $\lambda(H)$ \cite{AokiJPCM2022}.

    The present detailed specific heat measurements demonstrate 
    not only that there is a field-induced thermodynamic phase transition between two different superconducting states, but also that the pairing mechanisms driving these phases are likely different. 
    This is first seen from the phase diagram of Fig.~\ref{fig:PhaseDiagCp}, 
    where the appearance of the \hftransi phase appears very abruptly, marked by a sharp increase of \Tc(H) contrasting with the smooth continuation of the \lftransi phase.
    In addition, the specific heat anomaly for both phases is markedly different, with a very broad anomaly in the \hftransi phase whereas that of the \lftransi phase remains remarkably sharp (see Fig.~\ref{fig:RawCpLF_HF}).
    Both features differ strongly from the case of UPt$_3$ \cite {HasselbachPRL1989} or more recently of CeRh$_2$As$_2$ \cite{KhimScience2021, LandaetaPRX2022}, where \Tc(H) is always decreasing with field, and no change is observed on the shape of the specific heat transition along \Hc when switching from the low to the high field superconducting phases.
    
    In \UTe, there is clearly more than just a rotation of the \dvector between the \lftransi and \hftransi phases.
    Most likely, pairing is reinforced at high fields thanks to the emergence of a new pairing mechanism driven by the proximity to \Hm.
    Unfortunately, the nature of the magnetic fluctuations driving the metamagnetic transition is still unclear (appendix \ref{appendix:SuppCpHm}). 
    However, the phase diagram for \Hparab under pressure \cite{Knebel2020} shows that the \hftransi phase could well be the same as the pressure-induced higher \Tc superconducting phase \cite{Braithwaite2019,RanPRB2020,Aoki2020,LinNPJQuMat2020, ThomasPRB2021}.
    There are two main theoretical proposals for this "high-\Tc" pressure-induced phase.
    The first \cite{Shishidou2021} is that it is a $B_{2u}$ phase, having no component of the \dvector along the \baxis, with a pairing mechanism controlled by local ferromagnetic correlations. 
    The second is that it is a spin-singlet ($A_g$) phase \cite{Ishizuka2021}, induced by antiferromagnetic correlations becoming dominant over ferromagnetic fluctuations under pressure.
    
    The first proposal is the most natural one, explaining the different phases with a single pairing mechanism and transitions imposed by the Zeeman coupling of the \dvector with the applied field.
    In this framework, nothing should be changed to the procedure for the evaluation of the field dependent pairing strength $\lambda(H)$ between the \lftransi and \hftransi phases.
    The result is reported also in Fig.~\ref{fig:Lambda_H} (open red crosses), displaying a cusp at $\sim 15$~T, but a rather weak increase (30\% between zero field and \Hm) of the pairing strength, far from the factor $\approx 2$ observed on \CT at 1.8~K (Fig.~\ref{fig:Cp_T_FieldSweep}).

    Using the second proposal of a spin-singlet superconducting order parameter for the \hftransi phase seems paradoxical at first glance, but yields interesting results. 
    Once again, with a field-dependent pairing, the weak-coupling paramagnetic limit is easily exceeded thanks to the effective increase of the "zero-field" \Tc, and to strong coupling effects: 
    the ratio $\frac{H_{\rm{c2}}(0)}{T_{\rm{c}}}$ increases to a value of $\approx 4$ for $\lambda=1$, and exceeds 6 for $\lambda=2$ (see Fig. \ref{fig:FitsHc2LambdaFixed} above).
    
    This explains how a spin-singlet state could survive at $36$~T with a paramagnetic limitation set by a value $g=2$ of the gyromagnetic factor.
    Under pressure, this spin-singlet phase can explain the strong paramagnetic limitation observed along the \aaxis \cite{Knebel2019,Aoki2020a}.
    Here, it leads to a larger field dependence of the pairing strength (plain red crosses in Fig.~\ref{fig:Lambda_H}), required to overcome the saturation of \Hc (at fixed $\lambda$) due to the paramagnetic limitation. 
    Estimation of $\lambda(H)$ has been done for a gyromagnetic factor of 2, and (arbitrarily) with the same energy scale $\Omega$ than for the \lftransi phase, considering that both mechanisms should have similar characteristics in order to lead to similar critical temperatures.
    Using different values of $\Omega$ (but the same $g$-factor) changes little to the following analysis.
 
    Up to now, we have shown that the spin-singlet scenario for the HF phase is not unreasonable, requiring an increase of the pairing strength (from 1 to 2) which is compatible with the strong increase of electronic correlations observed on approaching \Hm as
    suggested by the strong enhancement of \CT in this field range.
    It is also in line with the change of pairing mechanisms between the \lftransi and \hftransi phases, supported by the difference in their respective specific heat anomalies , and by the marked positive value of \dHcT in the HF phase.
    We can however go a step further, showing that the spin-singlet state helps understanding quantitatively the change of the specific heat anomaly.
   
   \begin{figure}
	\centering
	\includegraphics[width=1\linewidth]{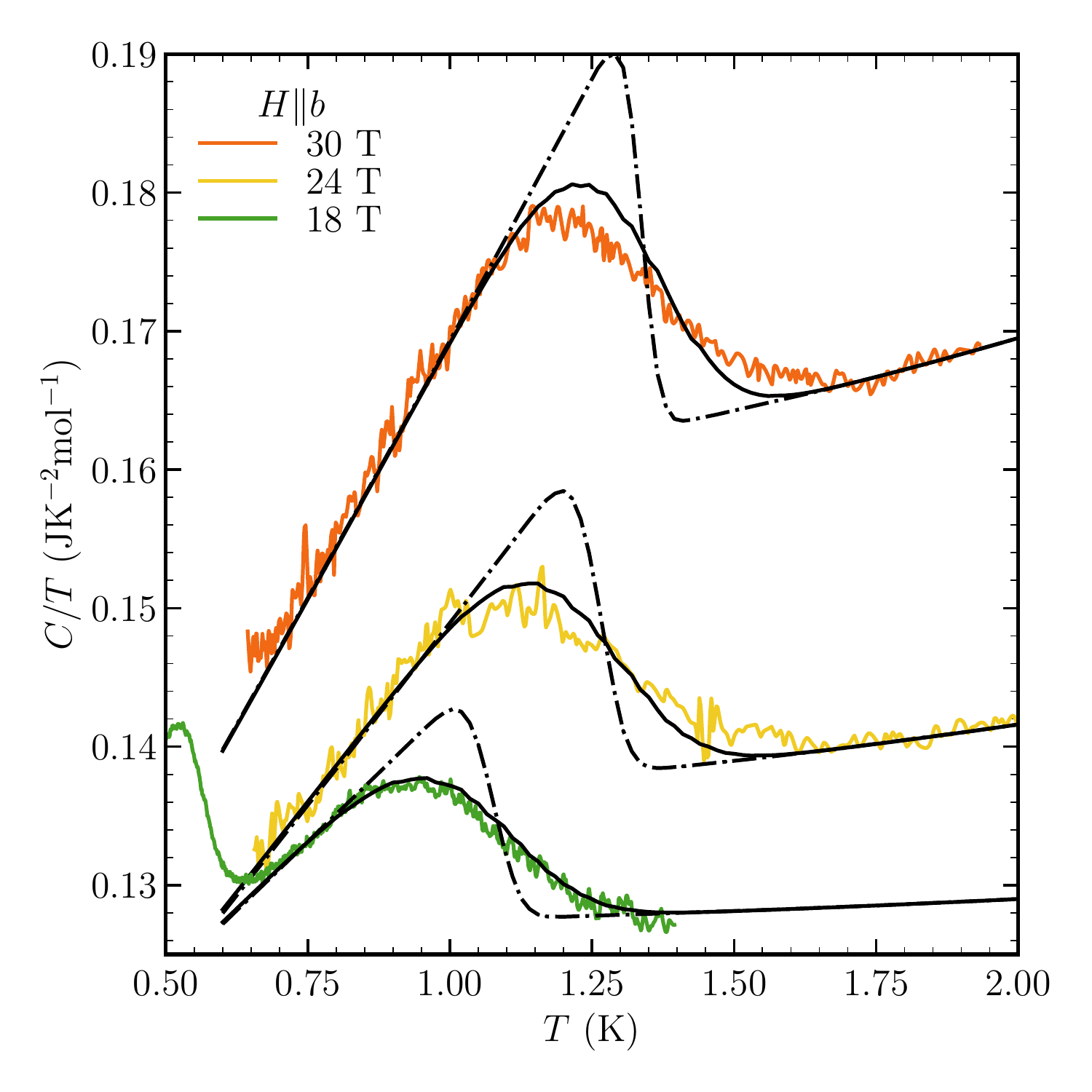}
	\caption{The \hftransi transition at 24 T, \Hparab, measured by specific heat. 
	Data (color) and fits (black lines) of the specific heat anomalies calculated for the spin-singlet (g=2 - continuous line) or spin triplet (g=0 - dashed line) superconducting state in the \hftransi phase.
	The broadening in the fits arises from the measured distribution of \Hm, however multiplied by a factor distribution 2.3. 
	For the spin-triplet state, even with a distribution of \Hm twice larger than given by our measurement, we fail to reproduce the broadening. 
	By contrast, for the same larger distribution of \Hm, the agreement is good for the spin-singlet scenario.}
	\label{fig:FitCpHF}
\end{figure}

    Indeed, with the field reinforced pairing the temperature of the superconducting transition in the HF phase depends on the applied field both through the usual orbital and possibly paramagnetic effects, and due to the field dependence of the pairing strength: $T_c = T_c(H,\lambda(H))$.
    So additional broadening of the superconducting transition may come from a field-dependent dispersion of $\lambda$.
    In the very likely hypothesis where the field increase of the pairing arises from the proximity to \Hm, a simple hypothesis is that $\lambda$ is a function of $\frac{H}{H_m}$. 
    
    Then, a dispersion of \Hm controlling the broadened specific heat anomaly reported in Fig.~\ref{fig:RawCpHm} for the metamagnetic transition, will translate into a distribution of \Tc, hence to a new mechanism for the broadening of the superconducting transition. 
    From the calculation of \Hc at fixed $\lambda$ used to extract the field dependence of the pairing, we can determine $T_c = \varphi\left(H, \lambda\left( \frac{H}{H_m}\right)\right)$. 
    This allows to determine the effect of the distribution of \Hm 
    on the specific heat anomalies of the \hftransi phase according to the two different determinations of $\lambda(H)$.
       
 	\begin{figure}[pt]
 		\centering
 		\includegraphics[width=1\linewidth]{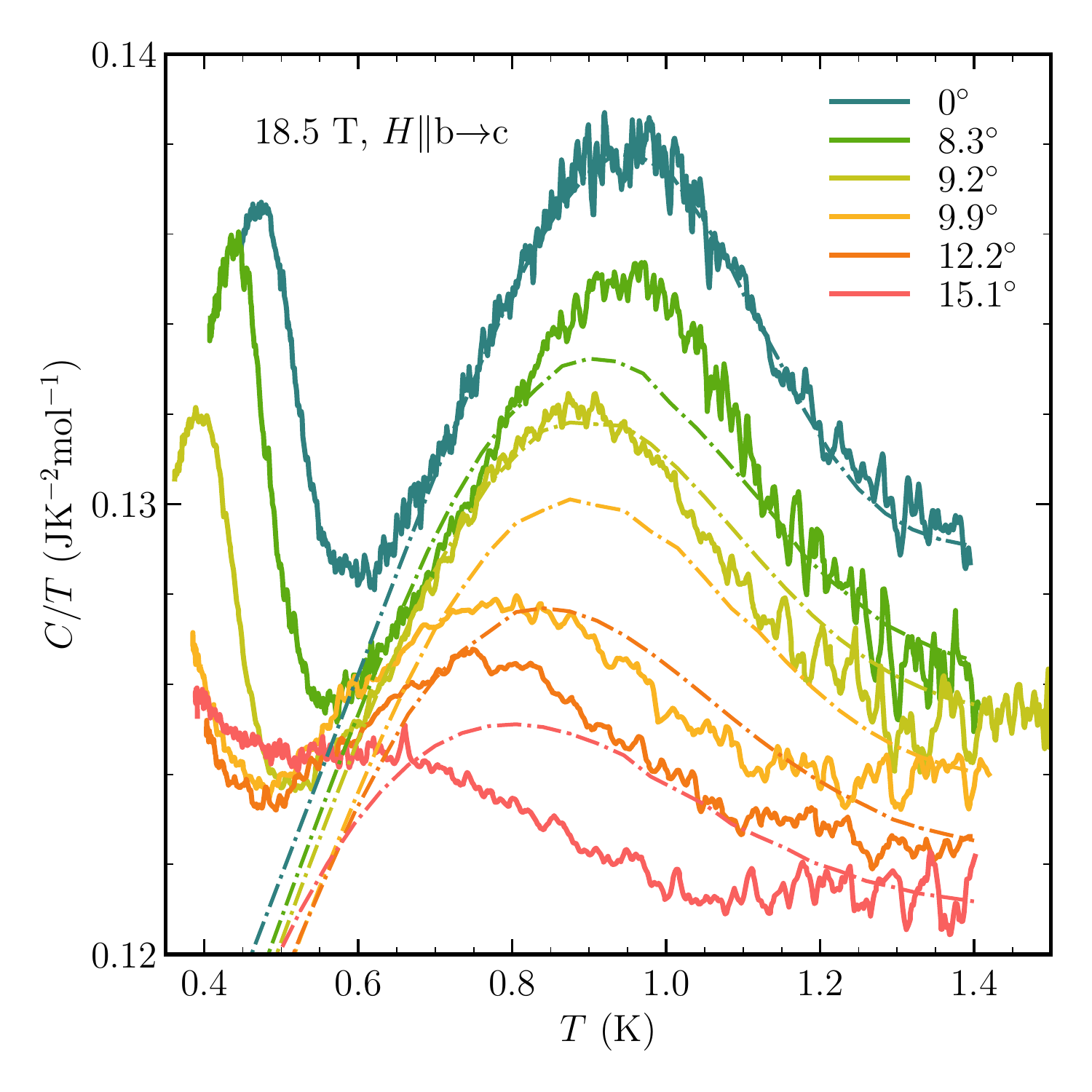}
 		\caption{Temperature dependence of $C/T$ measured on sample \diese3 \Hparab at 18.5~T for different angles in the ($b$,$c$) plane. Dashed lines are the transitions calculated from a distribution of \Hm, controlled by its angular dependence \cite{RanNatPhys2019} and a finite mosaicity of 3$^{\circ}$ in the sample.}
 		\label{fig:FitCp_theta}
 	\end{figure}
 	
    With this hypothesis in the spin-singlet case, the measured dispersion of \Hm of order 0.55$\%$, explains half the width of the superconducting transition. 
    As shown in Fig.~\ref{fig:FitCpHF}, the observed anomaly is well fitted all along the \Hc line of the \hftransi phase by doubling the measured \Hm dispersion. 
    By contrast,  it fails completely in the spin-triplet case. 
    A simple analysis (appendix \ref{appendix:Cp broadening}) reveals that the key advantage of the spin-singlet scenario is the much larger value of $\frac{\partial T_c}{\partial H}\vert_{\lambda}$ imposed by the paramagnetic limitation. 
    Within this scheme, the dispersion of \Hm is found to have a negligible influence on the \lftransi transition, so that it does give a first explanation for why the two superconducting phases could be marked by such different specific heat anomalies.
    
    Moreover, when the sample is misaligned in the ($b$,$c$) plane the transition shifts to lower temperatures and the amplitude of the jump decreases:
    see Fig.~\ref{fig:FitCp_theta} for an applied field of 18.5~T (more data in Appendix \ref{appendix:angular dependence}).
    At 15$^{\circ}$ the \hftransi transition almost disappears.
    Taking only into account the angular dependence of \Hm \cite{RanNatPhys2019}, and a hypothetical mosaicity of 3$^{\circ}$ in our crystal, we can roughly reproduce the huge broadening of the anomaly at finite angles, with the same dependence of \Tc on \Hm (dashed-dotted lines in Fig.~\ref{fig:FitCp_theta}), and otherwise a constant ideal specific heat jump. 
    This is another support for this explanation of the large broadening of the specific heat anomaly relying on the spin-singlet scenario.

\section{Perspectives}
\label{sec:discussion}
	A main result from this work is the requirement of a field-dependent pairing strength along all directions of the applied field, as shown by the anomalous temperature dependence of \Hc along the $a$, $b$ and $c$ axes already close to \Tc. 
	The strong decrease of the pairing strength along the \aaxis is reminiscent of the results on \ucoge along its easy magnetisation axis, and at first sight, it seems best compatible with a pairing mechanism involving true ferromagnetic fluctuations.
	Even subtle differences between the two systems are explained by such a mechanism. 
	For example in \ucoge, the slope of \Hc along the easy axis is strongly suppressed already at \Tc, showing that $\frac{d\lambda}{dH}$ is large and negative. 
	In \UTe for \Hparaa, comparison of \Hc and \Hcl showed that $\frac{d\lambda}{dH}\approx 0$.
    This 
    is consistent with the predictions for ferromagnetic and paramagnetic superconductors respectively,
    where $\frac{d\lambda}{dH}$ due to the suppression of "ferromagnetic" fluctuations is proportional to $M_z\frac{\partial M_z}{\partial H}$ ($M_z$ being the magnetisation along the easy axis) \cite{Mineev2017}. 
    So $\frac{d\lambda}{dH}$ at \Tc ($H=0$) should be zero in paramagnetic systems (like \UTe), and non-zero in ferromagnets below the Curie temperature, as long as the magnetization is not completely saturated.
	
	There are also several theoretical studies exploring other mechanisms leading also to spin triplet pairing, like finite momentum magnetic fluctuations \cite{KreiselPRB2022}, or only local ferromagnetic correlations within a unit cell \cite{Shishidou2021}.
	The field dependence of such mechanisms hasn't been explored.  
	However, the Fermi surface instability observed at 6~T along the easy axis \cite{NiuPRL2020} could play a key role if $\bm{Q}$-dependent pairing is important.
	Hence, even though ferromagnetic fluctuations are a likely mechanism for the \lftransi phase of \UTe, we cannot exclude that future investigations of these alternative mechanisms could also yield satisfying explanations of the present measurements.
	
	Concerning the results along the hard \baxis, the pertinence of the comparison of \UTe with the ferromagnetic superconductors becomes more suspicious. 
	For this field direction, the main result is
	the existence of two different bulk superconducting phases already at ambient pressure.

	In the course of revising this paper, we became aware of a new work on high-field NMR, recovering a similar phase diagram as reported here, but identifying the \hftransi phase as a spin-triplet $A_u+iB_{2u}$ state \cite{KinjoArXiv2022}. 
	This arises from Knight-shift measurements in the \hftransi phase, showing no detectable changes across \Tc. 
	We note that due to the field dependence of the pairing strength, these measurements in the \hftransi phase are all performed at values of $\frac{H}{H_{c2}^{eff}(0)}$ close to 1, where $H_{c2}^{eff}(0)$ is the effective value of \Hco, corresponding to the value of the pairing strength $\lambda(H)$ at the field $H$ of the measurement (see Fig.\ref{fig:FitsHc2LambdaFixed}). 
	At these large field values (with respect to $H_{c2}^{eff}(0)$), there is little change to expect for the Knight shift,  whatever the spin-state.
	
	More recently, a similar phase diagram was also reported from resistivity and ac susceptibility measurements \cite{SakaiArXiv2022}. 
	However, the anomalies revealing the transitions in both of these works could not be used to track the broadening of the transition in the  \hftransi phase as revealed by our specific heat measurements. 
	This change of the specific heat anomaly is a unique case showing that this new superconducting phase does not arise from a simple change of symmetry like in UPt$_{\rm{3}}$, or from a rotation of the \dvector: 
	it has to arise from a new pairing mechanism strongly reinforced on approaching \Hm.
	We have found support for a paramagnetically limited \Hc in the \hftransi phase, hence for a spin-singlet superconducting phase, as it can explain a large part of the strong broadening of the specific heat anomaly in the \hftransi phase, and the still increased broadening when turning away from the \baxis in the ($b$,$c$) plane.

	We became also aware of a theoretical work, proposing an alternate explanation for the phase diagram of \UTe, without field reinforced pairing \cite{YuPRB2022}: 
	admitting the existence of a transition line between a \lftransi and \hftransi superconducting phases, the "deep" of \Hc at 15T would be caused by thermal superconducting fluctuations boosted by a spatial distribution of critical temperatures in the sample. 
	This interesting scenario should now be explored against the present precise determination of the transition lines, and the change of the specific heat anomaly.
	
	Other open questions still remain. 
	A first one concerns the large difference between the irreversibility line observed on the magnetostriction and the specific heat anomaly: 
	it suggests a strong increase of the reversibility region in this phase, like the fact that the resistive transition only goes to zero when the bulk transition ends. 
	This is similar to previous observations in \ucoge \cite{Wu2018}. 
	It is consistent with the large reversibility region observed in the \hftransi phase between \Hc and the irreversibility field, and with an already weak pinning in the \lftransi phase \cite{Paulsen2021}.
	However an explanation for the additional suppression of the vortex pinning at high fields is still awaited.

	Another important point concerns the order of the different transition lines and the precise slopes of the lines at the tricritical point.
	Indeed, as for CeRh$_2$As$_2$ \cite{KhimScience2021}, in the case of a spin-triplet to spin-singlet transition, a first order transition is expected.
	In our specific heat measurements, we did not detect any hysteresis effects. 
	The only features visible in Fig.~\ref{fig:Cp_Jump_Width} are a smaller jump of \CT for the \lftransi to \hftransi transition than along \Hc in the \lftransi phase, as well as a slightly smaller width.
	This slight narrowing leaves open the possibility that the transition from \lftransi to \hftransi phases could be weakly first order.
	There are many cases in condensed matter physics, where first order transitions lead to negligible hysteresis:  see e.g. Ref.~\onlinecite{GuillouNatCom2018}, or the old example of the $^3$He melting curve. 
	This point requires however further experimental investigations.  

    If this transition is first order, of course, the question of the tricritical point is solved.
    If it is not, it remains an issue to determine if there is an additional transition line within the \lftransi superconducting phase, and whether or not the three transition lines determined in this work join with different slopes, or if the \Hc line has no change of slope (only a very strong positive curvature) at the tricritical point.

    The entrance into the \hftransi phase along \Hc cannot be done in a mixed singlet-triplet superconducting phase \cite{YanasePrivate}:
    it would require, like for the chiral superconducting state  \cite{Hayes2021}, a double transition which is not observed. 
    
   Theoretical works based on microscopic calculations have predicted that the interplay between ferromagnetic and antiferromagnetic fluctuations \cite{Xu2019, Ishizuka2021} could lead to competing pairing interactions \cite{Ishizuka2021}. 
   This competition could be central both for the pressure \cite{Ishizuka2021} and the field-induced phases of \UTe. 
   At ambient pressure, at the opposite of CeRh$_2$As$_2$, it could lead to a paradoxical spin-singlet phase at high fields, possibly driven by strong antiferromagnetic correlations on approaching the metamagnetic transition.
   Under pressure, this high-field phase would become the highest \Tc phase with the lowering of the metamagnetic field along the \baxis, whereas the pure spin-triplet phase would survive essentially for large enough fields along the easy \aaxis.
   \UTe is probably the first system where two competing pairing mechanism of similar strength exist, and can be arbitrarily tuned by field or pressure:
  it is an ideal case to challenge theoretical models and understand which conditions allow for the emergence of spin-triplet superconductivity. 

\section{Acknowledgements}
We thank Y. Yanase, K. Miyake, M. Houzet, M. Zhitomirsky and V. Mineev, for very fruitful discussions, and  A. Miyake and S. Imajo for kindly providing their data points. 
We got financial support from the CEA Exploratory program TOPOHALL, the French National Agency for Research ANR within the project FRESCO No. ANR-20-CE30-0020 and FETTOM ANR-19-CE30-0037,  
and from the JSPS programs KAKENHI (JP19H00646, JP20K20889, JP20H00130, JP20KK0061, JP20K03854, JP22H04933).
We acknowledge support of the LNCMI-CNRS, member the European
Magnetic Field Laboratory (EMFL), and from the Laboratoire d'excellence LANEF (ANR-10-LABX-51-01).

\appendix

\section{Thermal conductivity \texorpdfstring{\Hparab}{H parallel b}}
\label{appendix:thermalconductivity}

\begin{figure}
\includegraphics[width=0.85\columnwidth]{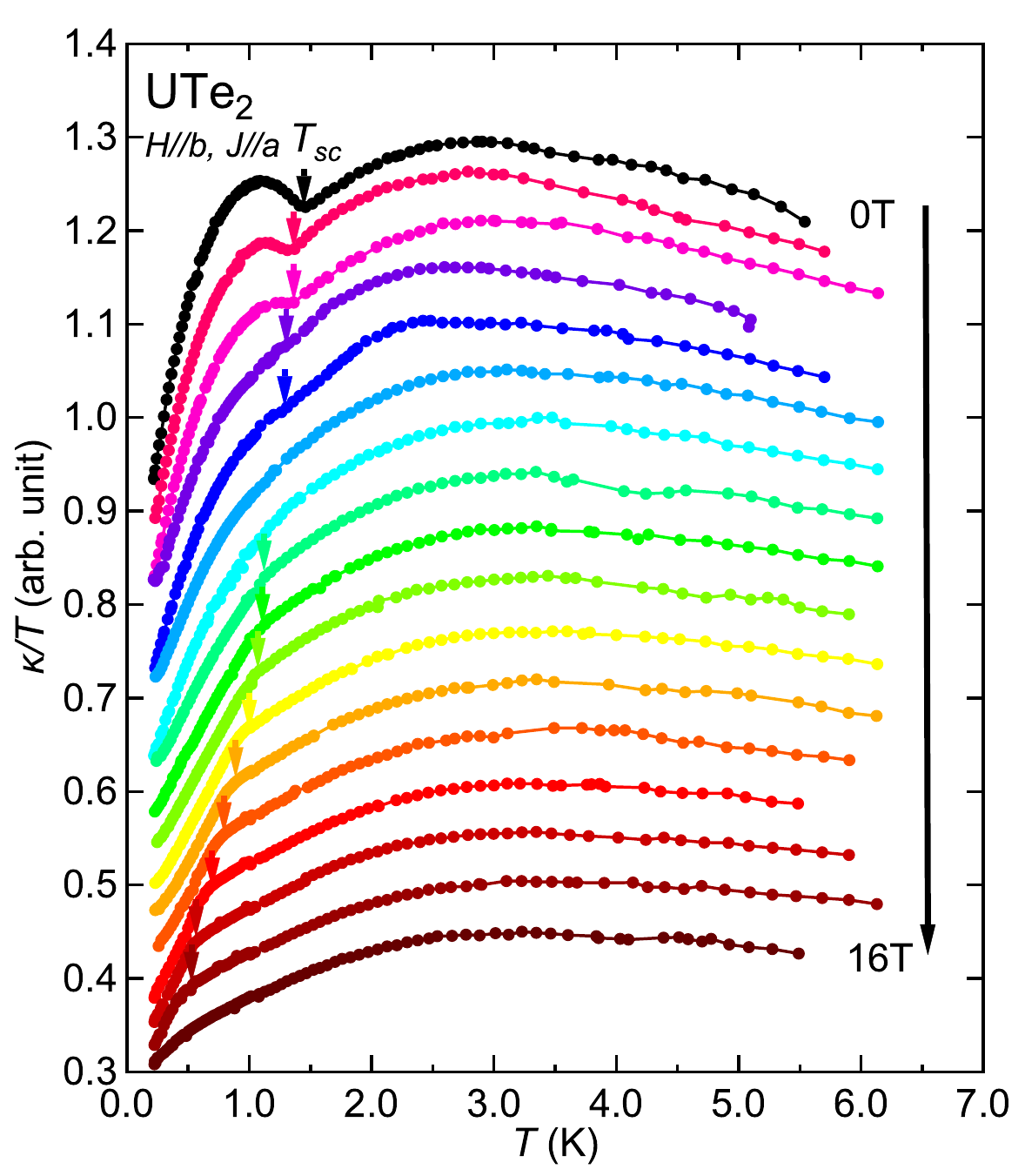}
\caption{Temperature dependence of $\kappa/T$ in \UTe with \Hparab between 0 and 16~T (every Tesla). Traces have been shifted for clarity. The superconducting temperature \Tc is represented by vertical arrows.}
\label{figS:KappaUTe2}
\end{figure}

Thermal conductivity $\kappa$ measurements have been performed on a sample from the same batch as sample \diese1.
The temperature sweeps have been measured on a home-made dilution refrigerator with a base temperature of 100~mK and a superconducting magnet with field up to 16~T using a standard "one heater-two thermometers" setup. The temperature dependence of $\kappa/T$ for different magnetic fields up to 16~T is represented in Fig.~\ref{figS:KappaUTe2}. $\kappa/T$ shows a broad maximum at around 3~K. At low field, there is a clear increase in $\kappa/T$ just below \Tc, which is suppressed by increasing the field. Such an increase below \Tc has also been observed in other systems such as CeCoIn$_5$ or YBCO, and attributed to a suppression of the inelastic scattering of heat carriers (electrons and phonons, respectively) by the opening of the superconducting gap. In the case of \UTe, the enhanced conductivity below \Tc is likely due to an increase of the electronic mean free path due to the opening of the superconducting gap. At higher field ($\mu_0 H>3$~T), entrance in the superconducting state is marked by a rapid decrease of the thermal conductivity, usually attributed to Andreev scattering on the vortex cores. 

\section{Thermal expansion for field \texorpdfstring{\Hparab}{H parallel b}}
\label{appendix:thermalexpansion}

As discussed in the main text, linear magnetostriction measurements are sensitive to pinning forces. 
Indeed, trapped flux imposes a field gradient at the sample surface, perpendicular to the applied field, controlled by the critical current density.
Pinning force trapping magnetic flux lines on the lattice should be balanced by equal but opposite forces acting on the lattice. 
Hence, it can be shown that the length change of the crystal $\Delta L_b/L_b$ is proportional to $\frac{H\Delta M}{E}C_\nu$, where $\Delta M$ is the non-equilibrium part of the magnetisation, $E$ the Young modulus, and $C_\nu$ a constant depending on the Poisson's ratio \cite{Eremenko1999}.

We also performed longitudinal thermal expansion  measurements (length change in the field direction) on sample \diese4 as a function of temperature at fixed magnetic fields, which also show similar effects due to vortex pinning. The measurements were performed up to 29.5~T along the \baxis in the high magnetic field laboratory LNCMI, and in addition using a superconducting magnet up to 13~T in the Pheliqs laboratory.  

\begin{figure}
\includegraphics[width=1.05\columnwidth]{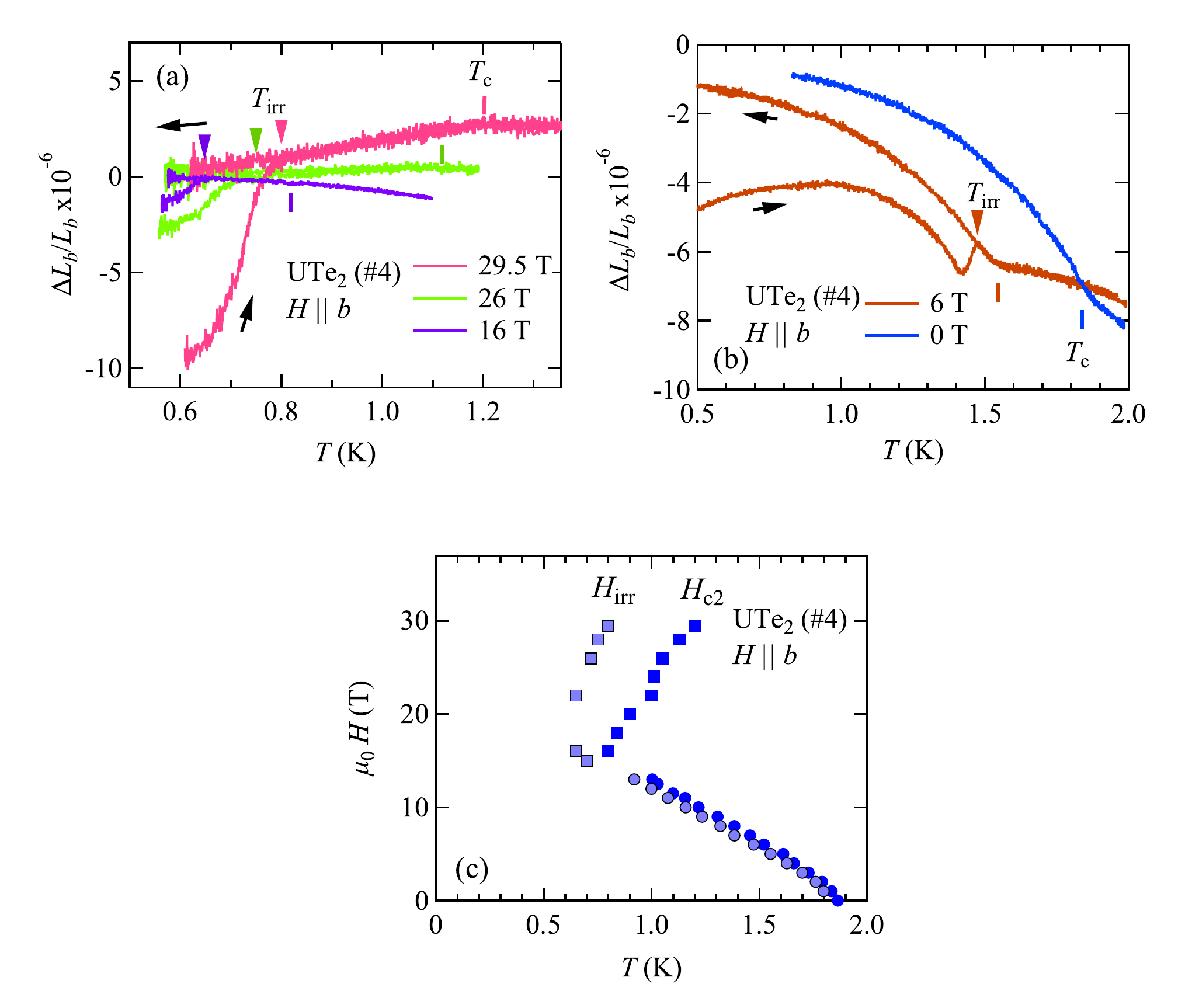}
\caption{Temperature dependence of the linear thermal expansion $\Delta L_b/L_b$ in \UTe with \Hparab at different magnetic fields measured in (a) the LNCMI Grenoble, and (b) using a superconducting magnet. The arrows indicate the direction of the temperature sweep, see text for details. The temperature of the irreversibility field $H_{\rm irr}$ and the upper critical field \Hc are indicated by arrows and vertical bars, respectively.
Panel (c) shows the phase diagram obtained from the thermal expansion measurements. While at low field ($H < 15$~T) $H_{\rm irr} \approx H_{\rm c2}$, at high fields the irreversibility line is much lower in temperature. }
\label{figS:RawDilatation_PhaseDiag}
\end{figure}

Figures~\ref{figS:RawDilatation_PhaseDiag} (a) and (b) show the temperature dependence of the relative length change along the $b$ axis $\Delta L_b/L_b$ for different magnetic fields. 
The data displayed in panel (a) are obtained by cooling from the normal state to the lowest temperature in a field 2~T below the target final field, then increasing the field at the lowest temperature up to the final field, heating at that field above $T_c$ and cooling again. 
Increasing the field at low temperature induces a non-equilibrium magnetisation inside the sample due to the flux pinning, while an equilibrium flux distribution occurs in the final-field cooled sweep. 
$H_{\rm irr}$ marks the onset of the irreversible magnetisation regime. 

The data in panel (b) are measured in a superconducting magnet using a dilution refrigerator. 
They are obtained from a similar procedure however starting by zero field cooling, then ramping the field up to its target value, heating up to the normal state and measuring the field-cooled length change. Again, the irreversibility line can be clearly determined. 

In panel (c) we show the temperature dependence of the upper critical field \Hc and of the irreversibility field $H_{\rm irr}$ determined from the cycles described above. Data for $H \leq 13$~T are from the experiments performed in a superconducting magnet, while for $H > 13$~T they stem from the high-field experiments in LNCMI. While in the LF superconducting phase $H_{\rm irr}$ is lower but very close to  $H_{\rm c2}$, in the HF superconducting phase $H_{\rm irr}$ is far lower in temperature than \Hc, indicating a wide reversible region (with low pinning ) behind \Hc in this state.  

\section{specific heat: normal phase}
\label{appendix:Cp normal phase}

 In general, the leading term of the specific heat of a heavy-fermion system at low temperatures ($T \ll T_F$, with $T_F$ being the effective Fermi temperature), far from any quantum criticality, is linear in temperature: $C \propto \gamma T$. 

\begin{figure*}[htb]
	\centering
	\includegraphics[width=2\columnwidth]{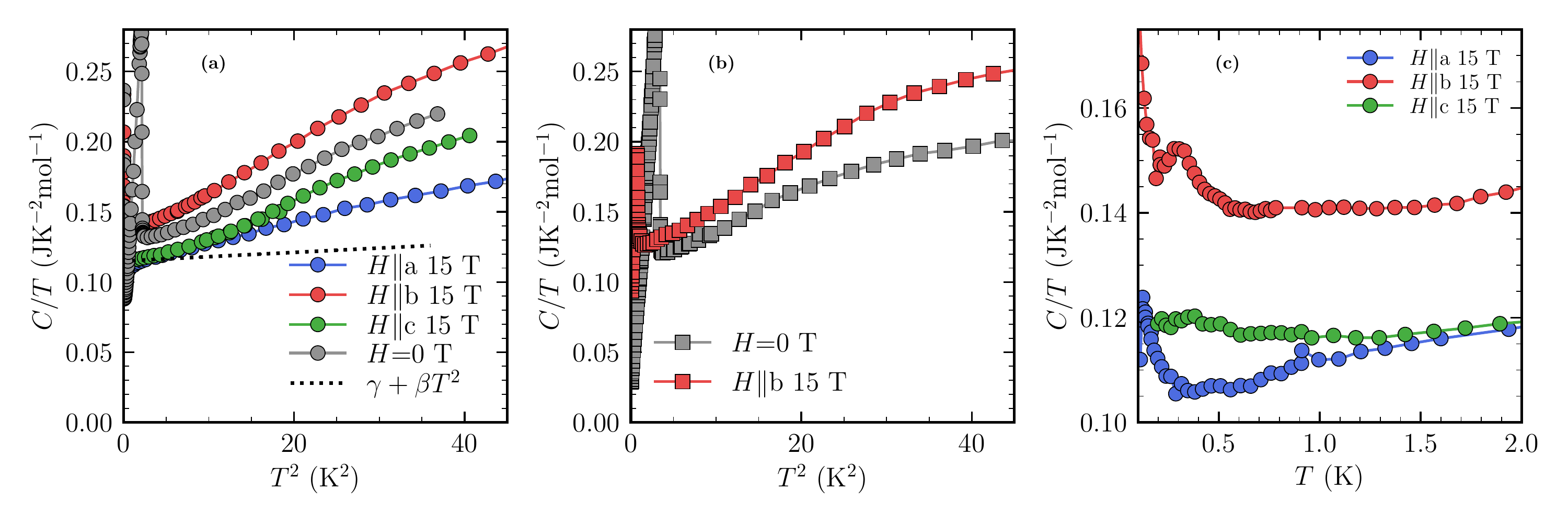}
 	\caption{Temperature dependence of $C/T$ (sample \diese1) for fields applied along the three crystallographic directions. 
 	(a) $C/T$ as a function of $T^2$, measured on sample \diese1:  
 	no linear behaviour is seen. 
 	At 15~T for \Hparaa, the temperature dependence is drastically suppressed compared to measurements at 0~T. 
 	whereas it is slightly larger for \Hparab, 
 	The doted line is the sum of a constant Sommerfeld term and a phonon contribution estimated from a Debye temperature deduced from high-temperature measurements \cite{Willa2021a}. 
 	(b) Same data for \Hparab on sample \diese2: 
 	the anomalous magnetic contribution seems more pronounced than for sample \diese1.
 	(c) $C/T$ at low temperatures at 15 T along the three axis measured on sample \diese1. The superconducting transition at $\sim 0.5$~K remains visible for \Hparab.}
	\label{figS:ct_vs_T2}
\end{figure*}

    The Sommerfeld coefficient $\gamma$  is proportional to the density of states at the Fermi level, which is, strongly renormalised compared to the free electron gas. 
    In UTe$_2$, at low temperature, the competition between different natures of magnetic fluctuations (ferromagnetic or antiferromagnetic) as well as the role of valence fluctuations due to the interplay between U$^{3+}$ and U$^{4+}$ configurations may occur. 
   The situation is even more complex in this system, because electronic correlations play the unusual role of driving the system from an insulating toward a metallic state \cite{Aoki2019, Xu2019, Shick2019, Ishizuka2019}.

\begin{figure}
   	\includegraphics[width=1\columnwidth]{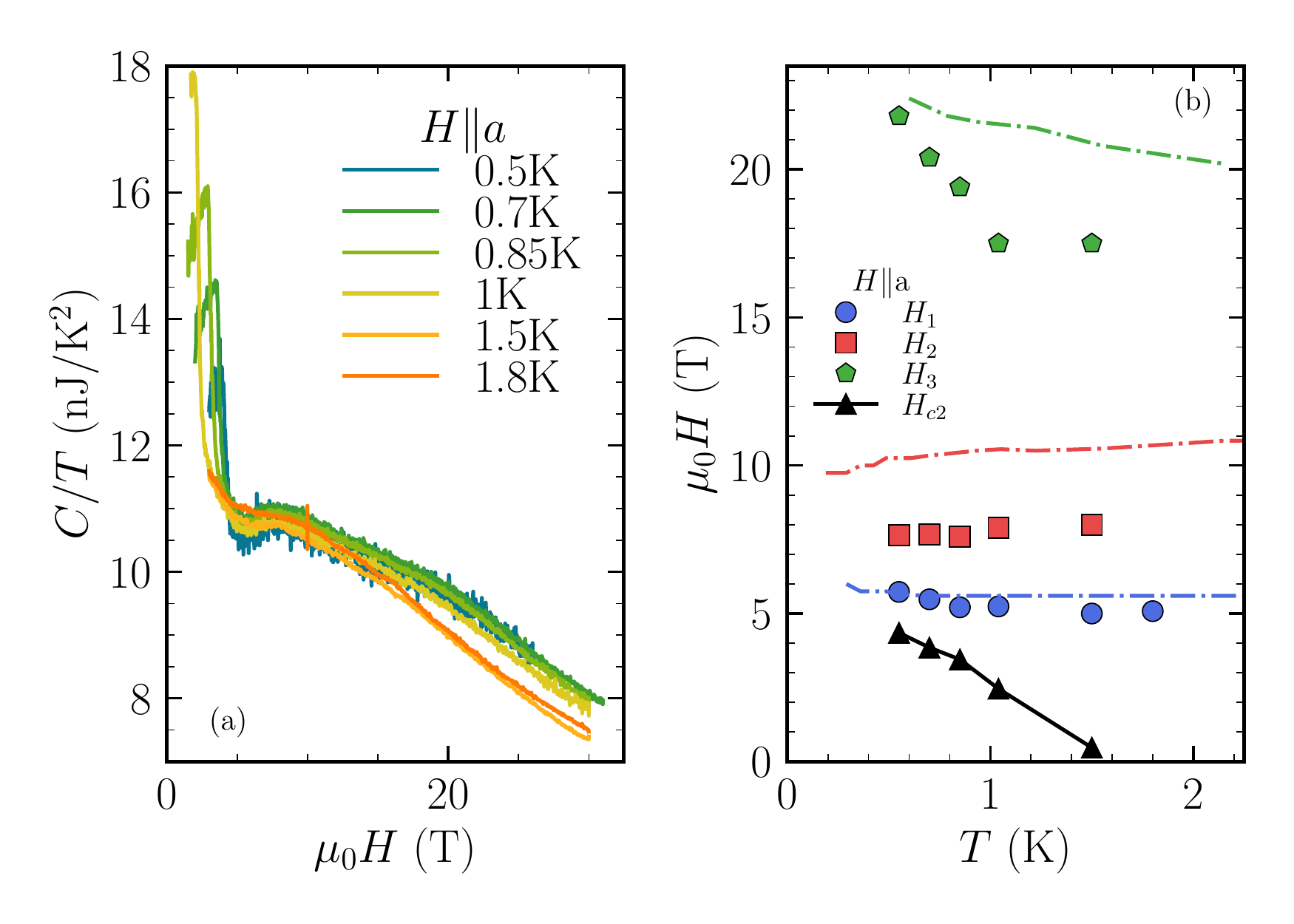}
	\caption{(a)$C/T$ measured for field sweeps at different temperatures with \Hparaa. 
	Measurements done on a sample coming from the same batch as sample \diese1, with a \Tc of 1.5 K.
	(b) Phase diagram \Hparaa up to 31 T. Black triangles represent the superconducting transitions, blue circles are the minima of $C/T(H)$. 
	Red squares represent the maxima of $C/T(H)$. 
	Green pentagons represent the inflexion point of $C/T$ observed on field sweeps.
	The dashed lines are the corresponding transitions measured by thermoelectric power in ref~\cite{NiuPRL2020}.}
	\label{figS:CT_vs_H_para_a}
   
\end{figure}

    As a consequence, even close  to \Tc,  $C/T$ is not the sum $\gamma + \beta T^2$ with $\beta T^2$  the phonon contribution far below the Debye temperature. An additional  contribution is observed, likely coupled to the "Schottky-like" anomaly detected at $T^\star \approx 12$~K \cite{Willa2021a}. 
    In Fig.~\ref{figS:ct_vs_T2}(a,b) shows $C/T$ vs $T^2$ for sample \diese1 and sample \diese2. 
    The phonon contribution has been calculated following the Debye model with a Debye temperature of $\theta_D = 180$~K \cite{Willa2021a}. 
    Clearly, the phonon contribution is low compared to the measured specific heat and cannot reproduce the strong temperature increase of \CT at zero magnetic field. 
    Under magnetic field, the anomaly at $T^\star$ shifts to higher temperatures for \Hparaa and lower temperatures for \Hparab. 
   Accordingly, \CT strongly depends on the magnitude of the applied field and on its direction. 
   Hence the low temperature Sommerfeld coefficient $\gamma$ cannot be determined properly from a direct analysis of the temperature dependence of \CT. 
   We tried, as done previously \cite{MiyakeA2019,Imajo2019}, to follow the field evolution of \CT at the lowest possible temperature, so as to be as close as possible to the value of $\gamma$.
    At 15~T for \Hparaa, $C/T$ decreases with temperature (down to 0.3~K), which is quite unusual (Fig.~\ref{figS:ct_vs_T2}(c)). 
    Thermoelectric power measurements have revealed the presence of several Lifshitz transitions in this field direction \cite{NiuPRL2020}. 
    In the main text, the insert of Fig.~\ref{fig:Cp_T_FieldSweep} 
    shows that the field dependence of $C/T$ at 1.8~K has a minimum followed by a maximum close to 5~T and 9~T respectively.
    Moreover, field sweeps have been performed up to 31~T for \Hparaa , on a sample with a \Tc of 1.45~K coming from the same batch as sample \diese1. Measurements of \CT are displayed in Fig.~\ref{figS:CT_vs_H_para_a}(a). 
    A minimum (around 6 T) followed by a maximum (around 8 T) are visible for fields above the superconducting transition. 
    At higher fields, a change of slope occurs for fields between 17~T and 22~T, depending on the temperature, where the magnetisation along the \aaxis starts to saturate \cite{MiyakeA2019}.
    We can follow these three anomalies in addition to the superconducting transition, and establish the phase diagram shown in Fig.~\ref{figS:CT_vs_H_para_a}(b). 
    The temperature dependence and the order of magnitude of the field where they occur are similar to those obtained from thermoelectric power measurements. The lower transition (at around 5-6~T) was clearly identified as a Lifshitz transition, the origin of the two others is less clear \cite{NiuPRL2020}. 
     Regarding the present specific heat data, the origin of the pronounced maxima of \CT observed at $\sim 1$~T for \Hparaa and $\sim 1.5$~T for \Hparac (Fig.~\ref{fig:Cp_T_FieldSweep} inset in the main text) is also not identified.

\section{Specific heat: metamagnetic transition}
\label{appendix:SuppCpHm}
\begin{figure}
        \centering
	\includegraphics[width=1\columnwidth]{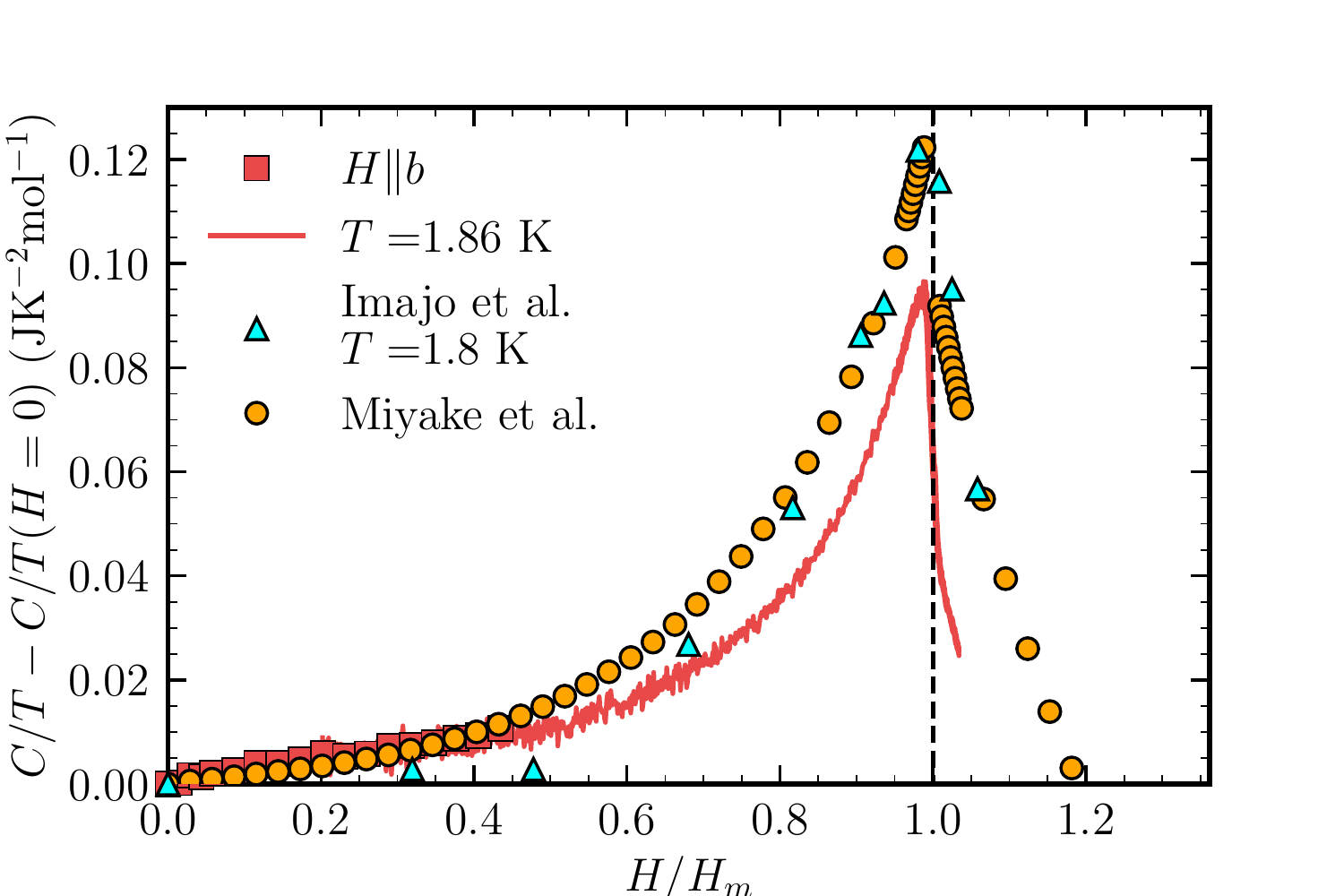}
\caption{Cyan triangles: specific heat measurements done in pulsed fields in ref~\cite{Imajo2019}. Orange circles: $\gamma(H)-\gamma(H=0)$ determined from the magnetisation measurements through thermodynamic relations in ref~\cite{MiyakeA2021}. 
Red line: our measurements.}
\label{figS:CompCT_Imajo}
\end{figure}

    As regards the metamagnetic transition, we could measure precisely the field dependence of the anomaly, and check that it did not depend on field sweep rate, varied between $\pm 350$ and $\pm 50$~Gauss/sec. 
    At this sweep rate, we also did not detect any magnetocaloric effect.
    Hence, the present continuous field sweep measurement show unambiguously that there is a jump at \Hm, slightly broadened by a distribution of \Hm. At 0.7~K, \CT decreases 55~mJK$^{-2}$mol$^{-1}$ and the width of the transition is 0.25~T. 
    This distribution of \Hm possibly comes from the strong sensitivity of \Hm to pressure \cite{Knebel2019,MiyakeA2022} and most likely to stress (crystal defects could then generate this \Hm distribution). 
    Figure~\ref{figS:CompCT_Imajo} shows the comparison of $C/T - C/T(H=0)$ near $T=1.8$~K for field along the \baxis determined from our experiment performed on sample \diese3 and experiments performed in pulsed field. 
    Ref.~\onlinecite{Imajo2019} reports specific heat experiments in highly stabilized fields, using the long pulsed fields facility at ISSP. 
    In Ref.~\onlinecite{MiyakeA2021} the Sommerfeld coefficient $\gamma$ has been determined from the magnetization measurements $M(T)$ under pulsed fields, using Maxwell's relation for $H \ne H_m$ as 
    $(\partial \gamma/\partial H)_T = (\partial ^2M/\partial T^2)_H$, and  using the Clausius-Clapeyron relation for the first order transition: 
    $\mu_0 dH_m/dT = -\Delta S/\Delta M$ \cite{MiyakeA2021} to get the jump 
    $\Delta\gamma = \Delta S/T$ at \Hm. 
    The last analysis indicated a discontinuous jump of  $\Delta \gamma = -30$~mJK$^{-2}$mol$^{-1}$ at $H_M$ for \Hparab, which is lower than that obtained in the present experiment. However, despite some quantitative differences (e.g. the absolute variation of $C/T - C/T(H=0)$ is larger in both pulsed field experiments) the general behaviour is similar: an increase with $H$ when approaching \Hm and a drop at \Hm followed by a strong decrease. A similar field dependence has been observed for the $A$ coefficient of the electrical resistivity, albeit without a clear jump above  \Hm\cite{Knafo2019, Knafo2021}. 

To extract the position and the width of the specific heat at the metamagnetic transition, it has been fitted like a broadened second-order phase transition, with the same Gaussian model as the superconducting transition.
Fields replace temperatures, and \Hm replaces \Tc.
\Hm for the up sweeps are roughly constant at 34.75 T, and for the down sweeps \Hm increases with the temperature. This goes along with the trend toward a closing of the hysteresis cycles with increasing temperatures, in agreement with the observations from resistivity measurements \cite{Knafo2019,NiuPhysRevRes2020}. 
The width of the transitions increases abruptly from 0.24 T to 0.45 T between 0.7 K and 0.97 K and then stays constant with temperature. 
The jump of \CT at \Hm strongly decreases on cooling from  0.97 K to 0.7 K, otherwise, above 0.97~K,  it decreases on warming. 
This anomaly at 0.7~K might be due to the presence of the \hftransi transition, which is wide enough in field and temperature to influence the drop of \CT at \Hm.

An important issue regarding this metamagnetic transition, notably for the discussion of the pairing mechanism responsible \hftransi superconducting phase is the nature of the magnetic correlations associated with \Hm. 
The question is presently open.
Indeed, Inelastic neutron experiments at such large fields are still not available. 
    If the metamagnetic transition would occur along the easy axis like in UCoAl \cite{AokiJPSJUCoAl2011}, the fluctuations would most likely be ferromagnetic, but it appears in \UTe along the hard axis. 
    Other criteria like the value of the Wilson ratio, claimed to support ferromagnetic fluctuations at low fields due to its large value \cite{Willa2021a} would be of no help close to \Hm: 
    calculating this ratio on approaching \Hm from raw data is certainly questionable in such a complex multiband system with local moment contributions. 
    Moreover, this calculation would yield much smaller values than along the \aaxis at low fields: 
    the susceptibility $\frac{\partial M}{\partial H}$, is at least six times smaller \Hparab than for \Hparaa, and the specific heat increases almost by a factor 2 between zero field and \Hm, suppressing the Wilson ratio deduced for \Hparaa by at least a factor 10.
    Arguments for antiferromagnetic fluctuations exist, but are far from rock solid:  
    besides the results from inelastic neutron measurements at low fields, we can note that the scaling relation found in many antiferromagnetic systems between the temperature of the maximum of the susceptibility $T_{\rm{\chi_{max}}}$ (35~K) and the value of \Hm \cite{Aoki2013} (33-35~T) is well obeyed in \UTe.

\section{Specific heat: superconducting phase}
\label{appendix:Cp superconducting phase}

\subsection{Measurements in zero field}
\label{appendix:Cp super zero field}
\begin{figure}
    \centering
	\includegraphics[width=1\columnwidth]{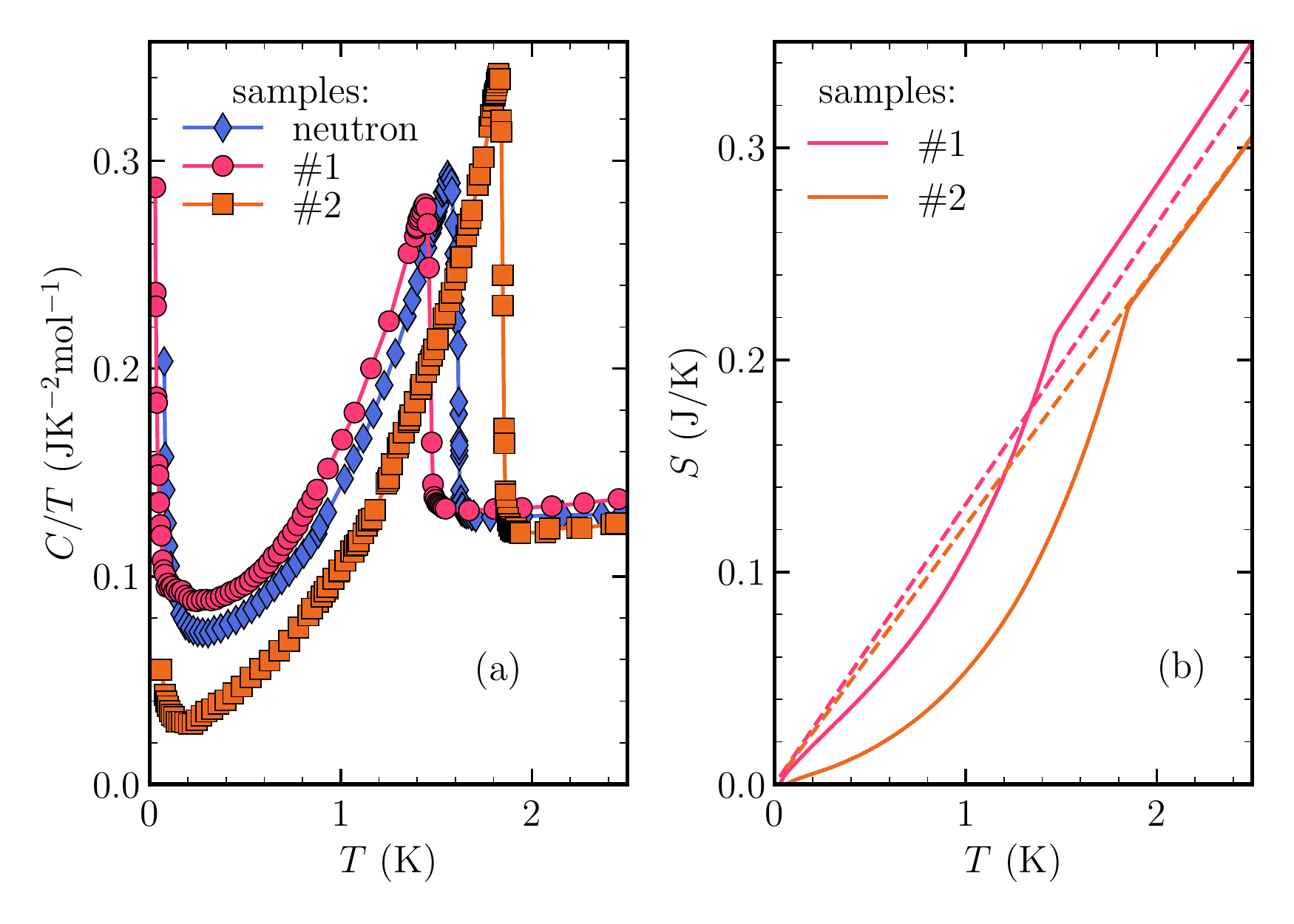}
	
 	\caption{(a) $C/T$ as a function of temperature at zero field and low temperatures for different samples. 
	Samples \diese1 and \diese2 are the one presented in this article. Sample neutron is a large sample of 241~mg used to perform the neutron diffusion experiments in ref~\cite{Knafo2021,RaymondJPSJ2021}.
	(b) Entropy calculated for samples \diese1 and \diese2, showing the bad balance for \diese1 and the very good one for \diese2.}
	\label{figS:CompCpSamplesBT}
\end{figure}

All measurements of \CT in \UTe display an upturn below 0.1K, and an extrapolated (from temperatures above the upturn) residual term at $T=0$ which was quite large in the first measurements \cite{RanScience2019,AokiRevJPSJ2019,Metz2019a}.
More recent studies are claiming that the residual term and the upturn are extrinsic to \UTe \cite{Cairns2020,AokiReview2022}.
Our measurements on different samples in Fig.~\ref{figS:CompCpSamplesBT}(a) show diverse behaviours at low temperatures.
The upturn is not monotonously correlated to \Tc, however, it is strongly reduced on our best samples.
The residual term seems to be more systematically decreasing with the \Tc increase, well in the trend reported in \cite{AokiReview2022}. 
In any case, these measurements do agree with an extrinsic nature of these anomalies. 
Note also that on sample \diese2, the entropy balance is perfectly satisfied at \Tc, within experimental errors (better than 1\%). 
The low temperature upturn plays a negligible role (see Fig.~\ref{figS:CompCpSamplesBT}(b)).

\subsection{Gaussian model for the specific heat anomaly}
\label{appendix:Gaussian fit}

A simple hypothesis is that broadening of the specific heat transition is controlled by a Gaussian $T_c$ distribution of the form: 
\begin{equation}
    p(T_c) = \frac{1}{\sigma \sqrt{2 \pi}} \exp \left( -\frac{1}{2}\left( \frac{T_c - T_{c0}}{\sigma} \right)^2\right)
\end{equation}

For the specific heat, or any additive quantity we can then write that:
\begin{equation}
	C/T = \int_{-\infty}^{\infty} p(T_c)  C/T(T,T_c)  dT_c
	\label{equS:CpDistrib}
\end{equation}

The simplest expression for $C/T(T,T_c)$ is a constant $\gamma$ term above \Tc, a jump at \Tc followed by a constant positive slope below \Tc. If both the slope and the jump are independent of \Tc, this amounts to:
\begin{equation}
	C/T(T,T_c)  = \gamma + \theta(T_c - T) \left( \deltaCT + \alpha (T-T_c) \right) 
	\label{equS:CpIdeal}
\end{equation}

Hence, for the total specific heat:
\begin{equation}
\begin{aligned}
	\dfrac{C}{T}(T)  &= \gamma + (\deltaCT + \alpha (T-T_{c0}))\left[\frac{1}{2}  - \frac{1}{2 } \text{erf} \left( \frac{T-T_{c0}}{\sigma} \right) \right] \\
	& -\alpha \frac{\sigma} {\sqrt{2 \pi}} \exp \left(-\frac{1}{2}\left( \frac{T - T_{c0}}{\sigma}  \right)^2 \right)  
	\label{equS:CpAnalytique}
\end{aligned}
\end{equation}

This is model is fine for the zero field transition, where $\deltaCT$ is independent of $T_c$. 
However, under magnetic field, the broadening of the transition may correspond to a distribution of slopes of \Hc (proportional to $T_c$ for clean type II superconductors). 
We can expect that $\deltaCT$ will be suppressed with field, with a decrease controlled by $H/H_{c2}(0)$. 
Hence $\deltaCT$ will not be constant within the broadened transition. 
More simply, we can assume that the jump will be suppressed like $T_c(H) / T_c(0)$. 
The problem is therefore to relate $T_c(H)$ and $T_c(0)$, or more precisely, to get the $T_c(0)$ corresponding to a given $T_c(H)$. 
Then we could take for a model of the transition that $\deltaCT$ is proportional, within the transition, to $T_c(H) / T_c(0)$. 
A simple way to find this relation is to assume a proportionality to the broadening so that : 
\begin{equation}
\begin{aligned}
	T_c(H) - T_{c0}(H) & = \frac{\sigma}{\sigma_0} \left( T_c(0) - T_{c0}(0) \right) \\
	\deltaCT(T_c) & = \deltaCT(T_{c0})\frac{T_c/T_{c0}}{1 + \frac{\sigma_0}{\sigma} \frac{ T_c - T_{c0} }{T_{c0}(0)}} 
\end{aligned}
\label{equS:dCT}
\end{equation}
In the last expression, we wrote $T_c = T_c(H)$ and $T_{c0} = T_{c0}(H)$.
As regards the slope, similarly, it should also depend on $T_c$. 
Indeed, in high fields for example, where the temperature dependence of $C/T$ is close to linear, the slope should depend both on $T_c$ and on $\deltaCT$. 
One way to keep some consistency within the transition is to assume that we have the same entropy balance for all the curves at different $T_c$ at a given field. 
At low field, where $C/T(T)$ has no specific reason to remain close to linear far below \Tc, there is no peculiar constraint on this entropy balance (the linear behaviour of $C/T$ below $T_c$ is valid only close enough to $T_c$). 
However, for fields closer to $H_{c2}(0)$, we can expect that this entropy balance should be more or less close to zero. 
Explicitly, we can enforce that : 
\begin{equation}
\begin{aligned}
	&\Delta S(T_c) = \int_0^{T_c} \left[ \deltaCT(T_c) + \alpha(T_c) \left( T - T_c \right) \right] dT = \beta T_c \quad \\
	&\text{with $\beta$ independent of $T_c$} \\
	&\alpha(T_c) = \frac{2}{T_c} \left[\deltaCT(T_c) - \beta \right]
\label{equS:alpha}
\end{aligned}
\end{equation}

Inserting equations (\ref{equS:dCT}) and (\ref{equS:alpha}) in the  equations (\ref{equS:CpIdeal}-\ref{equS:CpDistrib}), we obtain a final expression for $C/T(T)$, easily managed in its integral form by numerical calculations. 
It depends linearly on the parameters $\gamma$, $\deltaCT(T_{c0})$ and $\beta$ (close to zero in high fields), and non linearly on $\sigma$ and on $T_{c0}$. 
It has two additional inputs, taken from the zero field data: $\sigma_0$ and $T_{c0}(0)$.

\section{Measurements of \texorpdfstring{\Hc}{Hc2} for \texorpdfstring{\Hparaa}{H parallel to a}}
\label{appendix:Hc2 para a}

\begin{figure}
	\centering
	\includegraphics[scale=0.4]{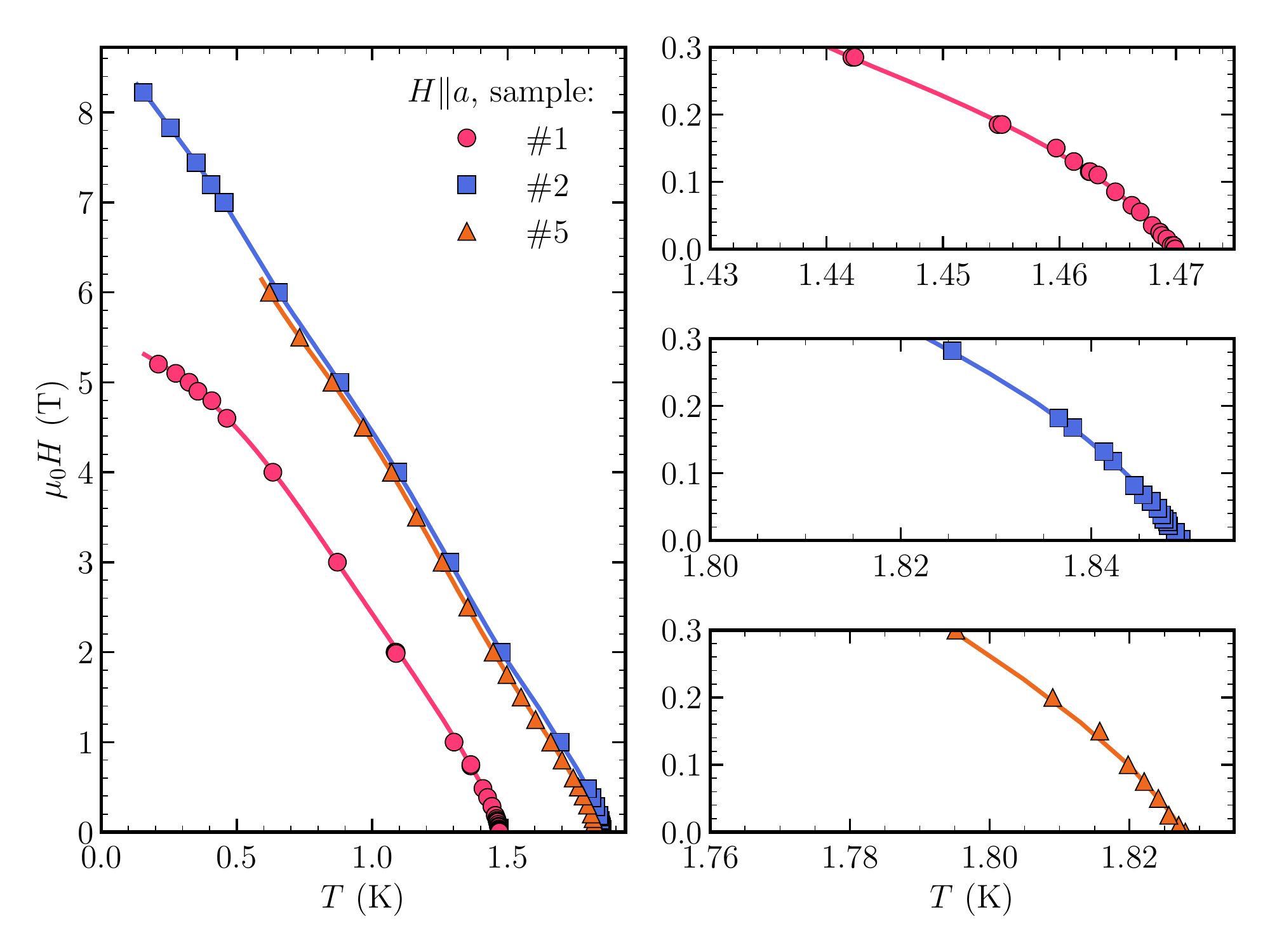}
	\caption{\Hc of 3 different samples for \Hparaa: the anomalous strong curvature near \Tc is reproducible, even for crystals with very different \Tc.}
	\label{figS:Curvature_H_a}
\end{figure}

Three different samples have been measured and their \Hc determined from the specific heat anomaly for \Hparaa including at very low fields.
The results are shown in Fig~\ref{figS:Curvature_H_a}.
The three samples come from different batches. Sample \diese1 and 2 have been measured with the same set up, and sample \diese5 with a different one.
They all exhibit a strong negative curvature near \Tc, proving that this feature is reproducible and intrinsic.

\added[id=OK]{To our knowledge, there are at least two other cases among heavy-fermion superconductors showing also an anomalous behaviour of \Hc at very low fields. 
The oldest one is UBe$_{13}$ \cite{RauchschwalbeZPhys1985}, however with a curvature which could be explained for example by a partial paramagnetic limitation \cite{ShimizuPRL2019}.
The other is the ferromagnetic superconductor URhGe, which even has a vertical \Hc along the easy axis, up to the field (of order 50~mT) where a single domain is induced in the sample \cite{HardyPRL2005}.
Such a mechanism is absent in \UTe which is paramagnetic and not ferromagnetic.}

\section{Measurements of \texorpdfstring{\Hc}{Hc2} for \texorpdfstring{\Hparab}{H parallel to b} - comparison with resistivity.}
\label{appendix:comparison Hc2 resistivity}

\begin{figure}
     \centering
	\includegraphics[width=1.05\columnwidth]{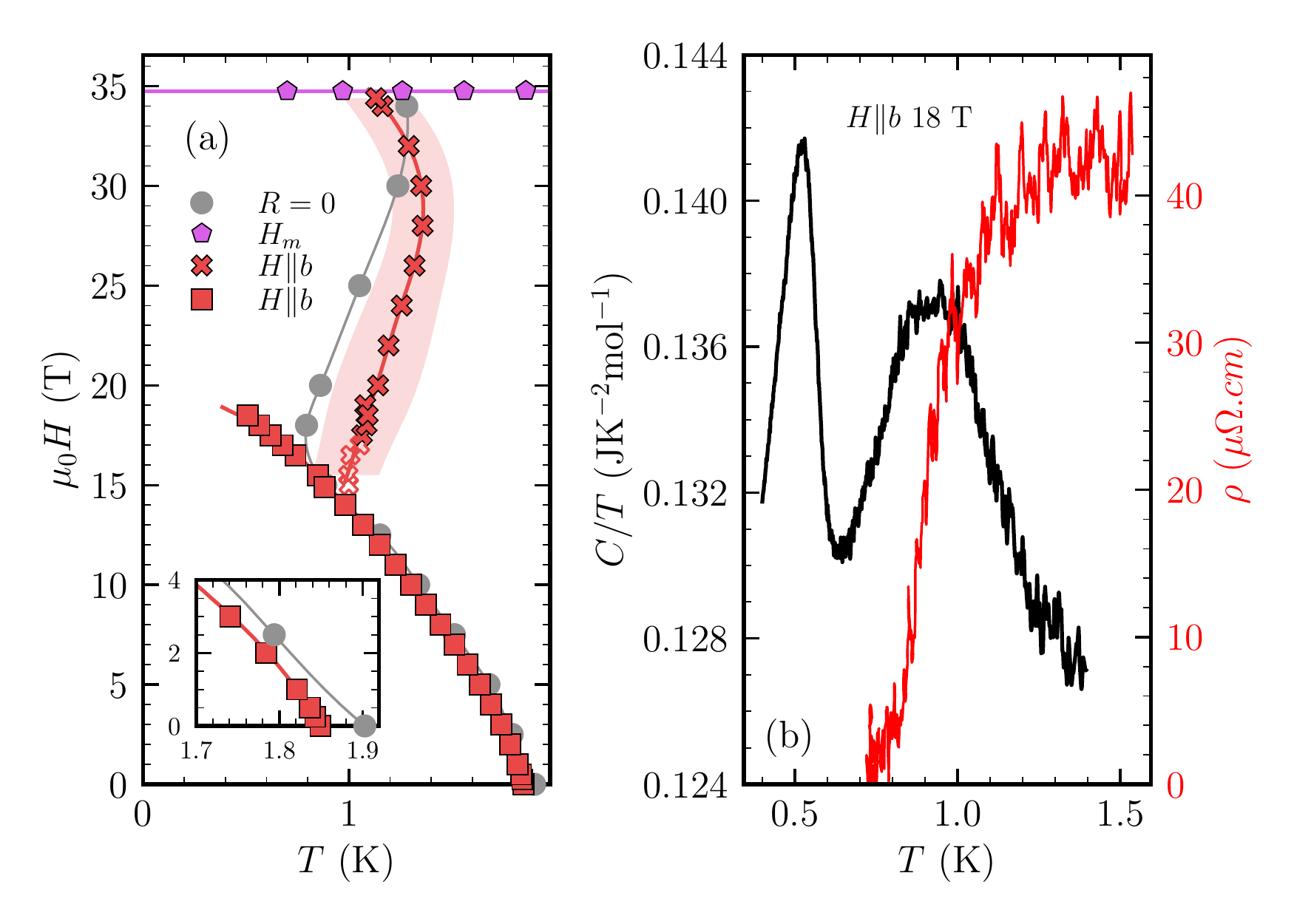}
	\caption{(a) \Hc for \Hparab determined by $C/T$ measurements (see main article). 
	Gray points correspond to the $R=0$ determined by resistivity measurements.
	The shaded region indicates the width of the \hftransi transition.
	The inset is a zoom for field below 4~T.
	(b) $C/T$ as function of temperature compared to $R$ as function of temperature, both measured at 18T.}
	\label{figS:PhaseDiagCpRho}
\end{figure}

Resistivity has also been measured on the sample from which we cut off sample \diese3. 
The critical field obtained with $R=0$ as criterion, is compared to \Hc determined by specific heat.
For the \lftransi phase, as expected, $R=0$ is above the specific heat transition. 
At low fields, there is a large difference between the initial slopes at \Tc for the determination from resistivity or specific heat anomaly, most likely due to the sensitivity of resistivity measurements to 
filamentary superconductivity, rapidly suppressed by (small) magnetic fields. 
For the \hftransi phase, $R=0$ is below the maximum of the specific heat transition, which is more unusual. 
This can be seen in Fig.~\ref{figS:PhaseDiagCpRho}(b), showing the temperature dependence of $C/T$ and of the resistivity at a fixed field of 18~T.

This discrepancy can arise from extrinsic inhomogeneities, like a continuous gradient of \Hm in the sample, or from more intrinsic phenomena like a weaker pinning of vortices in the \hftransi phase, which would induce a smaller critical current and possibly a shift of the resistive transition to lower temperatures.
This is well known in organic superconductors \cite{BelinJSuper1999} or in High-$T_c$ cuprates \cite{GrissonnancheNatCom2014}, where the resistivity remains non-zero in the vortex liquid state, favoured by the highly 2D anisotropy of their normal and superconducting properties. 
It has also been observed in iron-based superconductors \cite{KoshelevPRB2019}, and like in the organics or high-$T_c$ cuprates, with much stronger differences on the \Tc determination than observed in \UTe. 
The difficulty for such an explanation in \UTe is the same as faced for \ucoge \cite{Wu2018}:
the systems are 3D rather than 2D, hence superconducting fluctuations should be much less effective.
In addition, the discrepancy between resistivity and specific heat determination of \Hc occurs only at very high fields, whereas in the other systems, it arises very fast when entering the mixed state. 
It could be that the quantitative difference in the effect arises precisely because superconducting fluctuations are much less important in \UTe or \ucoge than in the quasi 2D systems.
However, it remains to be explained why this would happen only in the field-reinforced phase. 
This point remains a fully open question.

\section{Comparison of \texorpdfstring{\Hc}{Hc2} and \texorpdfstring{\Hcl}{Hc1}}
\label{appendix:comparison Hc2 Hc1}

In the Ginzburg-Landau regime near \Tc, well known relations exist between the lower, upper and thermodynamic critical fields. 
They are expressed through an anisotropic Ginzburg-Landau parameter $\kappa$:
\begin{equation}
\begin{aligned}
	H_{c1}&=\frac{H_c}{\sqrt{2} \kappa} \left(ln(\kappa)+0.49\right)\\
	H_{c2}&=\sqrt{2} \kappa H_c
\label{equS:GL}
\end{aligned}
\end{equation}

The thermodynamic critical field $H_c(T)$ is determined by double integration of the specific heat at 0~T, and we obtain a slope at \Tc of \added[id=OK]{$\frac{dH_c}{dT_c}$=-0.057~T/K for sample of Ref. \onlinecite{Paulsen2021} ($\frac{dH_c}{dT_c}$=-0.0685~T/K for sample \diese2).}
\Hcl has been measured on a crystal of the same batch as \diese1 in all field directions \cite{Paulsen2021}. 
Rescaling the values from Ref.\onlinecite{Paulsen2021} by the ratio of their respective \Tc, we can determine $\kappa$ in the three directions from the first equation (\ref{equS:GL}), and extract from the second a prediction for the value of \dHc of sample \diese2.
These values are reported in table~\ref{tabS:ParameterTable}.
The large value of \dHc at \Tc for \Hparaa is in very good agreement with the values of \dHcl as predicted by the Ginzburg-Landau relations. 
This is also true for \Hparac, but not for \Hparab.

\begin{table}
\caption{Table with the values of the slope \dHcl of \Hcl at \Tc from \cite{Paulsen2021}, the corresponding value of the calculated Ginzburg-Landau parameter $\kappa$, and the predicted value for \dHc for the sample of Ref.~\cite{Paulsen2021} (same batch as \diese1). For sample \diese2, \dHc is rescaled by the ratio of the \Tc of these samples. Last column is the initial slope measured on sample \diese2.}
    \label{tabS:ParameterTable}
    \begin{ruledtabular}
    \begin{tabular}{lccccr}
 & \Centerstack{-$\frac{H_{c1}}{dT_c}$($\nicefrac{\text{T}}{\text{K}}$)\\ ref~\cite{Paulsen2021} } & \Centerstack{$\kappa$\\   } & \Centerstack{-$\frac{H_{c2}}{dT_c}$($\nicefrac{\text{T}}{\text{K}}$) \\ ref~\cite{Paulsen2021}  }&  \Centerstack{-$\frac{H_{c2}}{dT_c}$($\nicefrac{\text{T}}{\text{K}}$) \\ rescaled} &  \Centerstack{-$\frac{H_{c2}}{dT_c}$($\nicefrac{\text{T}}{\text{K}}$) \\ \diese2 measured}\\
\colrule
\Hparaa & 0.00113 & 202.683 & 16.052 & 20.480 & 20\\
\Hparab & 0.00227 & 86.482 & 6.849 & 8.738 & 34.5\\
\Hparac & 0.00252 & 75.838 & 6.006 & 7.663 & 7.5 \\
    \end{tabular}
    \end{ruledtabular}
\end{table}

\begin{figure}
	\centering
	\includegraphics[width=1\linewidth]{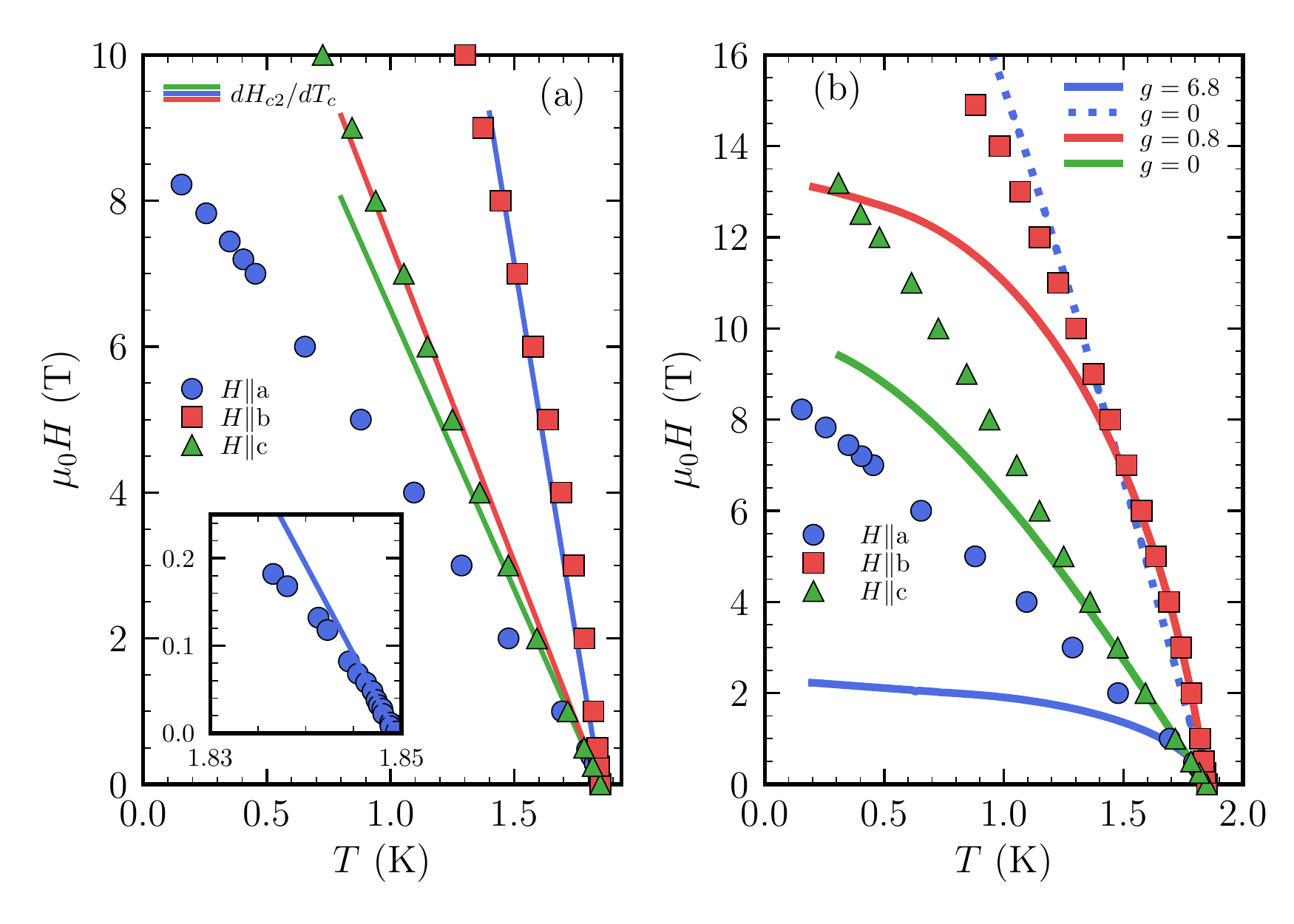}
	\caption{(a) \Hc measurements on sample \diese2: lines show \dHc calculated and rescaled from \dHcl determined in ref~\cite{Paulsen2021}. The insert is a zoom for fields \Hparaa below 0.25~T. There is a very good agreement between \Hcl and \Hc for \Hparaa and \Hparac, but a strong discrepancy for \Hparab.\\
	(b) \Hc determined by specific heat on sample \diese2. 
	The lines are the best adjustment of \Hc (orbital and paramagnetic limitations) to match the measured initial slopes and curvatures in each direction. 
	The dashed-dotted line corresponds to a pure orbital limitation of \Hc adjusted on its initial slope  for \Hparaa, evidencing the very strong negative curvature close to \Tc. $g$ is the gyromagnetic factor and $\lambda$ is set to 1.
    }
	\label{figS:CompHc2_Hc1}
\end{figure}

Figure~\ref{figS:CompHc2_Hc1}(a) shows \Hc determined by specific heat on sample \diese2 and \dHc at \Tc calculated from \Hcl. 
The insert of Fig.~\ref{figS:CompHc2_Hc1}(a) shows how the predicted \dHc matches the present measurement for \Hparaa very close to \Tc. 
The strong disagreement for \Hparab as well as the difference between the slope of \Hc at \Tc and the linear regime at lower temperatures for \Hparaa are clearly visible on the main Fig.~\ref{figS:CompHc2_Hc1}(a).
By contrast, the linear regime for \Hparac does match the Ginzburg-Landau prediction in a large temperature range.

\section{Strong coupling model for \texorpdfstring{\Hc}{Hc2}}
\label{appendix:strong coupling model} 
The model used for the calculation of \Hc in the strong coupling regime is fully described in Ref.\onlinecite{ThomasJLTP1996}. 
It extends that of Ref.\onlinecite{Bulaevskii1988} to include the paramagnetic limitation mechanism.
For completeness, we present here the basic equations.
This model is derived from the Eliashberg theory for electron phonon interaction in s-wave superconductors. 
We believe that it remains relevant for the estimation of strong-coupling effects on the upper critical field in unconventional superconductors (anyhow, we do not know of such calculation for p-wave superconductors).

A most simplified form of the Eliashberg interaction is used, sufficient to capture the most important properties of the strong coupling regime: 
the renormalization of the Fermi velocity and the pair-breaking effects arising from the presence of thermal phonons (or magnetic excitations) close to \Tc when the strong coupling constant $\lambda$ gets large.
The spectral density of interaction is taken as a simple delta function (Einstein spectrum):

\begin{equation}
\begin{aligned}
	\alpha^2F(\omega)&=\left(\frac{\lambda \Omega}{2}\right) \delta\left( \omega - \Omega\right)\\
\label{equS:densityInteraction}
\end{aligned}
\end{equation}
Where $\omega$ is the frequency, $\Omega$ the characteristic energy of the interactions (of order the Debye temperature for electron-phonon interaction), and $\lambda$ is the dimensionless strong-coupling constant.
\Hc is then determined by a system of linear equations for the gap: 

\begin{equation}
\begin{aligned}
	\Delta(i\tilde{\omega}_n)&=\left(\frac{\pi T}{\Omega}\right) \sum_{\vert \omega_m \vert<\omega_c} (\lambda(\omega_n-\omega_m) - \mu^\ast)\chi(\tilde{\omega}_m)\Delta(i\tilde{\omega}_m)\\
\label{equS:gap}
\end{aligned}
\end{equation}
Where $\omega_n=\pi T (2n+1)$ are Matsubara frequencies, $\mu^\ast$ is the screened Coulomb pseudo potential, $\omega_c$ is a frequency cut-off (8 to 10 times $\Omega$) and

\begin{equation}
\begin{aligned}
	\tilde{\omega}_n &= \omega_n + \pi T \sum_m \lambda(\omega_n - \omega_m) sgn(\omega_n) \\
	\lambda(\omega_n - \omega_m) &= \frac{\lambda \Omega^2}{\Omega^2 + (\omega_n - \omega_m)^2}\\
\label{equS:definitions}
\end{aligned}
\end{equation}
The function $\chi(\tilde{\omega}_n)$ in (\ref{equS:gap}) contains the effects of the field (B) on the gap equations through the orbital and paramagnetic effects:

\begin{equation}
\begin{aligned}
	\chi(\tilde{\omega}_n) &= \int_0^{\infty} dx \frac{\beta exp(-\beta x)}{\sqrt{\tilde{Q}^2 + x}} tan^{-1}\left( \frac{\sqrt{\tilde{Q}^2 + x}}{\frac{\vert \tilde{\omega}_n \vert + i g \mu_B B/2 sgn(\tilde{\omega}_n)}{\Omega}}\right)
\label{equS:chi}
\end{aligned}
\end{equation}
Here $\beta = \frac{2 \Omega^2}{\hbar e B \left({\bar{v}_F^{bare}}\right)^2}$ parameterizes the orbital effect: $\bar{v}_F^{bare}$ is a bare average Fermi velocity (meaning a Fermi velocity not renormalized by the pairing interaction), perpendicular to the applied magnetic field and e is the elementary charge. 
The paramagnetic limit is parametrized by the gyromagnetic factor $g$ in the direction of the applied field. 
Moreover, $\tilde{Q} = \frac{\hbar \bar{v}_F^{bare} Q}{2 \Omega}$ is the dimensionless amplitude of the (potential) Fulde-Ferrel, Larkin-Ovchinnikov wave vector , which has to be taken into account for dominant paramagnetic limit.
Hence for non vanishing $g$, the system of equations (\ref{equS:gap}) has to be solved (with the usual techniques of linear algebra) optimizing the solution with respect to $Q$ \added[id=OK]{for maximum \Hc: for dominant paramagnetic limitation, a finite $Q$ marking the entrance in the FFLO state can be found for temperatures below 0.55\Tc.}

\section{Field-dependence of the pairing strength modelled by a strong coupling parameter \texorpdfstring{\llambda}{lambda}}
\label{appendix:fitParameters} 

Figure~\ref{figS:CompHc2_Hc1}(b) shows \Hc along the 3 crystallographic directions, calculated with the strong coupling model for the upper critical field already used in ref.~\cite{Wu2017,Knebel2019} and summarized in Appendix \ref{appendix:strong coupling model}, at fixed pairing strength. 
The measured initial slopes \dHc at \Tc are controled by the orbital limit hence by \vF. 
The strong coupling constant $\lambda$ is set to 1, which seems a reasonable value for \UTe. 
The plain lines in Fig.~\ref{figS:CompHc2_Hc1}(b) are \HcT calculated with the orbital limit adjusted to match the measured \dHc, and the gyromagnetic factor $g$ adjusted to match the initial negative curvature along the $a$ and $b$ axes ($g=6.5$ along the \aaxis and $g=0.8$ along the \baxis). 
Deviations of the measured \Hc to such a usual combined orbital and paramagnetic limitation are observed for all applied directions of the magnetic field.

To explain these deviations, a field dependent pairing strength is assumed.
It is extracted from the data through a calculation of \HcT at fixed values of \llambda (see Fig.~\ref{fig:FitsHc2LambdaFixed} of the main text for the case of \Hparab). The typical energy controlling \Tc ($\Omega$), the Coulomb repulsion parameter $\mu^{\ast}$, and the bare average Fermi velocity for field along the $i$ axis ($\bar{v}_{F0}^{bare,i}$) controlling the orbital limit are taken independent of \llambda.
The effective Fermi velocity controlling the orbital limit and so \dHc (at fixed \llambda)  is renormalised as $\bar{v}_{F}^i = \frac{\bar{v}_{F0}^{bare,i}}{1+\lambda}$. 
If \llambda is field dependent, this effective Fermi velocity is also field dependent.
In the table~\ref{tabS:ParameterFitHc2}, we also calculate the Fermi velocity along each $i$-axis: $v_{F}^i$, deduced from the effective Fermi velocities through $v_{F}^i = \frac{\bar{v}_{F}^j\bar{v}_{F}^k}{\bar{v}_{F}^i}$, where $j$, $k$ are the axis perpendicular to $i$.

\begin{table}
     \caption{Parameter values of the fit. 
    We used a strong-coupling parameter $\lambda(H=0)=1$, with a typical energy (equivalent to the Debye energy) $\Omega=28.4$~K, $\mu^\ast=0.1$, pair-breaking impurity scattering rate $\Gamma=1.39$~K.
    Values of $\bar{v}_{F}$ used in the fit are reported for each field direction. Difference between $\bar{v}_{F}^i$ and $v_{F}^i$ is explained in the text. The corresponding coherence length are calculated from and $\xi_0 = 0.18 \frac{\hbar v_F}{k_B T_c}$ }
    \label{tabS:ParameterFitHc2}
    \begin{ruledtabular}
\begin{tabular}{lccccr}
 & \Centerstack{$\bar{v}_{F}^i(H=0)$\\(m/sec)} & \Centerstack{$g$\\ }  & \Centerstack{$\bar{\xi}_0^i$ ($\mathring{A}$)\\ } &  \Centerstack{$v_{F}^i(H=0)$ \\along $i$ axis} &  \Centerstack{$\xi_0^i$ ($\mathring{A}$) \\ along $i$ axis} \\
 \colrule
\Hparaa             & 5400 & 0 & 40 & 14400 & 106\\
\Hparab(\lftransi)  & 8600 & 0 & 64 & 5680 & 42\\
\Hparac             & 9044 & 0 & 67 & 5130 & 38\\
\Hparab(\hftransi)  & 8600 & 2 & 64 & 5680 & 42\\
    \end{tabular}
   \end{ruledtabular}
   
\end{table}

    It is to be noted that the lowest Fermi velocities (\vF) are along the $b$ and $c$ axes and highest along \aaxis.
    This matches qualitatively the anisotropy found in transport measurements between the different axes \cite{EoPRB2022}.
    Quantitatively, \Hc depends on an average of the Fermi velocities perpendicular to the applied field direction, weighted by the pairing strength.
     As a consequence for example, we never succeeded to compare quantitatively anisotropies of \Hc in UPt$_3$ or in URu$_2$Si$_2$ with the detailed determination of their respective Fermi surfaces by quantum oscillations \cite{McMullan_2008, Bastien2019}, even though the order of magnitude of the orbital limitation is consistent with measured effective Fermi velocities.
    This is probably even more acute if subtle $\bm{Q}$-dependent paring is responsible the specific pairing state realised in the system, as could well be the case in \UTe \cite{Xu2019,Ishizuka2021,KreiselPRB2022,ChenArXiv2021}.

For \Hparaa, as discussed in the main text, we have chosen the most natural hypothesis of an ESP (equal-spin-pairing) state along the \aaxis, hence no paramagnetic limitation of \Hc, an initial slope matching the measured one (agreement with \Hcl) implying a suppression of the pairing strength under field.
However, it is also possible to construct a model where the pairing strength would increase along the \aaxis. 
Indeed, maintaining an initial slope matching the measured one, if we suppose that the g-factor is in reality $\ge 6.8$, the same fitting procedure will lead to a field increase of $\lambda(H)$.

We rejected this scenario due to the very large and rather unrealistic value required for the g-factor, and the strong contradiction with NMR Knight-shift measurements observing no change at all for \Hparaa \cite{FujibayashiJPSJ2022}.
It should be stressed here that NMR measurements are performed at fixed field.
They yield the change of electronic spin susceptibility across \Tc from the temperature variation of the Knight-shift.
Hence, this measurement is not directly influenced by the field dependence of the pairing strength, as opposed to considerations on the violation of the paramagnetic limit on \Hc.

\section{Broadening of the specific heat transition by a distribution of \texorpdfstring{\Hm}{Hm}}
\label{appendix:Cp broadening}
As explained in Appendix \ref{appendix:SuppCpHm}, we could extract a standard deviation $\sigma = 0.19 $~T for the distribution of \Hm, hence a relative standard deviation $\frac{\sigma}{H_m} \sim 0.55 \%$

As explained in the main text (Fig.~\ref{fig:FitsHc2LambdaFixed}), from the calculation of \Hc at fixed values of the pairing strength \llambda, we can extract also the superconducting critical temperature under field as:
\begin{equation}
    \begin{aligned}
        & T_c  = \varphi\left( H, \Tilde{\lambda} \left( \frac{H}{H_m} \right) \right) \\
        & \Tilde{\lambda} \left( \frac{H}{H_m} \right)  = \lambda \left( H \frac{H_{m0}}{H_m} \right)
    \end{aligned}
    \label{equS:Tc_H_lambda}
\end{equation}
Where $H_{m0}$ is the centre of the distribution of metamagnetic fields \Hm, determined from the specific heat measurements of the metamagnetic transition. 
\llambdaH is the field dependent pairing strength deduced from the different models for \Hc and drawn in Fig.~\ref{fig:Lambda_H} of the main text.

From this relation, we can calculate the effect of a Gaussian distribution of \Hm on the specific heat anomaly at constant field of the superconducting transition, using for \CT (instead of Equ.~\ref{equS:CpDistrib}):
\begin{equation}
	C/T = \int p(H_m)  C/T\left(T,T_c\left(H,H_m\right)\right)  dH_m
	\label{equS:Cp_distrib_Hm}
\end{equation}
This is the way we could draw the broadening of the specific heat anomaly in Fig.~\ref{fig:FitCpHF} and Fig.~\ref{fig:FitCp_theta} of the main text, using the two different determinations of \llambdaHHm (with or without paramagnetic limitation of \Hc).

However, even without a full determination of the shape of the anomaly, requiring a numerical integration of Equ.~\ref{equS:Cp_distrib_Hm}, we can understand why the broadening is larger when there is a paramagnetic limitation of \Hc.
From Equ.~\ref{equS:Tc_H_lambda}, we can derive the derivative of \Tc with respect to \Hm at fixed H and for $H_m=H_{m0}$. 
It measures the sensitivity of \Tc to \Hm, hence the broadening of the \CT anomaly due to a distribution of \Hm:
\begin{equation}
    \left.\frac{\partial T_c}{\partial H_m}\right\vert_H =  \left.\frac{\partial T_c}{\partial \lambda}\right\vert_H \left(-\frac{H}{H_{m0}}\right) \left(\frac{d \lambda}{d H} \right) 
    \label{equS:dTc_Hm}
\end{equation}
When comparing both models, it is clear that one has a stronger field dependence of \llambda than the other, but this could be compensated by a different $\left.\frac{\partial T_c}{\partial \lambda}\right\vert_H$ which has to be computed at finite field (on the \HcT line).
Indeed, both models share the same \HcT. 
We can compute its temperature derivative \added[id=OK]{(at $H_m=H_{m0}$)} from Equ.~\ref{equS:Tc_H_lambda}:
\begin{equation}
    \begin{aligned}
        & dT  =  \left.\frac{\partial T_c}{\partial H}\right\vert_{\lambda} dH_{c2} +  
        \left.\frac{\partial T_c}{\partial \lambda}\right\vert_{H} \left(\frac{H_{m0}}{H_{m}}\right) \left(\frac{d \lambda}{d H} \right)dH_{c2} \\
        & \frac{dT}{d H_{c2}} - \left.\frac{\partial T_c}{\partial H}\right\vert_{\lambda}  =  -  \left(\frac{H_{m0}}{H}\right) \left.\frac{\partial T_c}{\partial H_m}\right\vert_H 
    \end{aligned}
    \label{equS:dTc_Hc2}
\end{equation}
The last equation shows that the difference between models for $\left.\frac{\partial T_c}{\partial H_m}\right\vert_H $ arises not directly from $\left(\frac{d \lambda}{d H} \right)$, but rather from  $\left.\frac{\partial T_c}{\partial H}\right\vert_{\lambda}$:
this term is much larger when \HcT at fixed \llambda becomes "flat" due to the paramagnetic limitation arising for singlet pairing, than for a pure orbital limit in case of a spin-triplet ESP state (see Fig.~\ref{fig:FitsHc2LambdaFixed} in the Main text).

\section{Angular dependence of the specific heat in the \texorpdfstring{($b$,$c$)}{(b,c)} plane}
\label{appendix:angular dependence}

\begin{figure}
     \centering
	\includegraphics[width=1\columnwidth]{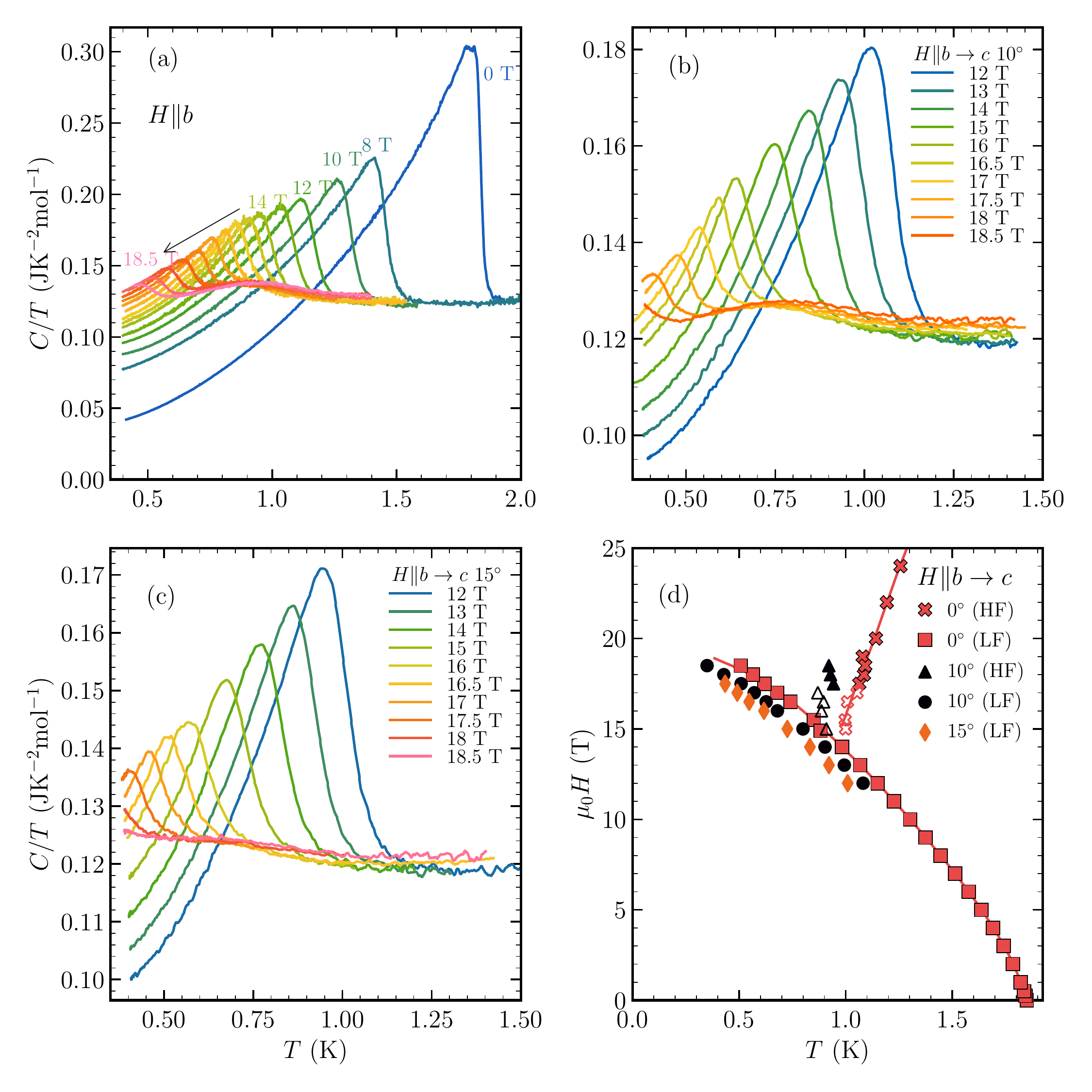}
\caption{$C/T$ as a function of temperature for several fields between 0~T and 18.5~T, measured on sample \diese3. 
(a) $C/T$ for \Hparab, 
(b) for an angle of 10$^\circ$ toward the \caxis and 
(c) for an angle of 15$^\circ$ toward the \caxis.
}
\label{figS:Cp_vs_H_3_angles}
\end{figure}

AC specific heat measurements have been performed on sample \diese2 up to 18.5 T for several angles in the ($b$,$c$) plane. 
Figure~\ref{figS:Cp_vs_H_3_angles}(a-b-c) show temperature sweeps for different fields for angles of 0$^\circ$, 10$^\circ$ and 15$^\circ$ from $b$ toward the $c$ axis. 
As the field is rotated toward the $c$ axis, the sharp transition of the \lftransi phase is shifted toward lower temperatures. 
The same behaviour is observed for the wide transition of the \hftransi phase. 
The corresponding critical temperature for the two transitions at the different angles are reported on the phase diagram of Fig.~\ref{figS:Cp_vs_H_3_angles}(d), using the same Gaussian analysis as in the main paper. 
For \Hc at 15$^\circ$, it was impossible to extract a reliable value of \Tc for the \hftransi transition.

\bibliography{RosuelMain2022}

\begin{thebibliography}{89}
\bibitem [{Not()}]{NoteResHc2}%
  \BibitemOpen
  \href@noop {} {}\bibinfo {note} {Generally speaking, in electrical transport
  measurements, filamentary superconductivity easily leads to slightly higher
  \Tc values. However this filamentary superconductivity is fragile under small
  fields, hence it is usual to observe a small positive curvature of \Hc
  determined by resistivity. In \UTe, it may mask the negative curvature of \Hc
  clearly visible in Fig.~\ref{fig:PhaseDiagCp} close to \Tc.}\BibitemShut
  {Stop}%
\end{thebibliography}

\end{document}